\RequirePackage{ifpdf}
\documentclass[11pt,letterpaper]{JHEP3}
\usepackage{epsfig}
\usepackage[latin1]{inputenc}
\usepackage{bbold,amsfonts}
\usepackage{graphicx}
\usepackage{amssymb,amsmath}
\usepackage{fancybox,framed}
\usepackage{dsfont}
\usepackage{mathtools}
\usepackage{braket}
\usepackage{slashed}
\usepackage{rotating}
\allowdisplaybreaks

\author{Marco S. Bianchi\\\\
Centre for Research in String Theory,
School of Physics and Astronomy\\
Queen Mary University of London,
Mile End Road, London E1 4NS, UK \\
\qquad\\
E-mail: \email{m.s.bianchi@qmul.ac.uk}
}

\abstract{We consider BPS Wilson loops in planar ABJM theory, wound multiple times around the great circle. We compute the expectation value of the 1/6-BPS and 1/2-BPS Wilson loops to three- and two-loop order in perturbation theory, respectively, dealing with the combinatorics of multiple winding via recursive relations.
For the 1/6-BPS Wilson loop we perform the computation at generic framing and at framing 1 we find agreement with the localization result.
For the 1/2-BPS Wilson loop we compute the expectation value at trivial framing and by comparison with the matrix model expression we extract the framing dependence of the fermion diagrams.
}

\preprint{May 2016 \\QMUL-PH-16-08
}

\title{A note on multiply wound BPS Wilson loops in ABJM
}


\csname @addtoreset\endcsname{equation}{section}

\def\bseq{\begin{subequation}}  
\def\eseq{\end{subequation}}
\def\bsea{\begin{subeqnarray}}  
\def\esea{\end{subeqnarray}}

\newcommand{\beq}{\begin{equation}}
\newcommand{\bea}{\begin{eqnarray}}
\newcommand{\eea}{\end{eqnarray}}
\newcommand{\eeq}{\end{equation}}

\newcommand {\non}{\nonumber}

\renewcommand{\a}{\alpha}
\renewcommand{\b}{\beta}

\renewcommand{\d}{\delta}
\newcommand{\pa}{\partial}
\newcommand{\g}{\gamma}
\newcommand{\G}{\Gamma}

\newcommand{\e}{\epsilon}

\newcommand{\s}{\sigma}

\def\beq{\begin{equation}}
\def\eeq{\end{equation}}
\def\bea{\begin{eqnarray}}
\def\eea{\end{eqnarray}}
\def\Tr{\mathrm{Tr}}

\def\a{\alpha}
\def\b{\beta}
\def\g{\gamma}

\def\d{\delta}
\def\e{\epsilon}

\def\G{\Gamma}

\declareslashed{}{\backslash}{0}{0}{\Omega}
\declareslashed{}{\backslash}{0}{0}{O}
\begin{document}

\allowdisplaybreaks

\section{Introduction}

During the last few years exact results in supersymmetric gauge theories have been providing interpolating functions from weak to strong coupling.
In particular, Wilson loops are basic observables in gauge theories, which, if constructed in such a way to preserve supersymmetry, can be computed exactly thanks to supersymmetric localization \cite{Pestun:2007rz}. This program has been applied extensively in four dimensions, and to a great degree in the context of superconformal ${\cal N}=4$ SYM theory, providing non-trivial tests of the AdS/CFT correspondence \cite{Maldacena:1997re,Wittenads,Gubser:1998bc}.\\ ABJM theory in three dimension is also a supersymmetric theory enjoying superconformal invariance and possessing an AdS/CFT dual \cite{ABJM,ABJ}. 
This is a gauge theory with Chern-Simons action for the gauge group $U(N_1)\times U(N_2)$ and matter transforming in the bifundamental representation thereof, preserving ${\cal N}=6$ supersymmetry.
Supersymmetric Wilson loop operators in ABJM have been defined in \cite{Berenstein:2008dc,DPY,Chen:2008bp,Rey:2008bh,Drukker:2009hy,Cardinali:2012ru} and it was shown that it is possible to compute the expectation value for circular ones as matrix model averages thanks to localization on $S^3$ \cite{Kapustin:2009kz,Drukker:2010nc}.

In this paper we study perturbatively the expectation value of such Wilson loops. 
In particular, we first consider the 1/6-BPS operator. This is the holonomy of the gauge connection of one of the two ABJM gauge groups, corrected by an adjoint bilinear constructed with the scalars of the theory, in such a way that the operator preserves locally two supersymmetries \cite{DPY,Chen:2008bp,Rey:2008bh}.
In order to preserve half of the supersymmetry, a superconnection for the two gauge groups has to be considered, which features a coupling to fermions too \cite{Drukker:2009hy}.

When the contour the Wilson loop operators is evaluated on is a circle, then supersymmetry can be preserved globally, the BPS Wilson loops expectation value is finite and non-trivial, and is amenable of an exact computation via localization.
This technique reduces the path integral of the ABJM field theory to a matrix model average.
The latter was derived in \cite{Kapustin:2009kz}
and studied using in particular insights from the topological string \cite{Marino:2009jd,Drukker:2010nc} and the Fermi gas approach \cite{Marino:2011eh}.
In particular, the matrix model average for the expectation value of the 1/6-BPS operators has been expanded perturbatively at weak coupling and the result coincides with the former two-loop evaluation of \cite{DPY,Chen:2008bp,Rey:2008bh}, which provides a strong test of the localization procedure.
Recently the three-loop term in the 't Hooft coupling expansion has been also successfully tested against a field-theoretical perturbative computation \cite{Bianchi:2016yzj}.
The matrix model of ABJM theory can be employed to compute the expectation value of the 1/2-BPS as well \cite{Drukker:2010nc}.
This allowed to derive a two-loop prediction (and higher order of course) for the 1/2-BPS Wilson loop at weak coupling, which later received a perturbative confirmation in \cite{Bianchi:2013zda,Bianchi:2013rma,Griguolo:2013sma}.
\bigskip

In this paper we study the expectation value of the ABJM 1/6-BPS and 1/2-BPS Wilson loops, in the generalization in which they wind $m$-times around the great circle of $S^3$.
The reason for considering such an object is two-fold.
On the one hand, from the point of view of localization this is a natural extension to be considered and its expectation value has been computed exactly in \cite{Klemm:2012ii}, via the Fermi gas approach.
This has never been given the support of a perturbative test and we want to provide such a backup in this paper.
On the other hand the multiply wound Wilson loops are relevant for computing other observables in ABJM theory, such as the entanglement entropy and the Bremsstrahlung function \cite{Lewkowycz:2013laa,Bianchi:2014laa}.

As a slightly more technical aside, we also use the multiple winding to investigate perturbatively the dependence of the 1/6-BPS and 1/2-BPS Wilson loops on the particular framing used in the computation.
It is known that the expectation value of Wilson loops in Chern-Simons theory suffers from ambiguities associated to the definition of the connections at coincident points.
In particular this ambiguity appears as a one-loop effect from a gluon exchange breaking topological invariance anomalously. The concept of framing provides a method to define the Wilson loop in a topologically invariant manner, at the price of introducing additional information on the contour. This consists in the choice of a nearby normal vector to the original path on which the second endpoint of the gauge propagator is allowed to run.
Pictorially this corresponds to thickening the path into a band of infinitesimal width.
Then the one-loop effect measures the linking number of the framing contour with respect to the original one, namely an integer counting how many times the former winds around the latter.
For pure Chern-Simons theory the net effect of framing is completely captured by a simple phase factor, the exponential of the one-loop contribution, multiplying the Wilson loop expectation value. This was argued non-perturbatively \cite{Witten} and then its emergence in perturbation theory was clarified up to three loops \cite{Guadagnini:1989am,Alvarez:1991sx}.

It is less known which the precise framing dependence in the case of Chern-Simons theories with matter is. The results from localization provide results at framing 1, whereas perturbative computations have been performed at trivial framing where computations simplify.
The comparison between the two entails identifying and removing the framing dependence from the localization result.
Conversely, in lack of solid arguments to determine the framing dependence, the comparison between the two may shed some light on the behaviour of diagrams at non-trivial framing, which is in principle amenable of a direct evaluation in perturbation theory.
Work in this direction has been recently performed for the 1/6-BPS Wilson loop. This indicates that a phase factor with a nontrivial function of the coupling constant may encapsulate the framing dependence of the Wilson loop \cite{Bianchi:2016yzj}.
For the 1/2-BPS Wilson loop the framing phase coincides with that of pure Chern-Simons with supergroup $U(N_1|N_2)$, but it has not been established yet how this arises perturbatively beyond one loop, and in particular which the role of fermionic diagrams in this is.

For multiply wound Wilson loops we expect the framing dependence to be more complicated. Indeed the expectation value of these is usually calculated as a combination of Wilson loops in different "hook" representations of the gauge group, each of which comes with a different framing phase \cite{Labastida:2000yw,Marino:2001re,Brini:2011wi,Hatsuda:2013yua}. Therefore their framing dependence is not in general a simple overall phase.
We study the effect of framing for the 1/6-BPS Wilson loop first, explicitly computing its expectation value perturbatively to three loops at generic winding and framing numbers.
We then compute the 1/2-BPS Wilson loop at multiple winding and trivial framing and argue from the comparison with the localization result, what the framing dependence of the individual fermionic diagrams looks like. 

\bigskip
The plan of the paper is as follows. In section \ref{sec:contours} we derive an algorithm to solve the combinatorial problem arising due to multiple winding.
In the section \ref{sec:winding1} we review the dynamics of the singly wound 1/6-BPS and 1/2-BPS circular Wilson loops in ABJM. We describe the computation of their expectation value up to three and two loops in perturbation theory and via localization emphasizing the role of framing. 
In sections \ref{sec:2loops} and \ref{sec:3loops} we apply the combinatorics and the dynamics of the Wilson loop to derive from perturbation theory the expectation value of the multiply wound Wilson loop up to three-loop order.
This computation shows agreement with the localization computation and extends it to generic framing number $f$.
Focussing on the pure Chern-Simons contribution we also find agreement with a formula for the expectation value of the Wilson loop at generic framing $f$ and winding $m$ \cite{Brini:2011wi}.
In section \ref{sec:halfbps} we compute the multiply wound 1/2-BPS Wilson loop at framing 0 and compare it with the localization result, thus inferring the putative framing dependence of the two-loop fermionic diagrams.

\section{General recursion relations}\label{sec:contours}

In this section we analyse in generality the problem of computing a multiply wound Wilson loops in perturbation theory. The following applies to any Wilson loop operator in any theory. 
We define the multiply wound Wilson loop as
\begin{equation}
W_m \equiv {\rm Tr}\, {\cal P}\, \exp\left(g \int_0^{2\pi m} d\tau\, {\cal L}(\tau) \right)
\end{equation}
for some coupling constant $g$ which we assume small in the following, so as to make perturbation theory possible. The Wilson loop is evaluated along a closed contour ${\cal C} = {\cal C}(\tau)$, with a certain parametrization in terms of the parameter $\tau \in [0,2\pi]$.
We shall focus on circular space-like contours, which we can parametrize as
\begin{equation}
{\cal C} \equiv \left( 0, \cos\tau, \sin\tau  \right)
\end{equation}
Expanding the exponential perturbatively and according to the path ordering, the computation of the expectation value of the Wilson loop involves contour integrals of the generic form
\begin{equation}
G_n(m) \equiv \int_0^{2\pi m} d\tau_n \int_0^{\tau_n} d\tau_{n-1} \dots \int_{0}^{\tau_2} d\tau_1\, F\left(\{\tau_i\}\right)
\end{equation}
for an integrand $F\left(\{\tau_i\}\right)$ which is the result of the relevant Wick contractions.
Such an integration domain can be decomposed as follows
\begin{align}
&\int_0^{2\pi m} d\tau_n \int_0^{\tau_n} d\tau_{n-1} \dots \int_{0}^{\tau_2} d\tau_1\, F\left(\{\tau_i\}\right) = 
\int_0^{2\pi (m-1)} d\tau_n \int_0^{\tau_n} d\tau_{n-1} \dots \int_{0}^{\tau_2} d\tau_1\, F\left(\{\tau_i\}\right) + \nonumber\\& +
\int_{2\pi(m-1)}^{2\pi m} d\tau_n \int_{2\pi(m-1)}^{\tau_n} d\tau_{n-1} \dots \int_{2\pi(m-1)}^{\tau_2} d\tau_1\, F\left(\{\tau_i\}\right)
\end{align}
On the last integral we can further massage the $\tau_{n-1}$ integration domain and arrive at
\begin{align}
&\int_0^{2\pi m} d\tau_n \int_0^{\tau_n} d\tau_{n-1} \dots \int_{0}^{\tau_2} d\tau_1\, F\left(\{\tau_i\}\right) = 
\int_0^{2\pi (m-1)} d\tau_n \int_0^{\tau_n} d\tau_{n-1} \dots \int_{0}^{\tau_2} d\tau_1\, F\left(\{\tau_i\}\right) + \nonumber\\& + 
\int_{2\pi(m-1)}^{2\pi m} d\tau_n \int_{0}^{2\pi(m-1)} d\tau_{n-1} \int_{0}^{\tau_{n-1}} d\tau_{n-2} \dots \int_{0}^{\tau_2} d\tau_1\, F\left(\{\tau_i\}\right) + \nonumber\\& + 
\int_{2\pi(m-1)}^{2\pi m} d\tau_n \int_{2\pi(m-1)}^{\tau_n} d\tau_{n-1} \int_{0}^{\tau_{n-1}} d\tau_{n-2} \dots \int_{0}^{\tau_2} d\tau_1\, F\left(\{\tau_i\}\right)
\end{align}
We can repeat this decomposition $n$ times until we arrive at the form
\begin{align}
&\int_0^{2\pi m} d\tau_n \int_0^{\tau_n} d\tau_{n-1} \dots \int_{0}^{\tau_2} d\tau_1\, F\left(\{\tau_i\}\right) = 
\int_0^{2\pi (m-1)} d\tau_n \int_0^{\tau_n} d\tau_{n-1} \dots \int_{0}^{\tau_2} d\tau_1\, F\left(\{\tau_i\}\right) + \nonumber\\& + 
\int_{2\pi(m-1)}^{2\pi m} d\tau_n \int_{0}^{2\pi(m-1)} d\tau_{n-1} \int_{0}^{\tau_{n-1}} d\tau_{n-2} \dots \int_{0}^{\tau_2} d\tau_1\, F\left(\{\tau_i\}\right) + \nonumber\\& + 
\int_{2\pi(m-1)}^{2\pi m} d\tau_n \int_{2\pi(m-1)}^{\tau_n} d\tau_{n-1} \int_{0}^{2\pi(m-1)} d\tau_{n-2} \int_{0}^{\tau_{n-2}} d\tau_{n-3} \dots \int_{0}^{\tau_2} d\tau_1\, F\left(\{\tau_i\}\right) + \dots + \nonumber\\& + 
\int_{2\pi(m-1)}^{2\pi m} d\tau_n \int_{2\pi(m-1)}^{\tau_n} d\tau_{n-1} \dots \int_{2\pi(m-1)}^{\tau_3} d\tau_2 \int_{0}^{2\pi(m-1)} d\tau_1\, F\left(\{\tau_i\}\right)
\end{align}
The integrand is a periodic function of the Wilson loop parameters and hence we can shift the extrema of the integration to obtain a simpler expression
\begin{align}\label{eq:recrelation}
&\int_0^{2\pi m} d\tau_n \int_0^{\tau_n} d\tau_{n-1} \dots \int_{0}^{\tau_2} d\tau_1\, F\left(\{\tau_i\}\right) = 
\int_0^{2\pi (m-1)} d\tau_n \int_0^{\tau_n} d\tau_{n-1} \dots \int_{0}^{\tau_2} d\tau_1\, F\left(\{\tau_i\}\right) + \nonumber\\& + 
\int_{0}^{2\pi} d\tau_n \int_{0}^{2\pi(m-1)} d\tau_{n-1} \int_{0}^{\tau_{n-1}} d\tau_{n-2} \dots \int_{0}^{\tau_2} d\tau_1\, F\left(\{\tau_i\}\right) + \nonumber\\& + 
\int_{0}^{2\pi} d\tau_n \int_{0}^{\tau_n} d\tau_{n-1} \int_{0}^{2\pi(m-1)} d\tau_{n-2} \int_{0}^{\tau_{n-2}} d\tau_{n-3} \dots \int_{0}^{\tau_2} d\tau_1\, F\left(\{\tau_i\}\right) + \dots + \nonumber\\& + 
\int_{0}^{2\pi} d\tau_n \int_{0}^{\tau_n} d\tau_{n-1} \dots \int_{0}^{\tau_3} d\tau_2 \int_{0}^{2\pi(m-1)} d\tau_1\, F\left(\{\tau_i\}\right)
\end{align}
We now introduce the convenient notation
\begin{align}
G_{n,i_1,i_2,\dots i_l}(m) &\equiv \int\limits_{2\pi>\tau_n>\tau_{n-1}>\dots>\tau_{i_1+1}>0} \prod_{j_1=i_1+1}^{n} d\tau_{j_1} \int\limits_{2\pi>\tau_{i_1}>\tau_{i_1-1}>\dots>\tau_{i_2+1}>0} 
\prod_{j_2=i_2+1}^{i_1} d\tau_{j_2} \dots \nonumber\\&
\dots \int\limits_{2\pi>\tau_{i_l}>\tau_{i_l-1}>\dots>\tau_{1}>0} \prod_{j_l=1}^{i_l} d\tau_{j_l} \, F\left(\{\tau_i\}\right)
\end{align}
with each index $n > i_1 > \dots >0$ corresponding to an integration over the whole circle followed by path ordered integrations.
According to this definition, relation \eqref{eq:recrelation} reads more simply
\begin{equation}
G_n(m) = \sum_{i=1}^{n-1}\, G_{ni}(m-1) + G_n(m-1) + G_n(1)
\end{equation}
The sum contains $G_n(m-1)$ and hence is a recursion relation for $G_n$.
In order to solve it we have to determine the functions $G_{n,1} \dots G_{n,n-1}$ as well.
Repeating the logic above we see that each of them can be given an independent recursion relation in terms of other functions $G_{n > i_1 > \dots >0}$ with ordered indices, producing the block triangular system of recursion relations
\begin{equation}\label{eq:system}
\bigwedge\limits_{k=0}^{n-1}\, \bigwedge\limits_{n>i_{1}>\dots i_{k}>0}\, \left( G_{n,\dots,i_k}(m) = \sum_{j=1}^{i_k-1}\, G_{n,\dots,i_k,j}(m-1) + G_{n,\dots,i_k}(m-1) + G_{n,\dots,i_k}(1) \right)
\end{equation}
Starting from $G_n(m)$ as step 0, at the $l^\text{th}$ step $\binom{n-1}{l}$ functions $G$ with $l+1$ indices are generated. Since $n-1$ steps are required to complete the decomposition, we have to solve a system of $2^{n-1}$ recurrence relations for as many functions.
At this stage this might look a little abstract and we provide a simple explicit example of its application in section \ref{sec:2loops}.

The procedure outlined above is algorithmic and hence it is straightforward to code a function generating such a system for generic $n$.
Whenever the functions $G$ have indices with 1 as last entry, their recursion relation is trivial and reads
\begin{equation}
G_{n,i_1\dots 1}(m) = m\, G_{n,i_1\dots 1}(1)
\end{equation}
Therefore, the system \eqref{eq:system} can be solved in general, provided the initial conditions are fixed for all the functions $G_{n,i_1,i_2\dots}(1)$.
These are contour integrals for the singly wound Wilson loop which we assume to be known.
Note, however, that the integration domains of these might not coincide with those of the original computation of the Wilson loop in that they are not path ordered.

In order to make contact with the original integrals one can algorithmically decompose the integration domains above into unions of ordered ones of the form $\left(2\pi > \tau_{i_1}> \tau_{i_2} > \dots > 0 \right)$.

One straightforward way to do this is for instance to consider all permutations of indices in the inequality $2\pi> \tau_n > \tau_{n-1} >\dots \tau_1 > 0$ and select only those which respect the constraints imposed by the indices of the relevant function $G$
\begin{equation}
G_{n,\dots,i_1,\dots,i_2\dots i_l} \to \left\{\begin{array}{l}
2\pi> \tau_n > \tau_{n-1} > \dots \tau_{i_1+1} > 0 \\
2\pi> \tau_{i_1} > \tau_{i_1-1} > \dots \tau_{i_2+1} > 0 \\
\dots \\
2\pi> \tau_{i_l} > \tau_{i_l-1} > \dots \tau_{1} > 0
\end{array}\right.
\end{equation}
At this point one can rename dummy variables in order to make all integration contours look like the original one at the price of reshuffling the arguments of the integrand $F(\{\tau\})$. This task can be optimized implementing the symmetry properties of the integrand under permutations of indices, which depends on its explicit form.
These are finally the same integrals as those appearing in the computation of the expectation value of the singly wound Wilson loop.
In other words, with the procedure outlined above we are able to separate the combinatorial problem of multiple winding (which can be treated by a computer) from the dynamics of the theory, encapsulated in the integrals for single winding.

We remark that in a non-abelian theory these integrals may come with a different color factor as in the $m=1$ case, though. In particular, integral topologies associated to non-planar particle exchanges may become relevant for the leading color expectation value of the winding $m$ case.

In the next sections we provide examples of this procedure, computing the expectation value of the 1/6-BPS circular Wilson loop in planar ABJM theory up to three loops and the 1/2-BPS Wilson loop up to two.

\section{The perturbative BPS Wilson loops}\label{sec:winding1}

In this section we briefly define the 1/6-BPS and 1/2-BPS Wilson loops for ABJM theory \cite{DPY,Chen:2008bp,Rey:2008bh} and review the perturbative computation of their expectation value up to three and two loops, respectively.
We stress the role of the framing dependence in these computations.
Finally we outline the exact computation of this Wilson loop via localization and provide its expansion at weak coupling.

\subsection{The 1/6-BPS Wilson loop}

We consider ABJM theory with unitary gauge groups $U(N_1)_k\times U(N_2)_{-k}$ and opposite Chern-Simons levels $k$, and the corresponding gauge fields $A$ and $\hat A$. The theory features a multiplet of scalar and fermionic degrees of freedom $C$, $\bar C$ and $\psi$, $\bar \psi$, transforming in the bifundamental representation of the gauge groups and in the (anti)fundamental of the R-symmetry group $SU(4)$. The theory possesses ${\cal N}=6$ supersymmetry and is superconformal.
The Chern-Simons level acts as an inverse coupling for the theory, which in the limit of $k\gg 1$ is amenable of a perturbative description.
Taking the ranks of the gauge groups to be large as well, one can also define a planar limit, where the perturbative expansion can be organised in terms of the effective 't Hooft couplings
\begin{equation}
\lambda_1 \equiv \frac{N_1}{k}\qquad\qquad \lambda_2 \equiv \frac{N_2}{k}
\end{equation}
We consider the theory in the weakly coupled planar regime where $\lambda_i \ll 1$.
In ABJM theory one can define a connection which preserves four superconformal supersymmetries out of the original 24 of the theory.
On particular contours such as a straight line or a circle the corresponding Wilson loop operator is also globally supersymmetric and its expectation value evaluates to 1 on the former, whereas on the latter it is a non-trivial, but finite, function of the coupling interpolating between the weak and strong regimes.
We focus on the computation of the expectation value of such a 1/6-BPS Wilson loop at weak coupling
\begin{equation}
\label{eq:WL}
\langle W^{1/6}[{\cal C}] \rangle = \frac{1}{N_1} \int D[A, \hat{A}, C, \bar{C}, \psi, \bar{\psi}] \; e^{-S_{\rm ABJM}}  \; \Tr \left[ {\cal P} \exp{
\left( - i \int_{\cal C} d\tau {\cal A}(\tau) \right) } \right]
\end{equation}
where $S_{\rm ABJM}$ is the euclidean action and the locally supersymmetric connection reads
\begin{equation}
\label{eq:connection}
{\cal A} =  A_\mu \dot{x}^\mu - \frac{2\pi i}{k}  |\dot{x}| (M_{1/6})_J^{\; I} C_I \bar{C}^J 
\end{equation}
containing a scalar matter bilinear governed by the matrix $M_{1/6} = {\rm diag}(+1,+1,-1,-1)$. Throughout this paper the trace is taken in the fundamental representation of the gauge group.
The path ${\cal C}$ is the unit circle parametrized as 
\begin{equation}
{\cal C} : \quad x^\mu(\tau) = (0, \cos{\tau}, \sin{\tau})
\end{equation}
Similarly, a 1/6-BPS Wilson loop $\hat W^{1/6}$ with a connection transforming in the adjoint of the second gauge group $U(N_2)$ can be defined.
\begin{equation}
\hat{\cal A} = \hat A_\mu \dot{x}^\mu - \frac{2\pi i}{k}  |\dot{x}| (M_{1/6})_J^{\; I} \bar C^J C_I 
\end{equation}
Its expectation value can be obtained from the $U(N_1)$ result by complex conjugation and exchange of the coupling constants.

\subsection{The 1/2-BPS Wilson loop}

A Wilson loop preserving 1/2 of the supercharges (up to a supergauge transformation) can be constructed in terms of a superconnection ${\cal L}$ of the supergroup $U(N_1|N_2)$. 
This allows to define the 1/2-BPS Wilson loop
\begin{equation}\label{eq:WLdef}
\langle W^{1/2}[{\cal C}] \rangle = \frac{1}{k}\, \int D[A, \hat{A}, C, \bar{C}, \psi, \bar{\psi}] \; e^{-S_{ABJM}}  \; {\rm Tr} \left[ {\cal P} \exp{
\left( - i \int_{\cal C} d\tau {\cal L}_{1/2}(\tau) \right) } \right]
\end{equation}
where ${\cal L}_{1/2}(\tau)$ is represented as the supermatrix  
\begin{equation}
{\cal L}_{1/2}(\tau) = \left( \begin{array}{ccc}
{\cal A} & -i \sqrt{\frac{2\pi}{k}} |\dot{x}| \eta_I \bar{\psi}^I \\
 -i \sqrt{\frac{2\pi}{k}}  |\dot{x}| \psi_I \bar{\eta}^I & \hat{\cal A}  \end{array}\right)  
\end{equation}
and ${\cal A}, \hat{\cal A}$ are the same connections as for the 1/6-BPS Wilson loops, though with different scalar matrix $M_{1/2}$. The coupling to fermions is governed by the commuting spinors and $\eta, \bar{\eta}$.
For a circular contour supersymmetry imposes that these have the form
\begin{align}
& M_{1/2} = {\rm diag}(-1,1,1,1)\\
& \eta_I^\a (\tau) = \left( e^{i\tau/2} \quad -i  e^{-i\tau/2} \right) \, \d^1_I  \quad , \quad \bar{\eta}_\a^I(\tau) = \left( \begin{array}{c}  
i  e^{-i\tau/2} \\
- e^{i\tau/2} 
\end{array}\right) \, \d_1^I \quad , \qquad (\eta \bar{\eta}) = 2i
\end{align}
For multiple winding gauge invariance requires the Wilson loop to have the form
\begin{equation}\label{eq:defhalf}
W_m^{1/2} = \frac{1}{k}\, {\rm Str} \left[ \left({\cal P} \exp{ \int d\tau {\cal L}_{1/2}(\tau) }\right)\, \left(\begin{array}{cc}
1 & 0\\
0 & (-1)^m
\end{array}\right) \right]
\end{equation}
Here and in \eqref{eq:WLdef} the supertrace on the supermatrix is necessary for enforcing supersymmetry conservation, as well as the additional matrix of \eqref{eq:defhalf} which basically interwines between supertrace and trace according to the parity of the winding number.
We finally comment on the unusual normalization of the Wilson loop.
This is due to the fact that the natural normalization by $N_1 - (-1)^m N_2$ is singular for even winding in the limit of $N_1=N_2$ and we want to avoid this singularity.
Moreover it is a convenient choice to make better contact with the localization result and keep formulae clean.

\subsection{Framing}

Wilson loops in gauge field theories may suffer from ultraviolet divergences caused by the integration of the connections along the contours. 
For Chern-Simons theories this is not the case and the Wilson loop, as a result, is finite.
Yet, the definition of connections at coincident points is ambiguous in the sense that it may produce different finite results, according to the specific choice.
In particular, this introduces in general a dependence on the metric which would spoil the  topological invariance of the Wilson loop expectation value.
The issue of defining a topologically invariant regularization of the Wilson loop has been investigated in the context of knot theory and its solution goes under the name of \emph{framing}.
It is reminiscent of a point-splitting regularization of correlation functions in quantum field theory.
Namely, one considers a vector field orthogonal to the original path on which the Wilson loop is evaluated
\begin{equation}
{\cal C}_f : \quad x^\mu(\tau) \rightarrow y^\mu(\tau) = x^\mu(\tau) + \epsilon \, n^\mu(\tau) \quad , \quad |n(\tau)| = 1
\end{equation}
Then the connections to be integrated over are defined on different contours, infinitesimally displaced by integer multiples of such a vector field
\begin{equation}
x^\mu(\tau_i) \rightarrow x^\mu(\tau_i) + (i-1)\, \epsilon \, n^\mu(\tau_i)
\end{equation}
This prescription has been shown to provide a topologically invariant manner of defining Wilson loops in pure non-abelian Chern-Simons theory, in the sense that their expectation values do not depend on the particular choice of the framing vector field, but only on the cotorsion, that is the number of times $f$ it winds around the original contour. 
Moreover, the effect of a non-trivial framing on the Wilson loop expectation value has been shown to be captured, non-perturbatively, by an overall phase factor \cite{Witten}
\begin{equation}
\langle W^{CS} \rangle_f = e^{i\,\pi\,\lambda\, f}\, \langle W^{CS} \rangle_0\qquad f\in \mathbb{Z}
\end{equation}
which is the exponential of the one-loop contribution, with $\lambda$ the (shifted by the quadratic Casimir of the gauge group) 't Hooft Chern-Simons coupling. This exponentiation has also been supported by perturbative computations up to three loops \cite{Guadagnini:1989am,Alvarez:1991sx} and an all-loop argument \cite{Alvarez:1991sx}.

Wilson loops with multiple winding can be decomposed in a combination of Wilson loops in different representations of the gauge group, each of which comes with its framing phase (e.g. \cite{Labastida:2000yw,Marino:2001re,Brini:2011wi}). We comment on this at the end of Section \ref{sec:3loops} from our result in ABJM theory, in a limit which selects its pure Chern-Simons component.

For ABJM BPS Wilson loops the effect of framing is much subtler. 
For the 1/6-BPS Wilson loop it was argued in \cite{Bianchi:2016yzj} that the effect of framing is likely to be still captured by a phase, which is however a non-trivial function of the 't Hooft couplings, instead of a simple exponential of the one-loop contribution (which coincides with the pure Chern-Simons framing phase).
For the 1/2-BPS Wilson loop it was pointed out that the framing factor is the same as for pure Chern-Simons with gauge supergroup $U(N_1|N_2)$ \cite{Drukker:2010nc}, since such a phase arises naturally in its exact computation via localization. The perturbative results of \cite{Bianchi:2013zda,Bianchi:2013rma,Griguolo:2013sma} for the expectation value at two loops confirm this expectation.
In both cases the removal of these phases leaves a real expectation value, as expected in a unitary theory.
For multiply wound Wilson loops only the matrix model results at framing one are available at the moment.
Part of the purpose of this paper is to get a perturbative handle on the behaviour of framing for multiply wound Wilson loops in ABJM.

\subsection{Perturbative computation}

The perturbative evaluation of the expectation value of the 1/6-BPS Wilson loop was carried out in \cite{DPY,Chen:2008bp,Rey:2008bh} up to two loops, building also on pure Chern-Simons results from \cite{Guadagnini:1989am}.

We review the outcome of such a computation diagrammatically for framing $f$
\begin{align}
\raisebox{-0.8cm}{\includegraphics[width=2.cm]{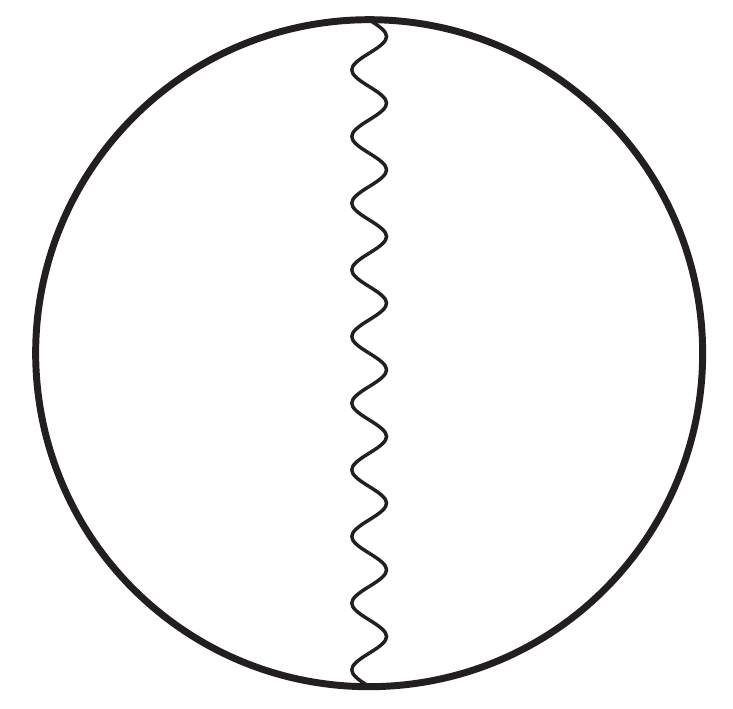}} &= i\, \pi\, f\, \lambda_1\\
\raisebox{-0.8cm}{\includegraphics[width=2.cm]{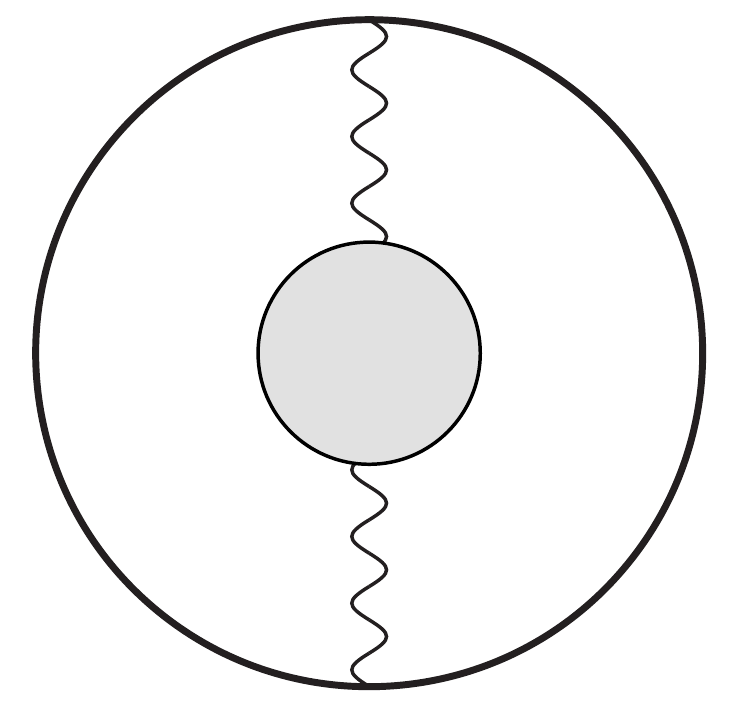}} + \raisebox{-0.8cm}{\includegraphics[width=2.cm]{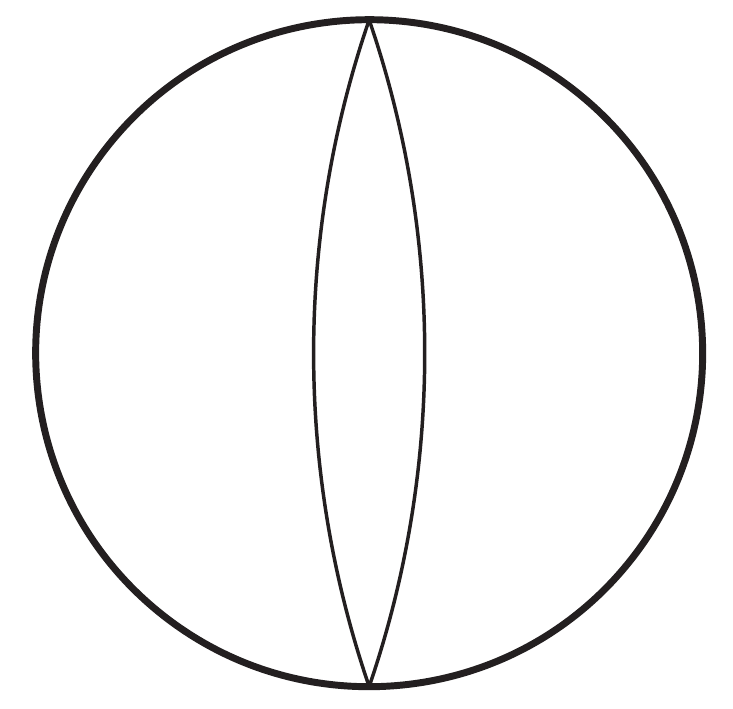}} &= \pi^2\, \lambda_1\lambda_2\\
\raisebox{-0.8cm}{\includegraphics[width=2.cm]{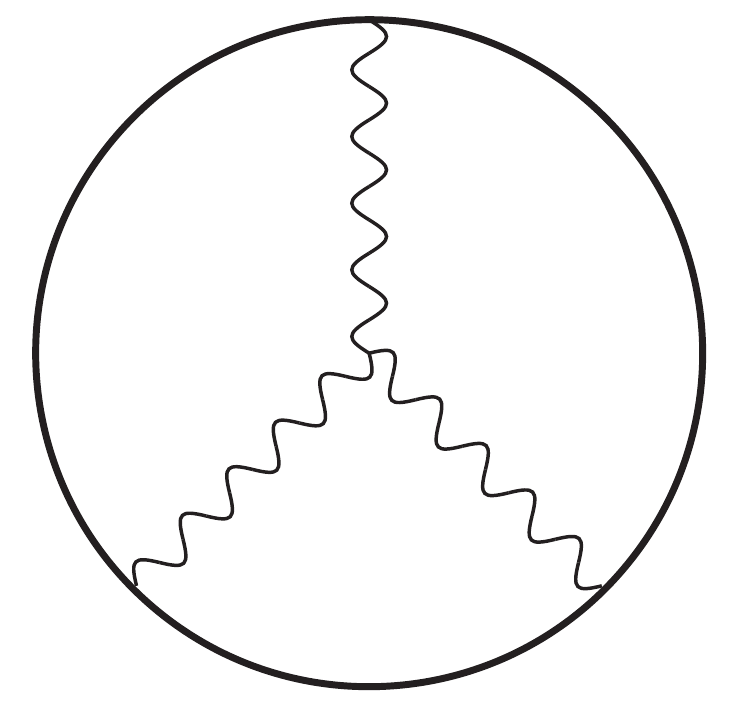}} &= -\frac{\pi^2}{6}\, \lambda_1^2 \label{eq:mercedesresult} \\
\raisebox{-0.8cm}{\includegraphics[width=2.cm]{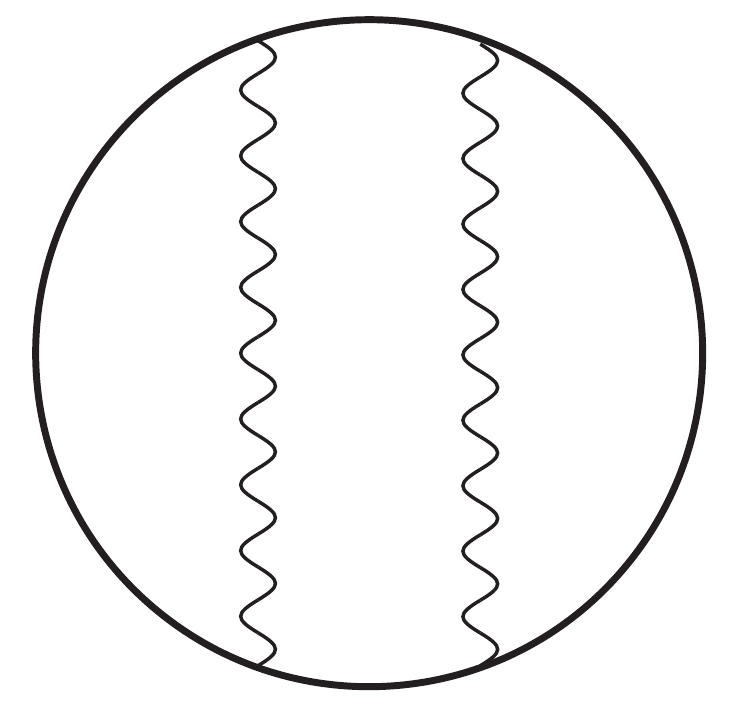}} + \raisebox{-0.8cm}{\includegraphics[width=2.cm]{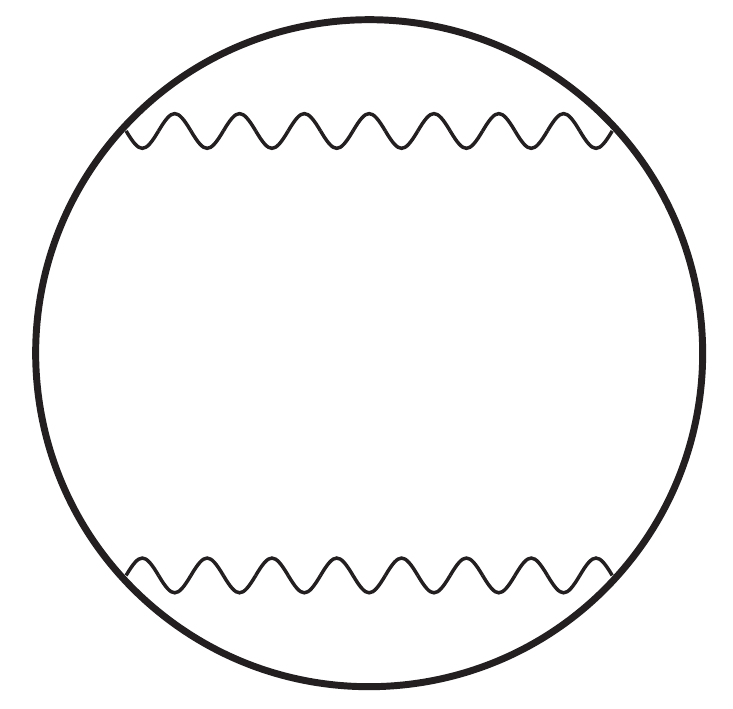}} + \raisebox{-0.8cm}{\includegraphics[width=2.cm]{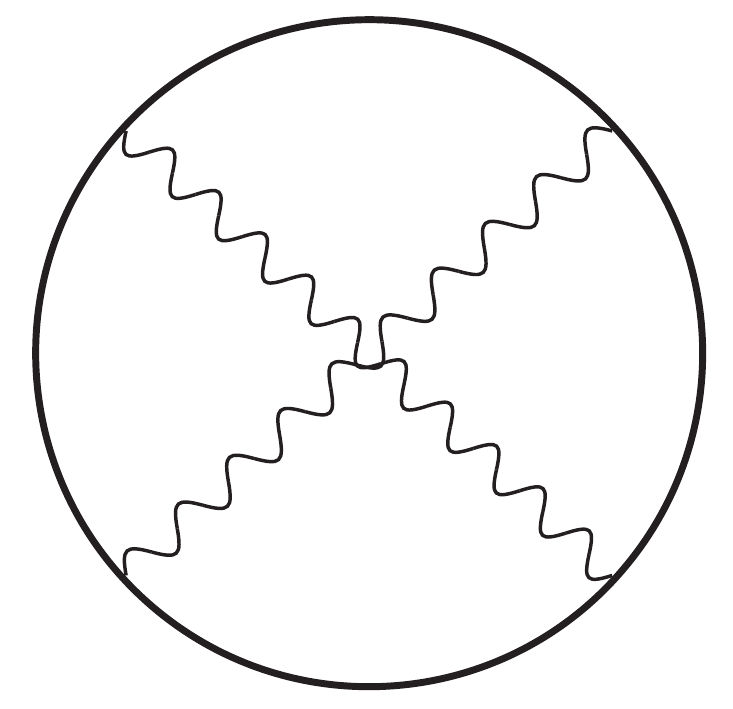}} &= \frac12\, \raisebox{-0.8cm}{\includegraphics[width=2.cm]{1Lgauge}} ^2 = -\frac{\pi^2\, f^2}{2}\, \lambda_1^2 \label{eq:double} \\
-\raisebox{-0.8cm}{\includegraphics[width=2.cm]{2Lcross}} &= 0 \label{eq:crossed}
\end{align}
where the pictorial integrals are defined in \ref{app:integrals}.
We recall in particular that 
\begin{itemize}
\item the one-loop contribution is framing dependent and vanishes for a planar contour;
\item the mercedes integral \eqref{eq:mercedesresult} is finite and framing independent \cite{Guadagnini:1989am};
\item the double exchange integral \eqref{eq:double} can be decomposed (in the planar limit) into a totally symmetric contribution and a "crossed" contribution;
\item the totally symmetric part is the square of the one-loop contribution, in particular it vanishes for a planar contour;
\item the crossed contribution \eqref{eq:crossed} is finite, framing independent and vanishes for a planar contour ({\em independently} of framing);
\item other integrals involving the scalar fields and not shown above vanish thanks to the condition $\Tr\, M = 0$.
\end{itemize}
At three loops the expectation value of the 1/6-BPS Wilson loop was investigated in \cite{Rey:2008bh}.
From inspection of the Feynman rules it was argued that at odd loops, only diagrams with an odd number $2l+1$ of $\varepsilon$ tensors are generated. By rewriting $2l$ of them in terms of products of metrics, one is left with a single Levi-Civita tensor, which is eventually contracted with three vectors which lie on a plane at trivial framing.
Therefore the three-loop expectation value vanishes for $f=0$. Actually the same holds true for any odd loop order.

With non-trivial framing the latter argument is no longer valid and framing dependent contributions may arise.
A study of them was performed for pure Chern-Simons theory (though with a different regularization scheme) in \cite{Alvarez:1991sx} and for ABJM theory with dimensional reduction scheme in \cite{Bianchi:2016yzj}.
The sum of these contributions reads
\begin{equation}
\langle W_1^{1/6} \rangle_f^{(3)} = i\, \pi^3 \left( -\frac16\, f^2(f+1)\, \lambda_1^3 + f\, \lambda_1^2\lambda_2 - \frac12\, f\, \lambda_1\lambda_2^2 \right)
\end{equation}
Then we obtain for the 1/6-BPS Wilson loop at framing $f$
\begin{align}
\langle W_1^{1/6} \rangle_f = e^{i\, \pi\, f\, \lambda_1 - i\, \frac{\pi^3}{2}\, f\, \lambda_1\lambda_2^2 }\left[1 + \pi^2 \left(\lambda_1\lambda_2-\frac{1}{6}\, \lambda_1^2\right)\right] + {\cal O}\left(k^{-4}\right)
\end{align}
In particular, at framing 0 we recover the result \cite{DPY,Chen:2008bp,Rey:2008bh}
\begin{equation}\label{eq:framing0}
\langle W_1^{1/6} \rangle_0 = 1 + \pi^2 \left(\lambda_1\lambda_2-\frac{1}{6}\, \lambda_1^2\right) + {\cal O}\left(k^{-4}\right)
\end{equation}
whereas at framing 1 the Wilson loop expectation value reads
\begin{equation}\label{eq:framing1}
\langle W_1^{1/6} \rangle_1 = 1 + i\, \pi\, \lambda_1 + \pi^2 \left(\lambda_1\lambda_2-\frac{2}{3}\, \lambda_1^2\right) + i\, \pi^3 \left( -\frac13\, \lambda_1^3 + \lambda_1^2\lambda_2 - \frac12\, \lambda_1\lambda_2^2 \right) + {\cal O}\left(k^{-4}\right)
\end{equation}
The latter is relevant for comparison with the localization result, as we explain below.

The perturbative computation of the expectation value of the 1/2-BPS Wilson loop at framing 0 has been carried out in \cite{Bianchi:2013zda,Bianchi:2013rma,Griguolo:2013sma}.
The bosonic diagrams are effectively the same as for the 1/6-BPS case, albeit the different scalar coupling matrix $M_{1/2}$ and we refer to the previous results.
The new diagrams appearing in the 1/2-BPS case are those involving fermions. 
At one loop there is a fermion exchange diagram, whose contribution can be shown to be subleading in dimensional regularization (at trivial framing).
At two loops there are three new diagrams: a single exchange of a one-loop corrected fermion, the double fermion exchange and the fermionic mercedes.
At zero framing and hence discarding all contributions that vanish thanks to the planarity of the contour they evaluate
\begin{align}
\raisebox{-0.8cm}{\includegraphics[width=2.cm]{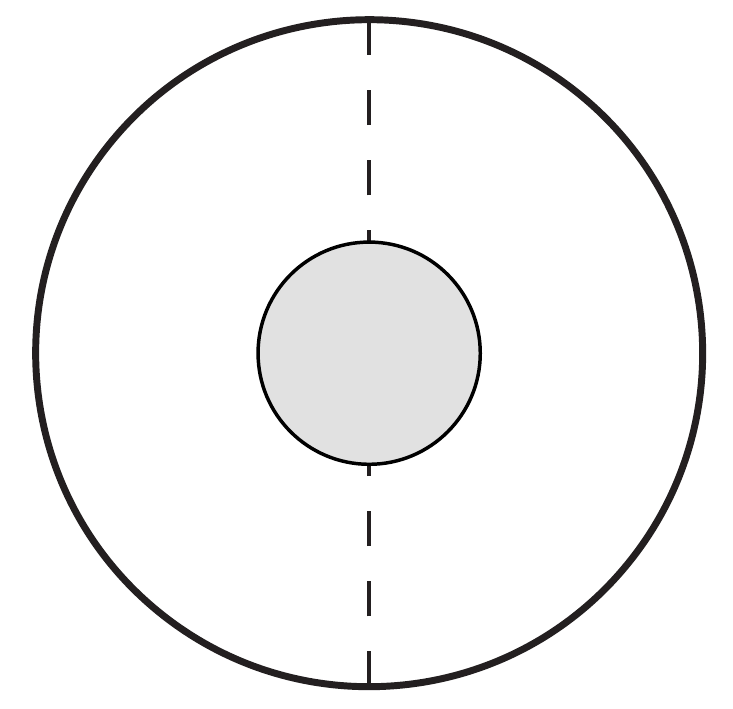}} &= 0 \label{eq:fermionex}\\
\raisebox{-0.8cm}{\includegraphics[width=2.cm]{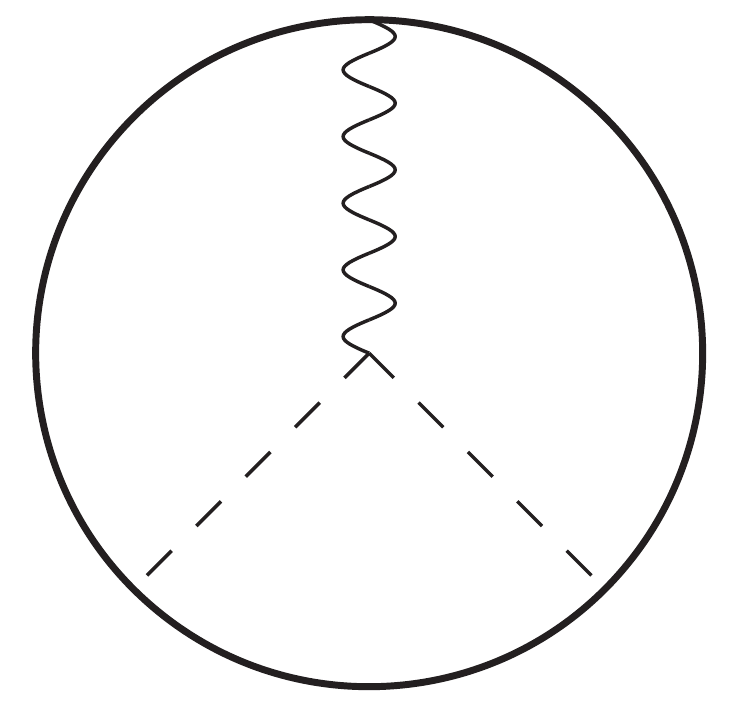}} &= -2\pi^2\, \lambda_1\lambda_2 \left(\lambda_1 + \lambda_2\right) \label{eq:fermionmercedes0} \\
\raisebox{-0.8cm}{\includegraphics[width=2.cm]{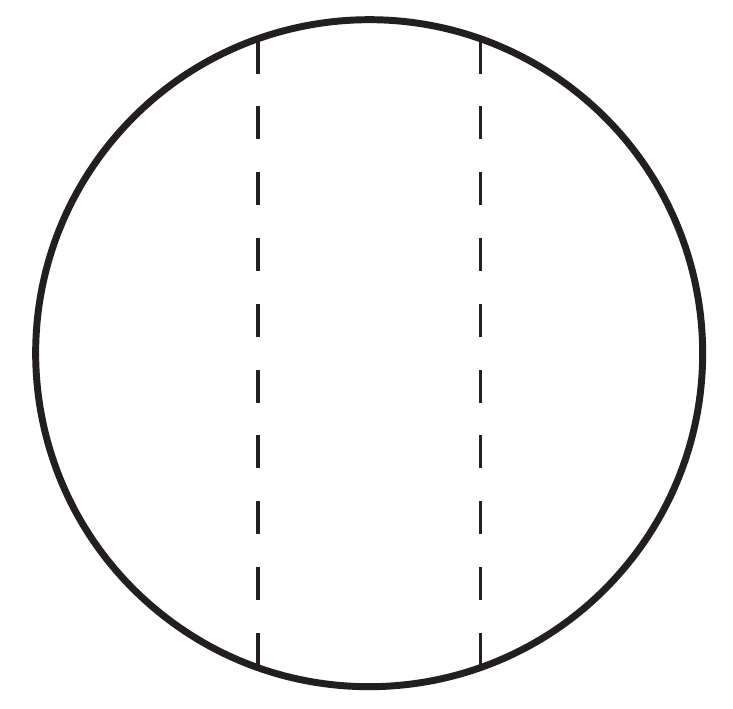}} + \raisebox{-0.8cm}{\includegraphics[width=2.cm]{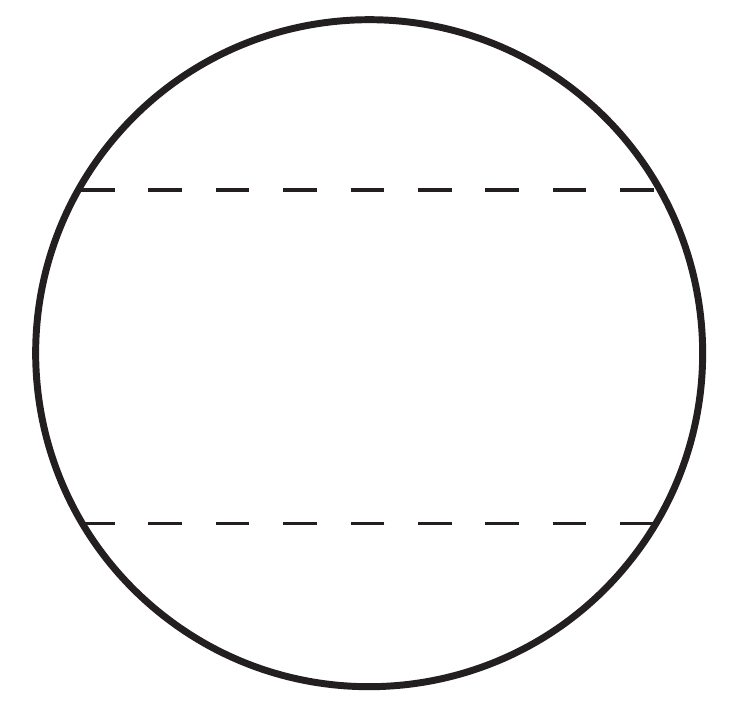}} &= \frac{3\pi^2}{2}\, \lambda_1\lambda_2 \left(\lambda_1 + \lambda_2\right) \label{eq:fermionexchange0} 
\end{align}
We remark that the one-loop corrected fermion exchange \eqref{eq:fermionex} vanishes as a result of the subtraction between the two blocks in \eqref{eq:defhalf}, but each does not individually.
The final result for the 1/2-BPS Wilson loop reads
\begin{equation}\label{eq:halfsingle}
\langle W^{1/2}_1 \rangle_0 = \left(\lambda _1 + \lambda _2\right) \left[
1 - \frac{\pi^2}{6} \left(\lambda _1^2+\lambda _2^2-4\lambda _1 \lambda _2\right)\right] + {\cal O}\left(k^{-4}\right)
\end{equation}

\subsection{Localization result}

The expectation value of the 1/6-BPS Wilson loop in ABJM theory can be computed exactly by supersymmetric localization on $S^3$.
Thanks to this technique the path integral of the theory collapses to the matrix model \cite{Kapustin:2009kz,Marino:2009jd,Drukker:2010nc}
\begin{align}
Z(N_1, N_2, k)&=\int\prod_{i=1}^{N_1} d\mu_i \prod_{j=1}^{N_2} d \nu_j \prod_{i<j}  \sinh^2 \left( \frac{\mu_i -\mu_j}{2}\right)  \sinh^2 \left( \frac{\nu_i -\nu_j}{2}\right) \nonumber\\&
\times \prod_{i,j}  \cosh^2 \left( \frac{\mu_i -\nu_j}{2}\right)\,  e^{-\frac{k}{4\pi\, i}\left(  \sum_i \mu_i^2 +\sum_j \nu_j^2\right)}
\end{align}
It was shown in \cite{Marino:2009jd} how to solve for the expectation value of the Wilson loop in the planar limit in terms of the integral formula
\begin{equation}\label{eq:mmwl}
\langle W_1^{1/6} \rangle_1 = \frac{1}{2\pi^2\, i\, \lambda_1}\, \int_{-a}^a\, e^{x}  {\rm arctan} {\sqrt\frac{ \alpha - 2 \cosh x}{\beta + 2\cosh x }}\, d x 
\end{equation}
In the integrand the parameters $\alpha$ and $\beta$ are related to the endpoints of the cuts of the Chern-Simons lens space $\left(1/a,a\right)$ and $\left(-1/b,-b\right)$, which the eigenvalues condense around at large gauge group ranks, via
\begin{equation}
\alpha \equiv a + \frac{1}{a} \qquad \beta \equiv b + \frac{1}{b}
\end{equation}
where in turn $a$ and $b$ are functions of the couplings $\lambda_1$ and $\lambda_2$ which can be obtained at weak coupling inverting perturbatively the relations spelled out in \cite{Marino:2009jd}.

The computation of the multiply wound Wilson loop is performed evaluating the matrix model average of the operator $e^{m x}$.
The first three perturbative orders, which we check perturbatively in this paper, read
\begin{align}\label{eq:localization}
\langle W_m^{1/6} \rangle_1 &= 1+i\, \pi\, m^2\, \lambda _1 + 
\pi^2\left[\left(-\frac{m^2}{3}-\frac{m^4}{3}\right) \lambda _1^2 + m^2\, \lambda _1 \lambda _2\right] + \nonumber\\&
+ i\,\pi^3\left[ \left(-\frac{m^2}{18} -\frac{2\,m^4}{9} -\frac{m^6}{18} \right) \lambda _1^3 + \left(\frac{m^2}{3} +\frac{2\, m^4}{3}\right) \lambda _1^2 \lambda _2 - \frac{m^2}{2}\, \lambda _1 \lambda _2^2 \right] + {\cal O}\left(k^{-4}\right)
\end{align}
A few more orders are displayed explicitly in \eqref{eq:localization8loops}.

The coefficients at odd loop orders are imaginary. Such terms are ubiquitous in localization based computations in three-dimensional Chern-Simons theories as pointed out in \cite{Closset:2012vg}. The localization procedure in fact implies that some background fields are imaginary in order to preserve rigid supersymmetry on $S^3$. This causes a loss of unitarity, indeed signalled by an imaginary contribution to the expectation value of a physical observable.
For Wilson loop expectation values in pure Chern-Simons theory this ambiguity has the interpretation of a framing dependence. We expect the same phenomenon to occur in Chern-Simons-matter theories as well.
In particular, as explained in \cite{Kapustin:2009kz}, localization by construction assumes that a framing of the Wilson loop is introduced in a supersymmetry preserving way. This in turn can be achieved if the framing contours are the Hopf fibers of $S^3$. 
As these circles have linking number 1, the matrix model average \eqref{eq:mmwl} yields naturally a framing 1 result, hence the index in the expectation value of \eqref{eq:localization}.
Indeed, for single winding, the expression \eqref{eq:localization} agrees with the perturbative computation at framing 1 \eqref{eq:framing1} reviewed before. 

For $m=1$ the framing dependence can be shown to exponentiate into a phase \cite{Bianchi:2016yzj}, which in the presence of matter is itself a function of the coupling with a non-trivial perturbative expansion. 
Such a phase can then be removed, for instance by taking the modulus of \eqref{eq:framing1}, reproducing \eqref{eq:framing0}.

In the following sections we ascertain agreement between the matrix model and the field-theoretical computations for general winding as well.

Finally, the 1/2-BPS Wilson loop expectation value at general winding $m$ can be obtained via localization as a combination of the 1/6-BPS expectation values
\begin{equation}\label{eq:equiv}
\langle W^{1/2}_m \rangle_1 = \lambda_1\, \langle W^{1/6}_m \rangle_1 - (-1)^m\, \lambda_2\, \langle \hat W^{1/6}_m \rangle_1 
\end{equation}
where we have used again the index 1 to remark that the localization result is derived at framing 1. The factors of $\lambda$ are present for consistency between normalization conventions.
Up to two-loop order it reads
\begin{align}
\langle W^{1/2}_m \rangle_1 &= \lambda _1 - (-1)^m \lambda _2 + i\, \pi\,  m^2 \left(\lambda _1^2+ (-1)^m\lambda _2^2\right) + \nonumber\\&
- \frac{\pi ^2}{3}\, m^2\, \left(\lambda _1^2 \left((m^2+1)\lambda _1-3 \lambda _2\right) - (-1)^m\lambda _2^2 \left(-3 \lambda _1+(m^2+1)\lambda _2\right)\right) + {\cal O}\left(k^{-4}\right)
\end{align}
For single winding, and after removing the framing phase, the result is in agreement with the perturbative expression \eqref{eq:halfsingle}.


\section{Multiply wound Wilson loops} 

In this section we compute the ABJM Wilson loops perturbatively at generic framing $f$ and winding $m$.
We start with the 1/6-BPS Wilson loop up to two loops, where the combinatorics of multiple winding is sufficiently simple to be explained in full detail, providing an explicit example of the general procedure outlined in section \ref{sec:contours}.
We then complete the evaluation of the 1/6-BPS Wilson loop expectation value at three loops and compare the result to previous literature.
Finally we compute the 1/2-BPS Wilson loop to two loops.

\subsection{1/6-BPS Wilson loop to two loops}\label{sec:2loops}

We provide the details of the computation of the diagrams for the 1/6-BPS Wilson loop at $m$ windings.
At one loop the relevant integral for a gluon exchange is symmetric under the exchange of the two integration variables and hence can be symmetrized. The resulting integral has a domain $[0,2\pi m]\times[0,2\pi m]$ and therefore is trivially equal to $m^2$ times the original contribution at single winding. Hence at framing $f$ it gives 
\begin{equation}
\langle W_m^{1/6} \rangle_f^{(1)} = i\, \pi\, m^2\, f\, \lambda_1
\end{equation}
At two loops the diagrams with matter can be treated in the same way as the one-loop contributions and seen to equal $m^2$ times their result at winding 1, namely \eqref{eq:matterm}.
The mercedes diagram is a bit more complicated and needs the combinatorics spelled out in section \ref{sec:contours}.
In order to provide an explicit and simple example of how it works, we derive it in full detail.
The integral relevant for this diagram reads explicitly
\begin{equation}
G_{3}[f](m) \equiv \raisebox{-0.8cm}{\includegraphics[width=2.cm]{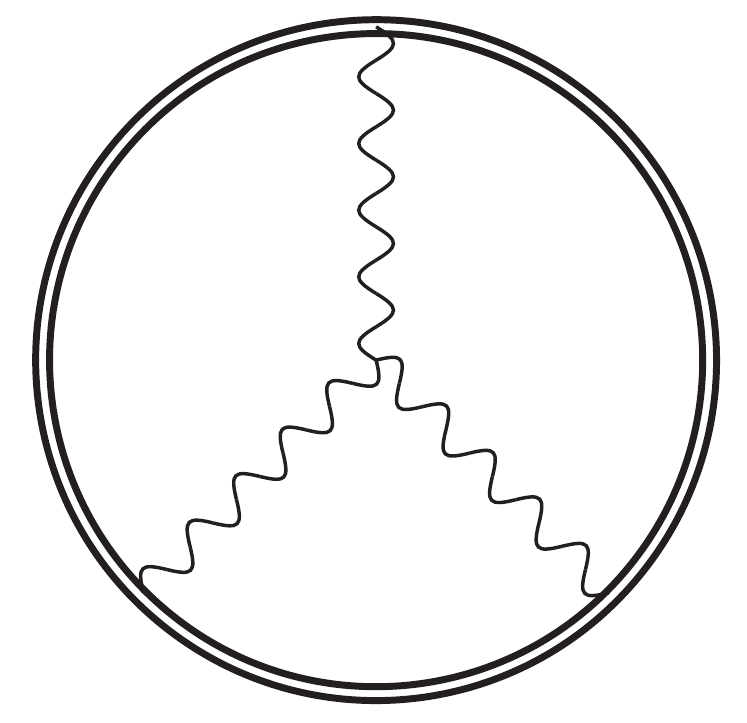}} = \int_{0}^{2\pi m}d\tau_1 \int_{0}^{\tau_1}d\tau_2\, \int_{0}^{\tau_2}d\tau_3\, f(\tau_1,\tau_2,\tau_3)
\end{equation}
where the function $f$ is specified in \eqref{eq:mercedes}, but in dealing with the generalization to $m$ windings we only need to know that it is antisymmetric under the exchange of any pair of variables and periodic in each.
Following the general analysis of section \ref{sec:contours}, we can decompose the integration as follows
\begin{align}
G_{3}[f](m) = & \equiv \int_{0}^{2\pi m}d\tau_1 \int_{0}^{\tau_1}d\tau_2\, \int_{0}^{\tau_2}d\tau_3\, f(\tau_1,\tau_2,\tau_3) = \nonumber\\& = \int_{2\pi (m-1)}^{2\pi m}d\tau_1 \int_{0}^{2\pi (m-1)}d\tau_2\, \int_{0}^{\tau_2}d\tau_3\, f(\tau_1,\tau_2,\tau_3) + \nonumber\\& ~~+ 
\int_{2\pi (m-1)}^{2\pi m}d\tau_1 \int_{2\pi (m-1)}^{\tau_1}d\tau_2\, \int_{2\pi (m-1)}^{\tau_2}d\tau_3\, f(\tau_1,\tau_2,\tau_3)
+  \nonumber\\& ~~+ 
\int_{2\pi (m-1)}^{2\pi m}d\tau_1 \int_{2\pi (m-1)}^{\tau_1}d\tau_2\, \int_{0}^{2\pi (m-1)}d\tau_3\, f(\tau_1,\tau_2,\tau_3)
+  \nonumber\\& ~~+ 
\int_{0}^{2\pi (m-1)}d\tau_1 \int_{0}^{\tau_1}d\tau_2\, \int_{0}^{\tau_2}d\tau_3\, f(\tau_1,\tau_2,\tau_3)
\end{align}
Using periodicity of $f$, we can construct the recursive relation
\begin{align}\label{eq:mercedesstep}
G_{3}[f](m) & = G_{3}[f](m-1) + \int_{0}^{2\pi}d\tau_1 \int_{0}^{2\pi (m-1)}d\tau_2\, \int_{0}^{\tau_2}d\tau_3\, f(\tau_1,\tau_2,\tau_3) + \nonumber\\& ~~+ 
\int_{0}^{2\pi}d\tau_1 \int_{0}^{\tau_1}d\tau_2\, \int_{0}^{\tau_2}d\tau_3\, f(\tau_1,\tau_2,\tau_3)
+ \nonumber\\& ~~+ 
\int_{0}^{2\pi}d\tau_1 \int_{0}^{\tau_1}d\tau_2\, \int_{0}^{2\pi (m-1)}d\tau_3\, f(\tau_1,\tau_2,\tau_3)
\end{align}
The third term is by definition $G_{3}[f](1)$, i.e. the first step of the iteration, which evaluates \eqref{eq:mercedesresult}
\begin{equation}
G_{3}[f](1) = -\frac{\pi^2}{6}\, \lambda_1^2
\end{equation}
The last term can be handled as follows: first, thanks to periodicity it equals $(m-1)$ times the integral
\begin{equation}
G_{3,1}[f](1) = \int_{0}^{2\pi}d\tau_1 \int_{0}^{\tau_1}d\tau_2\, \int_{0}^{2\pi}d\tau_3\, f(\tau_1,\tau_2,\tau_3)
\end{equation}
Such an integral can be decomposed into three pieces according to the position of $\tau_3$ on the circle, namely
\begin{equation}
\left(\int_{0}^{2\pi}d\tau_{1>2>3} + \int_{0}^{2\pi}d\tau_{1>3>2}+ \int_{0}^{2\pi}d\tau_{3>1>2} \right) f(\tau_1,\tau_2,\tau_3)
\end{equation} 
These can in turn be handled changing dummy integral variables
\begin{equation}
\int_{0}^{2\pi}d\tau_1 \int_{0}^{\tau_1}d\tau_2\, \int_{0}^{\tau_2}d\tau_3\, \left( f(\tau_1,\tau_2,\tau_3) + f(\tau_1,\tau_3,\tau_2) + f(\tau_2,\tau_3,\tau_1) \right)
\end{equation} 
Using the antisymmetry of $f$ under exchanges one concludes that such a term is again equal to $G_{3}[f](1)$.

The first term in \eqref{eq:mercedesstep} can be simplified as follows:
\begin{align}\label{eq:mercedesstep2}
G_{3,2}[f](m-1) &\equiv \int_{0}^{2\pi}d\tau_1 \int_{0}^{2\pi (n-1)}d\tau_2\, \int_{0}^{\tau_2}d\tau_3\, f(\tau_1,\tau_2,\tau_3) = \nonumber\\& = \int_{0}^{2\pi}d\tau_1 \int_{2\pi (n-2)}^{2\pi (n-1)}d\tau_2\, \int_{0}^{2\pi (n-2)}d\tau_3\, f(\tau_1,\tau_2,\tau_3) +
\nonumber\\&~~ +
\int_{0}^{2\pi}d\tau_1 \int_{2\pi (n-2)}^{2\pi (n-1)}d\tau_2\, \int_{2\pi (n-2)}^{\tau_2}d\tau_3\, f(\tau_1,\tau_2,\tau_3) +
\nonumber\\&~~ +
\int_{0}^{2\pi}d\tau_1 \int_{0}^{2\pi (n-2)}d\tau_2\, \int_{0}^{\tau_2}d\tau_3\, f(\tau_1,\tau_2,\tau_3)
\end{align}
As above we use periodicity to rewrite it as
\begin{equation}
G_{3,2}[f](m-1) = \int_{0}^{2\pi}d\tau_1 \int_{0}^{2\pi}d\tau_2\, \int_{0}^{\tau_2}d\tau_3\, f(\tau_1,\tau_2,\tau_3) + G_{3,2}[f](m-2)
\end{equation}
The first term in \eqref{eq:mercedesstep2} can in fact be discarded since it evaluates to 0, because it is the integral of an antisymmetric function over a symmetric domain.
The recursion relation can be rewritten as 
\begin{equation}
G_{3,2}[f](m-1) = G_{3}[f](1) + G_{3,2}[f](m-2)
\end{equation}
since the integrals are equivalent thanks to the symmetry properties of the integrand.
The solution of the iteration is obviously
\begin{equation}
G_{3,2}[f](m-1) = (m-1) G_{3}[f](1)
\end{equation}
Plugging these results into \eqref{eq:mercedesstep}, we find the recursion relation
\begin{align}
\left\{\begin{array}{l}
G_{3}[f](m) = (2m-1)\, G_{3}[f](1) + G_{3}[f](m-1)\\
G_{3}[f](1) = -\frac{\pi^2}{6}\, \lambda_1^2
\end{array}\right.
\end{align}
which we can easily solve to obtain
\begin{equation}
G_{3}[f](m) = -\frac{\pi^2}{6}\, m^2\, \lambda_1^2
\end{equation}
We now move to the last diagram at two loops, namely the double gluon exchange.
The symmetric part of the double exchange diagrams behaves simply in the multiple winding case.
Namely, since the contour is totally symmetric, it is enhanced by a power of $m^4$, leading to 
\begin{equation}
\raisebox{-0.8cm}{\includegraphics[width=2.cm]{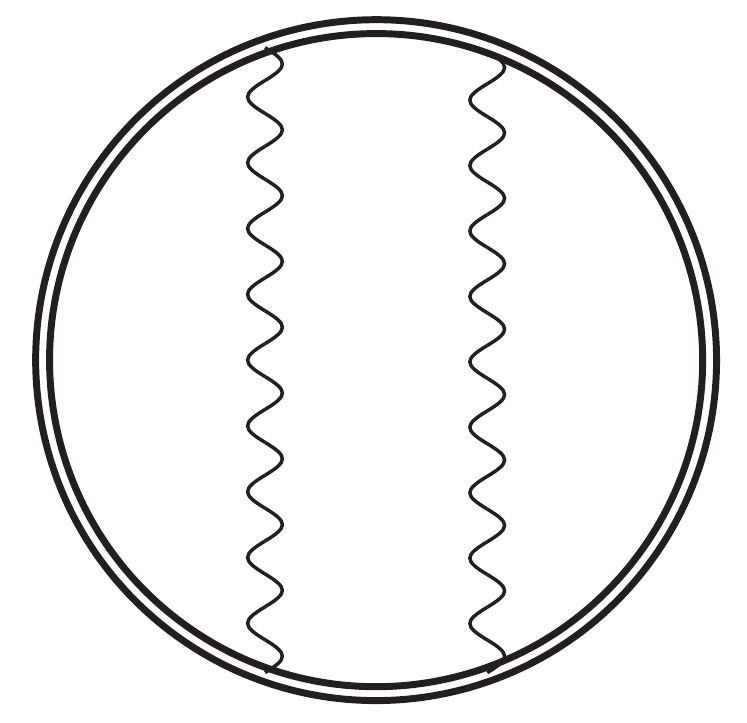}} + \raisebox{-0.8cm}{\includegraphics[width=2.cm]{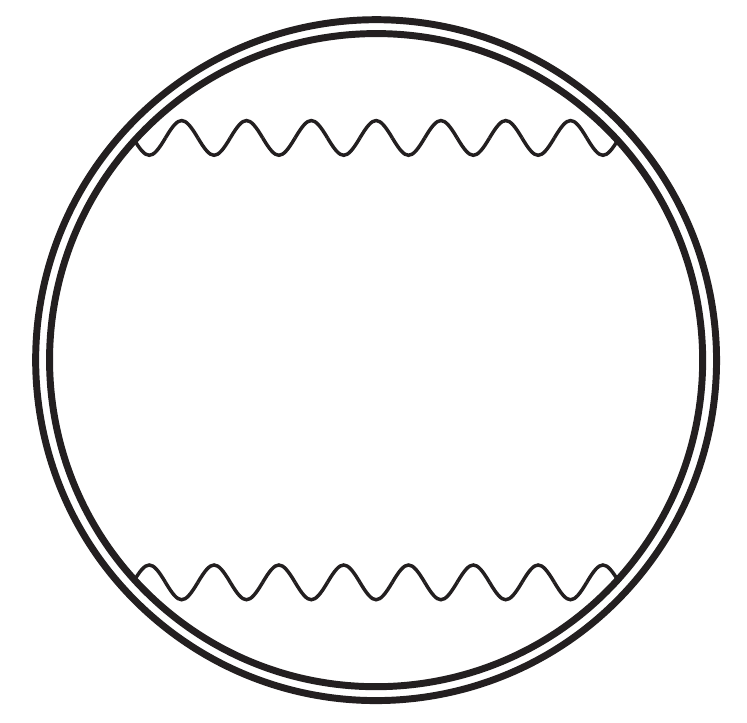}} + \raisebox{-0.8cm}{\includegraphics[width=2.cm]{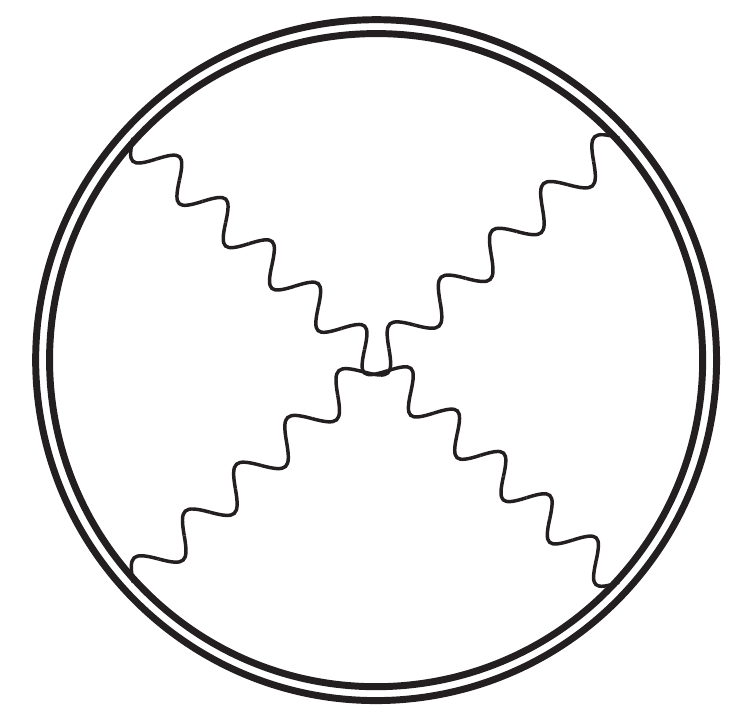}} = -\frac{\pi^2}{2}\, m^4\, f^2\, \lambda_1^2
\end{equation}
The novel feature of multiply wound Wilson loops is the contribution from the crossed diagram, as we now demonstrate.
We start with the integral
\begin{equation}
G_4[g](m) \equiv \raisebox{-0.8cm}{\includegraphics[width=2.cm]{2Lcrossm}} = \int_{0}^{2\pi m}d\tau_1 \int_{0}^{\tau_1}d\tau_2\, \int_{0}^{\tau_2}d\tau_3\, \int_{0}^{\tau_3}d\tau_4\, g(\tau_1,\tau_3)\, g(\tau_2,\tau_4)
\end{equation}
The novelty, as sketched at the end of section \ref{sec:contours}, consists in the fact that in the integration contour the endpoints are reshuffled in such a way that not only "crossed" contributions occur, but also the other terms, which have singularities at coinciding points and give contributions when regularized with framing.
We analyse the integral similarly as for the mercedes diagram, handling contours.
The analysis is a little more intricate, as a system of recursion relations is needed to solve for the combinatorics. Therefore we take this as an exemplary case to show how the algorithm of section \ref{sec:contours} works in a slightly non-trivial situation.
We start writing down the first recursion relation for the integral we are interested in
\begin{align}\label{eq:recursiondouble}
G_4[g](m) & = 
G_4[g](m-1) + G_4[g](1) + G_{4,3}[g](m-1) + G_{4,2}[g](m-1) + G_{4,1}[g](m-1)
\end{align}
where we encounter two new integrals to deal with.
Before tackling them we observe that the integral $G_4[g](1)$ is the crossed double gluon exchange for single winding which was proven to be framing independent and vanishing.
Therefore this term can be discarded.
Moreover, the integral $G_{4,1}[g](m-1)$, as all iterative integrals with a 1 index, is just a multiple of a single winding integral, more precisely
\begin{equation}
G_{4,1}[g](m-1) = (m-1)\, G_{4,1}[g](1)
\end{equation}
The latter integral can be dealt with decomposing the contour in four ordered pieces, according to the position of $\tau_4$ on the circle. Then we can relabel integration variables to rewrite everything in terms of the canonical integration contour $2\pi>\tau_4>\tau_3>\tau_2>\tau_1>0$, changing the integrand.
We obtain pictorially
\begin{equation}
G_{4,1}[g](1) = 2(m-1)\, \left[
 \raisebox{-0.55cm}{\includegraphics[width=1.5 cm]{2Lgauge2}} + \raisebox{-0.55cm}{\includegraphics[width=1.5cm]{2Lcross}} \right]
\end{equation}
Next we analyse the other contributions to the recursive relation.
In the second step of the contour decomposition we can write down the recursion relations
\begin{equation}
\left\{ \begin{array}{l}
G_{4,3}[g](m) = G_{4,3}[g](m-1) + G_{4,3}[g](1) + G_{4,3,2}[g](m-1) + G_{4,3,1}[g](m-1)\\
G_{4,2}[g](m) = G_{4,2}[g](m-1) + G_{4,2}[g](1) + G_{4,2,1}[g](m-1)
\end{array}\right.
\end{equation}
The latter relation is irreducible, for $G_{4,2,1}[g](m-1)$, as all iterative integrals with a 1 index, is just a multiple of a single winding integral, more precisely
\begin{equation}
G_{4,2,1}[g](m-1) = (m-1)\, G_{4,2,1}[g](1)
\end{equation}
Again, the latter integral in the first winding can be approached by dividing the contour into ordered ones, relabelling variables and using the symmetry properties of the integrand. In the end its recursion relation reads pictorially
\begin{align}
G_{4,2}[g](m) = & G_{4,2}[g](m-1) + 4(m-1) \left[ \raisebox{-0.55cm}{\includegraphics[width=1.5 cm]{2Lgauge}} + \raisebox{-0.55cm}{\includegraphics[width=1.5 cm]{2Lgauge2}} + \raisebox{-0.55cm}{\includegraphics[width=1.5cm]{2Lcross}} \right] + \nonumber\\& + 2 \left[ 2 \raisebox{-0.55cm}{\includegraphics[width=1.5 cm]{2Lgauge}} + \raisebox{-0.55cm}{\includegraphics[width=1.5cm]{2Lcross}} \right]
\end{align}
The solution of this relation yields
\begin{align}
G_{4,2}[g](m) =& 4 m \raisebox{-0.55cm}{\includegraphics[width=1.5 cm]{2Lgauge}} + 2 m \raisebox{-0.55cm}{\includegraphics[width=1.5cm]{2Lcross}} + \nonumber\\& + 2m(m-1) \left[ \raisebox{-0.55cm}{\includegraphics[width=1.5 cm]{2Lgauge}} + \raisebox{-0.55cm}{\includegraphics[width=1.5 cm]{2Lgauge2}} + \raisebox{-0.55cm}{\includegraphics[width=1.5cm]{2Lcross}} \right] 
\end{align}
The final term of the recursive relation \eqref{eq:recursiondouble} reads
\begin{equation}
G_{4,3}[g](m) \equiv \int_{0}^{2\pi}d\tau_1 \int_{0}^{2\pi (m-1)}d\tau_2\, \int_{0}^{\tau_2}d\tau_3\, \int_{0}^{\tau_3}d\tau_4
\end{equation}
Again by managing contours it can be given an iterative relation
\begin{align}\label{eq:recursiveL}
G_{4,3}[g](m) =& G_{4,3}[g](m-1) + 4(m-1) \left[ \raisebox{-0.55cm}{\includegraphics[width=1.5 cm]{2Lgauge}} + \raisebox{-0.55cm}{\includegraphics[width=1.5 cm]{2Lgauge2}} + \raisebox{-0.55cm}{\includegraphics[width=1.5cm]{2Lcross}} \right] + \nonumber\\& + 2 \left[ \raisebox{-0.55cm}{\includegraphics[width=1.5 cm]{2Lgauge2}} + \raisebox{-0.55cm}{\includegraphics[width=1.5cm]{2Lcross}} \right] + G_{4,3,2}[g](m-1)
\end{align}
where, as the final step of the decomposition, $G_{4,3,2}[g](m)$ is defined itself by a recursive relation
\begin{align}
G_{4,3,2}[g](m) =& G_{4,3,2}[g](m-1) + 8(m-1) \left[ \raisebox{-0.55cm}{\includegraphics[width=1.5 cm]{2Lgauge}} + \raisebox{-0.55cm}{\includegraphics[width=1.5 cm]{2Lgauge2}} + \raisebox{-0.55cm}{\includegraphics[width=1.5cm]{2Lcross}} \right] + \nonumber\\& + 4 \left[ \raisebox{-0.55cm}{\includegraphics[width=1.5 cm]{2Lgauge}} + \raisebox{-0.55cm}{\includegraphics[width=1.5 cm]{2Lgauge2}} + \raisebox{-0.55cm}{\includegraphics[width=1.5cm]{2Lcross}} \right]
\end{align}
Hence we can first solve
\begin{equation}
G_{4,3,2}[g](m) = 4m^2 \left[ \raisebox{-0.55cm}{\includegraphics[width=1.5 cm]{2Lgauge}} + \raisebox{-0.55cm}{\includegraphics[width=1.5 cm]{2Lgauge2}} + \raisebox{-0.55cm}{\includegraphics[width=1.5cm]{2Lcross}} \right]
\end{equation}
then plug it into \eqref{eq:recursiveL} and solve it to obtain
\begin{align}
G_{4,3}[g](m) =& 2m \left[ \raisebox{-0.55cm}{\includegraphics[width=1.5 cm]{2Lgauge2}} + \raisebox{-0.55cm}{\includegraphics[width=1.5cm]{2Lcross}} \right] + \nonumber\\&
+ \frac43\, (m+1)m(m-1) \left[ \raisebox{-0.55cm}{\includegraphics[width=1.5 cm]{2Lgauge}} + \raisebox{-0.55cm}{\includegraphics[width=1.5 cm]{2Lgauge2}} + \raisebox{-0.55cm}{\includegraphics[width=1.5cm]{2Lcross}} \right]
\end{align}
Plugging all these results into \eqref{eq:recursiondouble} we can rewrite the recursive relation as
\begin{equation}
G_4[g](m) = G_4[g](m-1) + \frac23\, m(m-1)(m-2) \left[ \raisebox{-0.55cm}{\includegraphics[width=1.5 cm]{2Lgauge}} + \raisebox{-0.55cm}{\includegraphics[width=1.5 cm]{2Lgauge2}} + \raisebox{-0.55cm}{\includegraphics[width=1.5cm]{2Lcross}} \right]
\end{equation}
after discarding extra vanishing crossed contributions.
We can solve it to have
\begin{equation}
G_4[g](m) = \frac13\, m^2(m^2-1) \left[ \raisebox{-0.55cm}{\includegraphics[width=1.5 cm]{2Lgauge}} + \raisebox{-0.55cm}{\includegraphics[width=1.5 cm]{2Lgauge2}} + \raisebox{-0.55cm}{\includegraphics[width=1.5cm]{2Lcross}} \right]
\end{equation}
Summarizing, up to two loops the diagrams evaluate for general winding $m$ and framing $f$
\begin{align}
\raisebox{-0.8cm}{\includegraphics[width=2.cm]{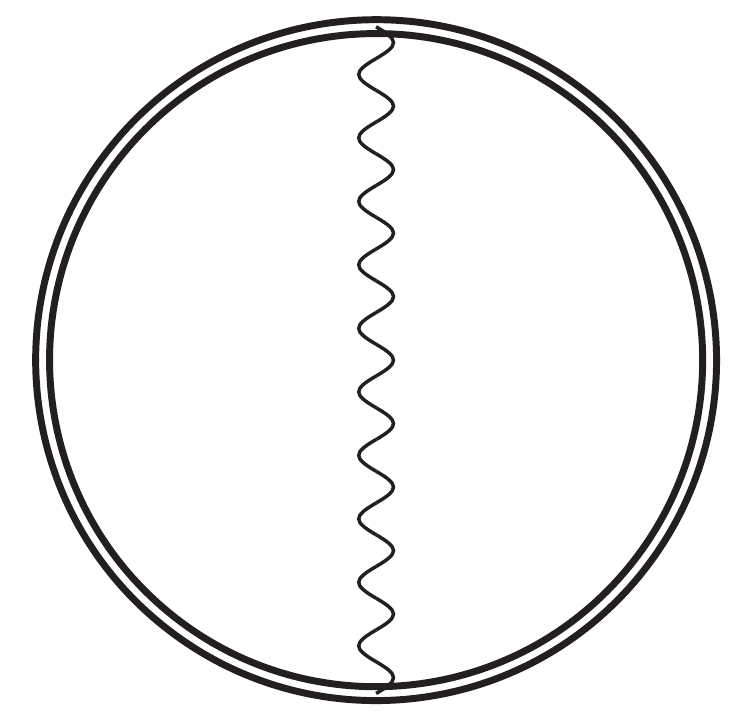}} &= i\, \pi\, m^2\, f\, \lambda_1 \label{eq:1loopn} \\
\raisebox{-0.8cm}{\includegraphics[width=2.cm]{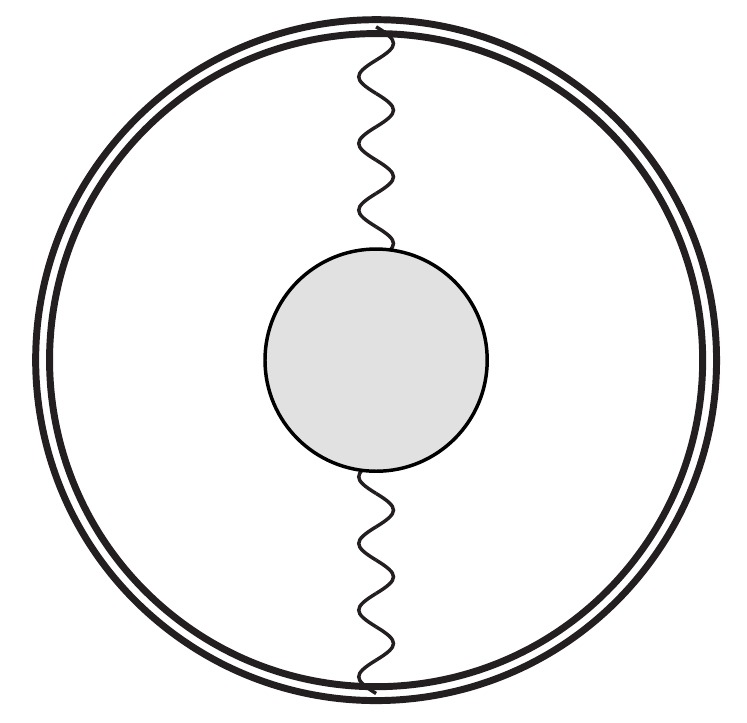}} + \raisebox{-0.8cm}{\includegraphics[width=2.cm]{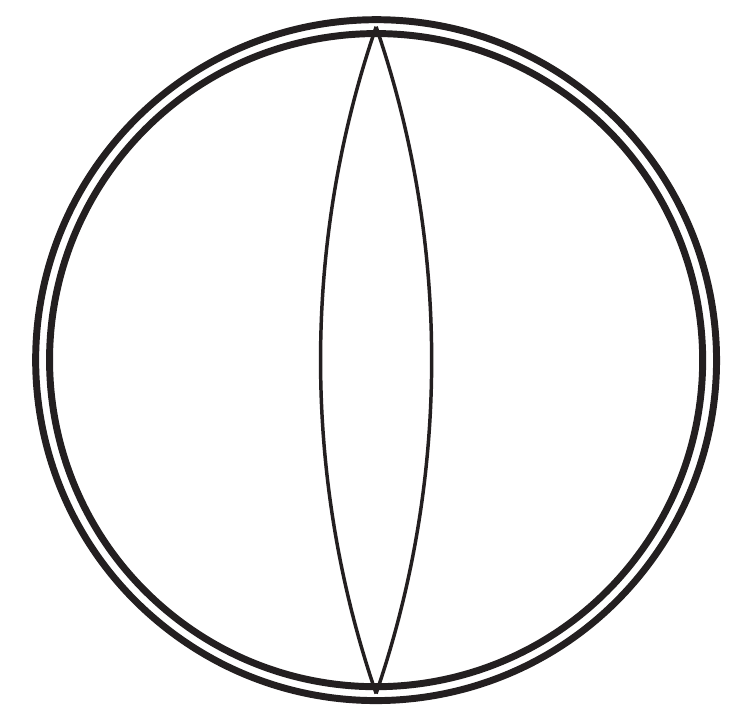}} &= \pi^2\, m^2\, \lambda_1\lambda_2 \label{eq:matterm}\\
\raisebox{-0.8cm}{\includegraphics[width=2.cm]{2LgaugeMBm}} &= -\frac{\pi^2}{6}\, m^2\, \lambda_1^2 \label{eq:mercedesm} \\
\raisebox{-0.8cm}{\includegraphics[width=2.cm]{2Lgaugem}} + \raisebox{-0.8cm}{\includegraphics[width=2.cm]{2Lgauge2m}} + \raisebox{-0.8cm}{\includegraphics[width=2.cm]{2Lcrossm}} &= \frac12\, \raisebox{-0.8cm}{\includegraphics[width=2.cm]{1Lgaugem}} ^2 = -\frac{\pi^2}{2}\, m^4\, f^2\, \lambda_1^2 \label{eq:symmn} \\
-\raisebox{-0.8cm}{\includegraphics[width=2.cm]{2Lcrossm}} &= \frac{\pi^2}{6}\, m^2(m^2-1)\, f^2\, \lambda_1^2 \label{eq:crossedn}
\end{align}
where here and in the rest of the paper we use the double line on the contour to indicate multiple winding, as opposed to the single line notation.
Summing up we obtain
\begin{equation}
\langle W_m^{1/6} \rangle_f = 1 + i\, \pi\, m^2\, f\, \lambda_1 + \frac{\pi^2}{6}\left(-m^2(1+f^2+2\,f^2\,m^2)\, \lambda_1^2 + 6\, m^2\, \lambda_1\lambda_2 \right) + {\cal O}\left(k^{-3}\right)
\end{equation}
For $f=1$ it is in agreement with the localization result \eqref{eq:localization}.

\subsection{1/6-BPS Wilson loop at three loops}\label{sec:3loops}

In this section we push the computation to three loops in order to verify agreement with the localization prediction.
It is easier to consider ABJM theory with different ranks and separate contributions according to their color factor.
Building on previous results for the usual 1/6-BPS Wilson loop, we start with the color structure $N_1 N_2^2$. As shown in \cite{Bianchi:2016yzj} the only contribution to this part comes from the two-loop matter corrections to the gluon self-energy.
This has the same structure as the tree level propagator and hence the same $m^2$ behaviour at $m$ windings
\begin{equation}
\raisebox{-0.8cm}{\includegraphics[width=2.cm]{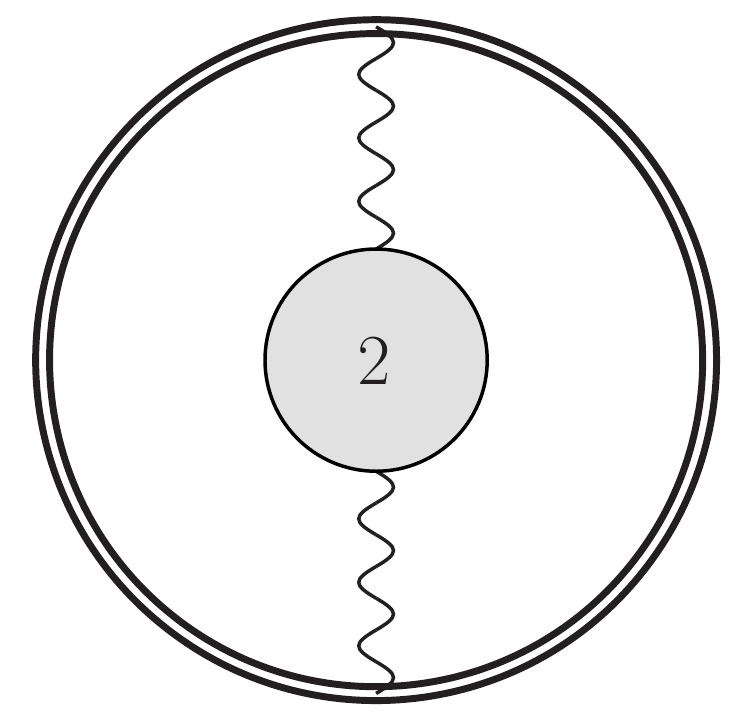}} = -i\, \frac{\pi^3}{2}\, f\, m^2\, \lambda_1\lambda_2^2
\end{equation}
For the singly wound Wilson loop the contribution to the $N_1^2N_2$ structure only comes from the factorization of the one-loop diagram and the two-loop matter diagrams whose combination we represent pictorially
\begin{equation}
\raisebox{-0.8cm}{\includegraphics[width=2.cm]{2Lgauge1}} + \raisebox{-0.8cm}{\includegraphics[width=2.cm]{2Lscalar}} \equiv \raisebox{-0.8cm}{\includegraphics[width=2.cm]{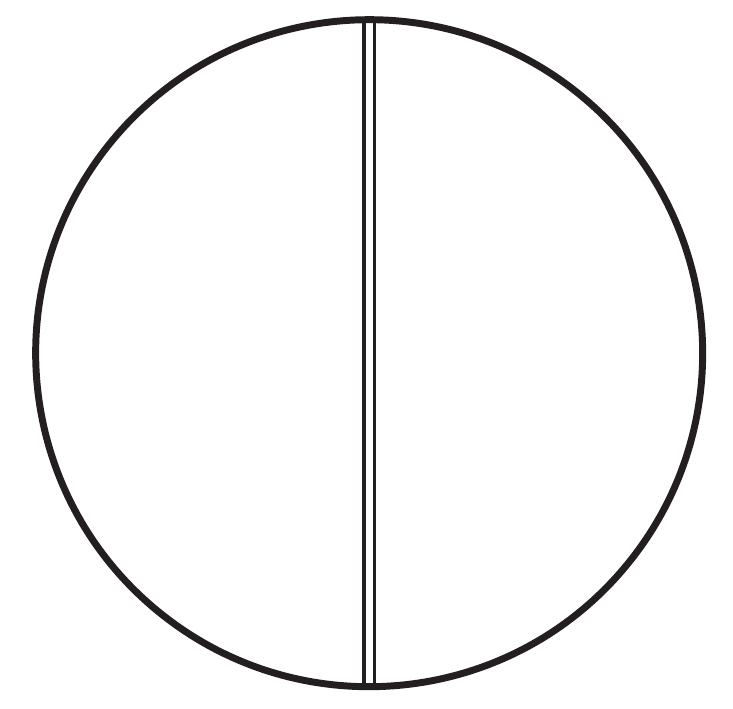}}
\end{equation}
We assume that this holds true also for multiple windings. Then the same analysis as for the two-loop double exchange of gauge vectors can be carried out. Namely, adding and subtracting (twice) the crossed exchange (which would be subleading in color), the sum can be symmetrized giving rise to the factorization
\begin{align}
& \phantom{+2} \raisebox{-0.8cm}{\includegraphics[width=2.cm]{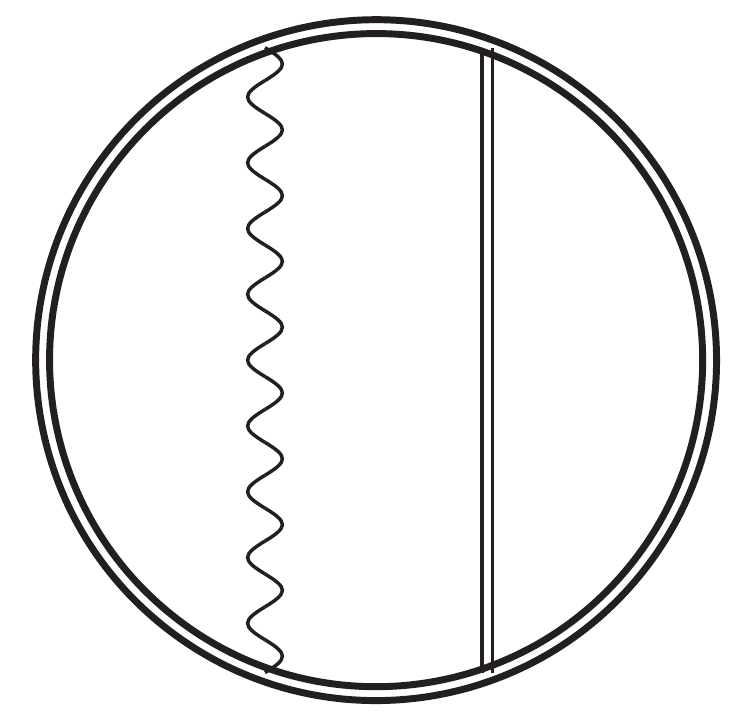}} + \raisebox{-0.8cm}{\includegraphics[width=2.cm]{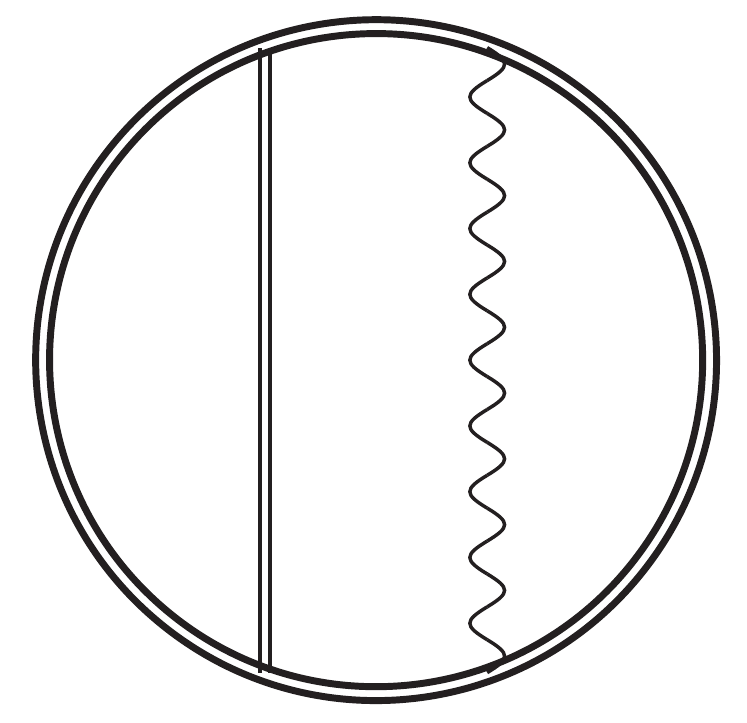}} + \raisebox{-0.8cm}{\includegraphics[width=2.cm]{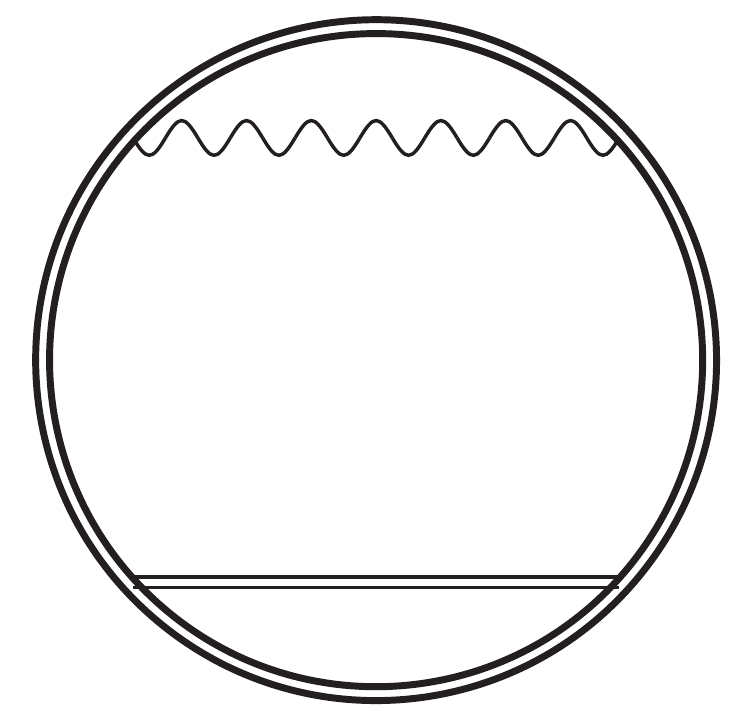}} + \raisebox{-0.8cm}{\includegraphics[width=2.cm]{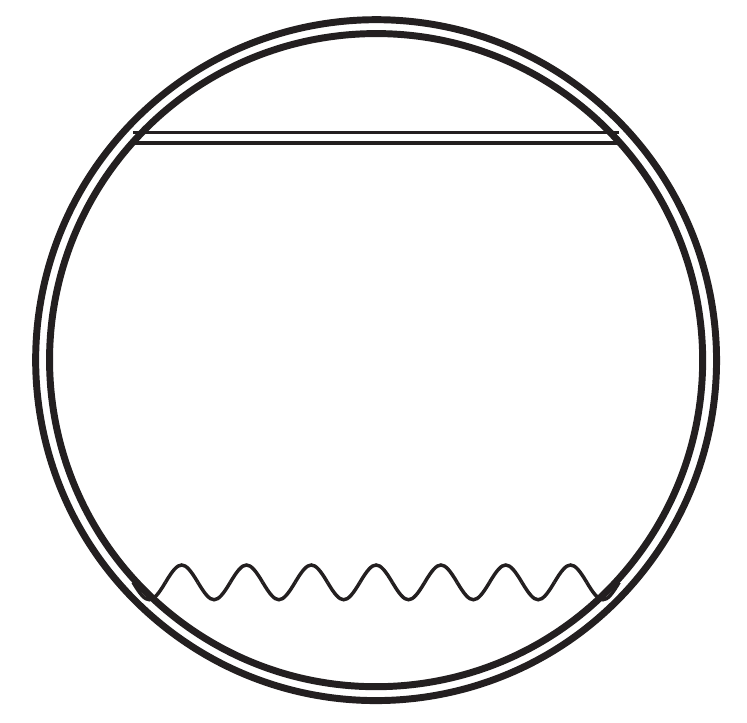}} + \nonumber\\& + \raisebox{-0.8cm}{\includegraphics[width=2.cm]{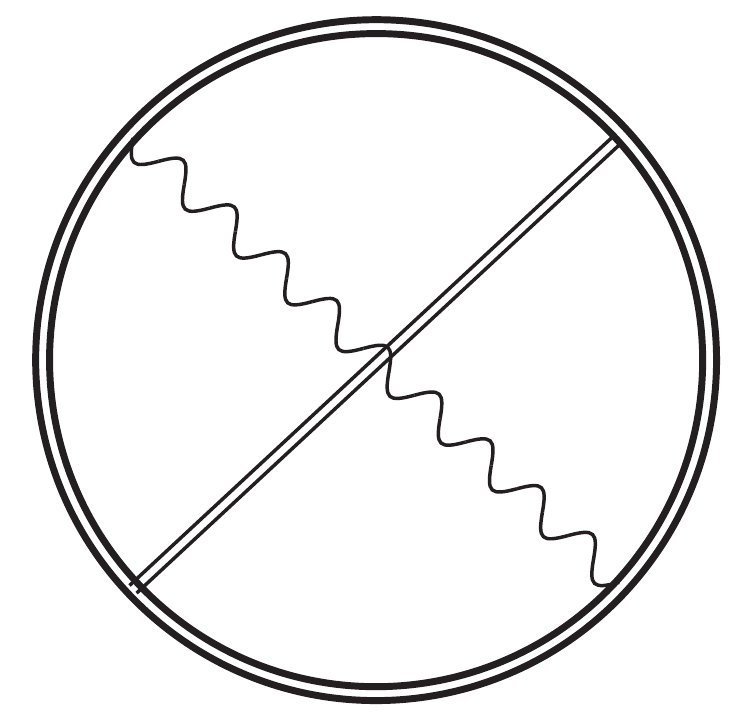}} + \raisebox{-0.8cm}{\includegraphics[width=2.cm]{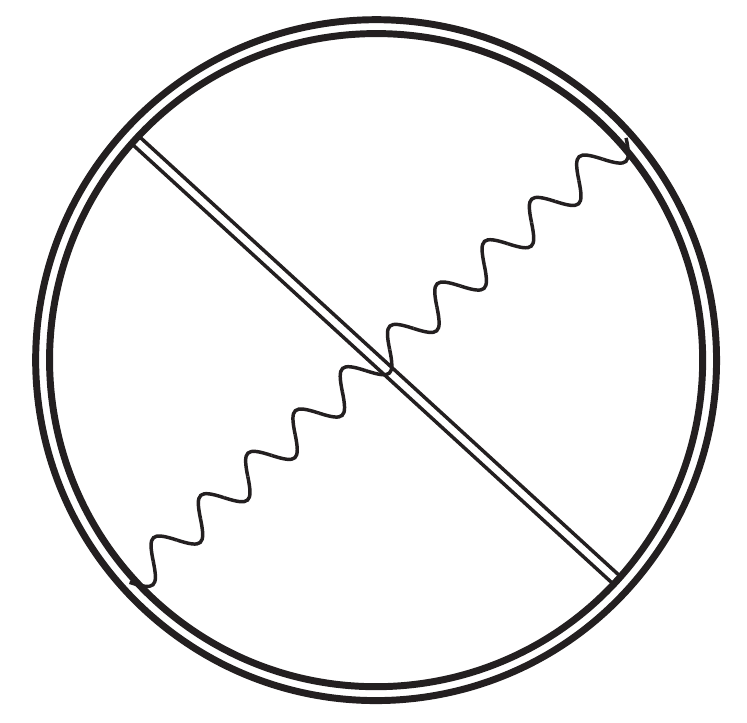}} =  \raisebox{-0.8cm}{\includegraphics[width=2.cm]{1Lgaugem}} \times \raisebox{-0.8cm}{\includegraphics[width=2.cm]{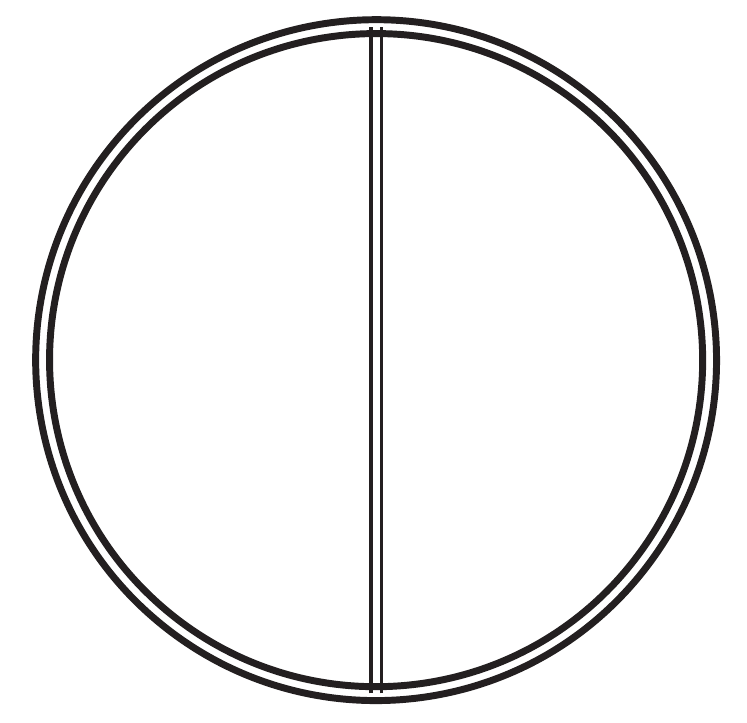}} = i\, \pi^3\, m^2\,f\, \lambda_1^2\lambda_2
\end{align}
whereas the subtracted crossed contribution can be given the same recursive relation as for the gluon double exchange, pictorially
\begin{equation}
- \raisebox{-0.8cm}{\includegraphics[width=2.cm]{mixed5m}} - \raisebox{-0.8cm}{\includegraphics[width=2.cm]{mixed6m}} = - i\, \frac{\pi^3}{3}\, m^2 (m^2-1)\, f\, \lambda_1^2\lambda_2
\end{equation}
The total contribution for this diagram reads
\begin{equation}
\langle W_m^{1/6} \rangle_f^{(3)} \bigg|_{N_1^2N_2} = i\, \frac{\pi^3}{3}\, m^2 (2m^2+1)\, f\, \lambda_1^2\lambda_2
\end{equation}
and agrees with the $N_1^2N_2$ contribution to the localization three-loop term. This implies that all other potential contributions to this color structure, namely the two-loop corrections to the gauge propagator with color $N_1 N_2$ and framing dependent interaction diagrams have to cancel each other and drop in the final answer, as suggested for single winding in \cite{Bianchi:2016yzj}.

Finally, the most involved three-loop contribution is the one proportional to $N_1^3$. This is obviously given by pure Chern-Simons interactions, and relying on previous analysis we expect it to be produced entirely by the triple gluon exchanges and the mercedes diagram with an additional free gluon propagator inserted.
For these cases we implemented the algorithm of section \ref{sec:contours} on \emph{Mathematica} to construct and solve systems of recursion relations.
The result for the mercedes diagram with an additional insertion of a gauge propagator reads
\begin{equation}\label{eq:3L5}
\raisebox{-0.8cm}{\includegraphics[width=2.cm]{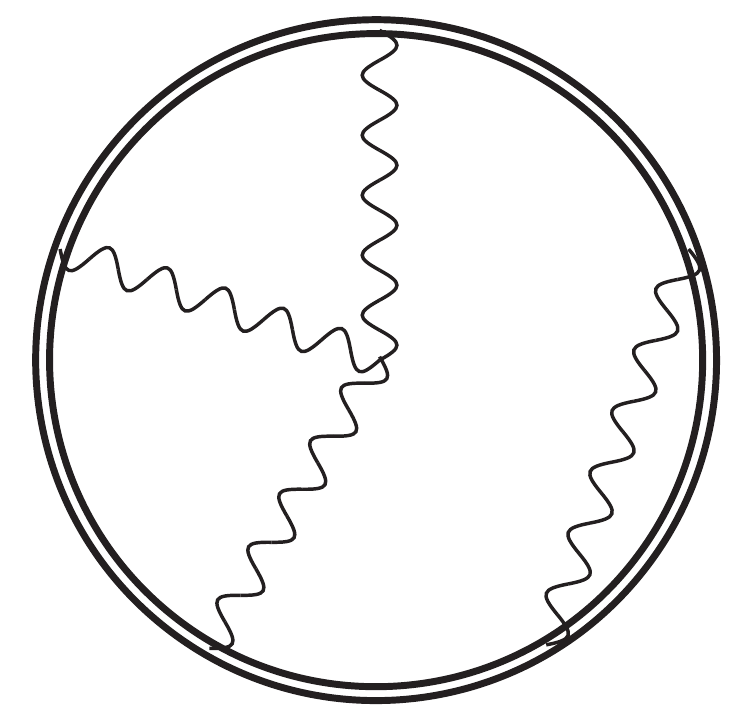}} + perms = -\frac{1}{18} i f m^2 \left(1+2 m^2\right) \pi ^3\, \lambda_1^3
\end{equation}
whereas that for the triple gluon exchange gives
\begin{equation}\label{eq:3L6}
\raisebox{-0.8cm}{\includegraphics[width=2.cm]{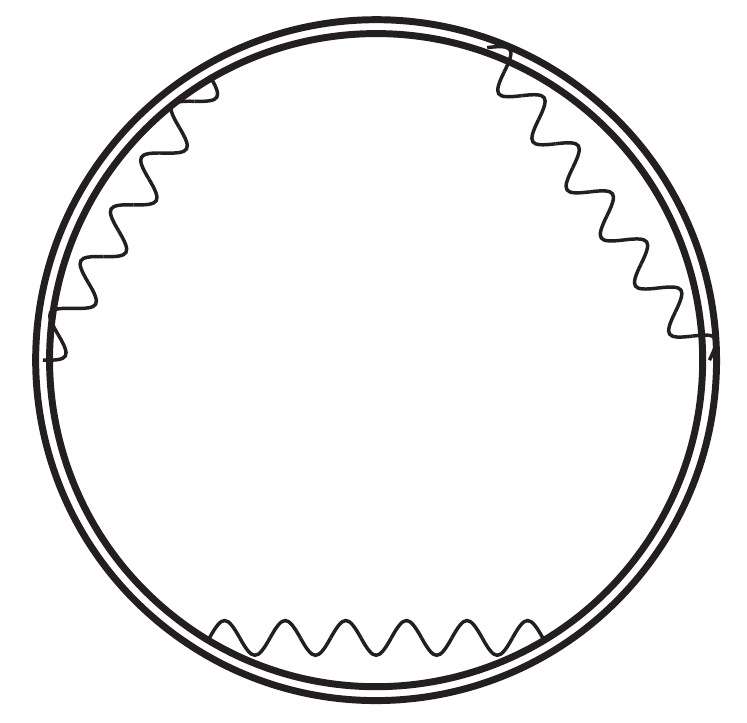}} + perms = -\frac{1}{18} i f^3 m^4 \left(2+m^2\right) \pi ^3\, \lambda_1^3
\end{equation}
The details of this computation and the explicit systems of recursive relations are spelled out in appendix \ref{app:3loop}.
Altogether we find the three-loop expectation value
\begin{align}\label{eq:result}
\langle W_{m}^{1/6} \rangle_f &= 1 + i\, \pi\, m^2\, f\, \lambda_1 + \frac{\pi^2}{6}\left(-m^2(1+f^2+2\,f^2\,m^2)\, \lambda_1^2 + 6\, m^2\, \lambda_1\lambda_2 \right) + \nonumber\\& -  i\, \frac{\pi^3}{18}\, f\, m^2 \left[\left(1+m^2 \left(2+f^2 \left(2+m^2\right)\right)\right) \lambda_1^3 -6(1+2m^2) \lambda_1^2\lambda_2 + 9\, \lambda_1\lambda_2^2\right] + {\cal O}\left(k^{-4}\right)
\end{align}
We remark that at framing 1 \eqref{eq:result} is in agreement with the localization result \eqref{eq:localization} and provides the highest order perturbative check of it.
We stress that for multiple winding the framing dependence of the 1/6-BPS Wilson loop is not captured by an overall phase and in particular a comparison to the perturbative computation at framing 0 cannot be performed by simply taking the modulus of \eqref{eq:localization}, on the contrary it entailed deriving the perturbative results directly at framing 1.

Focussing on the pure Chern-Simons part, which can be clearly obtained by sending $\lambda_2\to 0$
\begin{align}\label{eq:limitCS}
\langle W^{CS}_{m} \rangle_f &= 1 + i\pi\, f\, m^2\, \lambda_1 - \frac{\pi^2}{6}\, f^2\, m^2(2m^2+1)\, \lambda_1^2 + \nonumber\\& - i\, \frac{\pi ^3}{18}\, f\, m^2 \left(1+2 \left(1+f^2\right) m^2+f^2 m^4\right)\, \lambda_1^3 + {\cal O}\left(\lambda_1^4\right)
\end{align}
we can check that it coincides with the general result of the unknot in pure $U(N)$ Chern-Simons at winding $m$ at framing $f$ obtained in \cite{Brini:2011wi}
\begin{equation}
\langle W^{CS}_{m} \rangle_f = \frac{1}{N}\, \sum_{l=0}^m\, e^{i\pi \frac{N}{k} (2l+mf-m) }\, (-1)^{m+l}\, \frac{[mf+l-1]!}{[m-l]![l]![mf-f+l]!}
\end{equation}
where $[n]$ is the $q$-number
\begin{equation}
[n] \equiv e^{q/2} - e^{-q/2}
\end{equation}
and $[n]!$ is the $q$-factorial
\begin{equation}
[n]! \equiv [n][n-1]\dots[1]
\end{equation}
For our purposes we set $q=e^{\frac{2\pi i}{k}}$, without the shift of the Chern-Simons level by the quadratic Casimir of the gauge group \cite{Witten}, as this is absent for supersymmetric Chern-Simons theories \cite{Kapustin:2009kz}.
Indeed expanding this formula perturbatively to three loops we find agreement with \eqref{eq:limitCS}.

\subsection{1/2-BPS Wilson loop to two loops}\label{sec:halfbps}

We compute the expectation value of the 1/2-BPS Wilson loop perturbatively at framing 0, following the analysis of \cite{Bianchi:2013zda,Bianchi:2013rma,Griguolo:2013sma}.
The bosonic diagrams are effectively the same as for the 1/6-BPS case, albeit the different scalar coupling matrix $M_{1/2}$
\begin{align}
\raisebox{-0.8cm}{\includegraphics[width=2.cm]{1Lgaugem}} &= i\, \pi\, m^2\, f\, \left(\lambda_1^2 + (-1)^m \lambda_2^2\right) \label{eq:1loopn2} \\
\raisebox{-0.8cm}{\includegraphics[width=2.cm]{2Lgauge1m}} + \raisebox{-0.8cm}{\includegraphics[width=2.cm]{2Lscalarm}} &= \pi^2\, m^2\, \lambda_1\lambda_2 \left(\lambda_1 - (-1)^m \lambda_2\right) \label{eq:matterm2}\\
\raisebox{-0.8cm}{\includegraphics[width=2.cm]{2LgaugeMBm}} &= -\frac{\pi^2}{6}\, m^2\, \left(\lambda_1^3 - (-1)^m \lambda_2^3\right) \label{eq:mercedesm2} \\
\raisebox{-0.8cm}{\includegraphics[width=2.cm]{2Lgaugem}} + \raisebox{-0.8cm}{\includegraphics[width=2.cm]{2Lgauge2m}} & = -\frac{\pi^2}{6}\, m^2(2m^2+1)\, f^2\, \left(\lambda_1^3 - (-1)^m \lambda_2^3\right) \label{eq:}
\end{align}
The fermionic diagrams have to be treated in a slightly more delicate way, because of the additional matrix in \eqref{eq:defhalf}, enforcing gauge invariance.
For the double fermion exchange, as for the bosonic diagrams, the effect of this is just a $(-1)^m$ additional factor in the coupling constants and the combinatorics associated to multiple winding of the integrals
\begin{equation}
\raisebox{-0.8cm}{\includegraphics[width=2.cm]{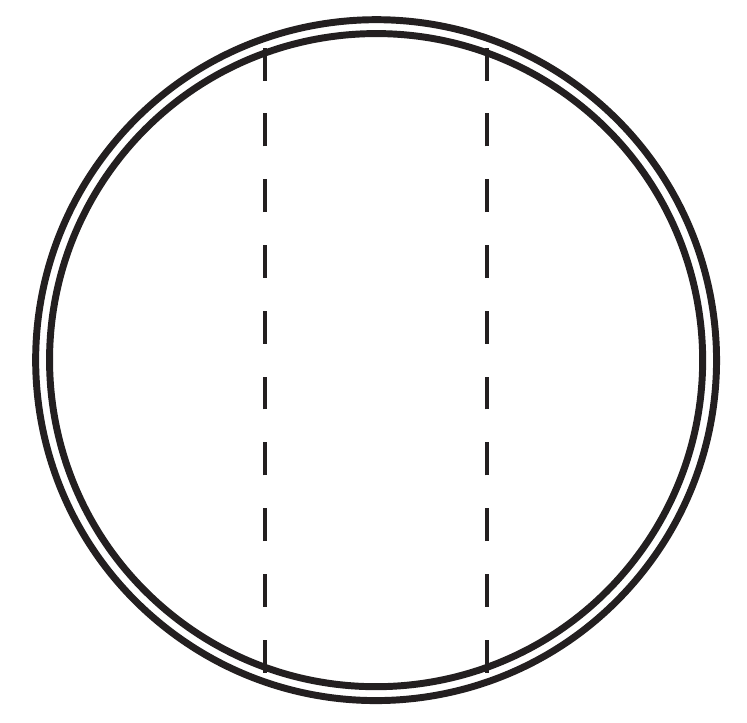}} + \raisebox{-0.8cm}{\includegraphics[width=2.cm]{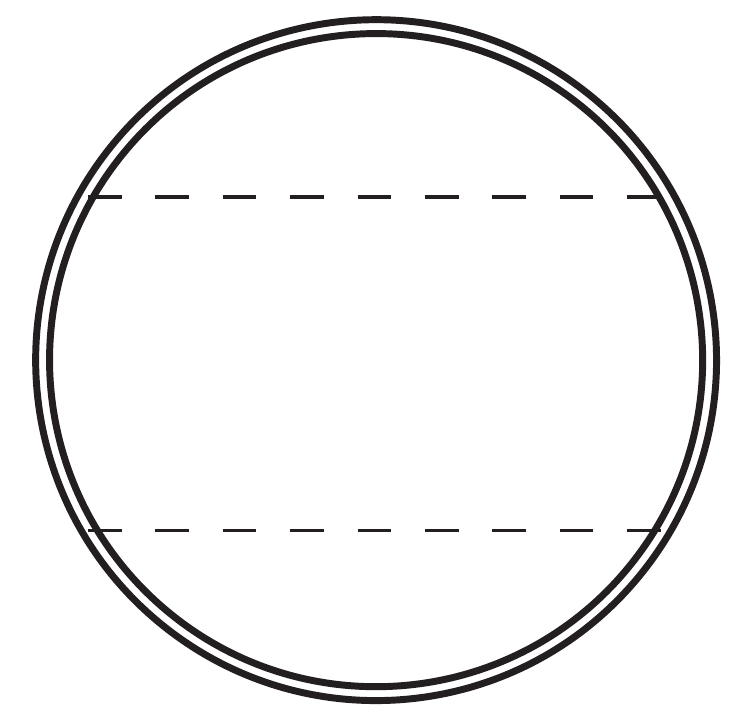}} = \frac{\pi^2}{2}\, m^2(2m^2+1)\, \lambda_1\lambda_2 \left(\lambda_1 - (-1)^m \lambda_2\right) + {\cal O}(f)
\end{equation}
For the remaining diagrams, the one-loop corrected fermion and the mercedes, the effect of that matrix is more intricate. 
The algebra of the former diagram reads
\begin{equation}\label{eq:fermioncorrected}
\raisebox{-0.8cm}{\includegraphics[width=2.cm]{2Lfermion}} = \frac{2\pi}{k^2} \,  \int_0^{2\pi} d\tau_1 \int_0^{\tau_1} d\tau_2\,  {\rm Tr} \left[ \eta_{1 I} \bar{\eta}_2^J 
\langle  \bar{\psi}_1^I \psi_{2 J} \rangle^{(1)} - (-1)^m \bar{\eta}_1^I \eta_{2 J} \langle \psi_{1\, I} \bar{\psi}_2^J \rangle^{(1)} \right] \, |\dot{x}_1| |\dot{x}_2|
\end{equation}
whereas that of the latter yields
\begin{align}
\label{eq:3expansion}
\raisebox{-0.8cm}{\includegraphics[width=2.cm]{2Lfermionmercedes}} &= - i\, \frac{2\pi}{k^2}\, \int d\tau_{1>2>3}\,
{\rm Tr} 
\Big\{ \left[ \eta_{2  I} \bar{\eta}_3^J  \, \langle A_{1\mu} \bar{\psi}_2^I \psi_{3  J} \rangle \, \dot{x}_1^\mu  
~+~ \bar{\eta}_3^I \eta_{1 J}  \, \langle \bar{\psi}_1^J \hat{A}_{2 \mu} \psi_{3 I}  \rangle \, \dot{x}_2^\mu + \right.\nonumber\\& \left.
~+~ \eta_{1 I} \bar{\eta}_2^J  \, \langle  \bar{\psi}_1^I \psi_{2 J} A_{3 \mu} \rangle \, \dot{x}_3^\mu  \right] ~-~  (-1)^m \left[  \bar{\eta}_2^I \eta_{3 J}    \, \langle   \hat{A}_{1 \mu}  \psi_{2 I} \bar{\psi}_3^J    \rangle \, \dot{x}_1^\mu   + \right.\nonumber\\& \left. 
~+~ \eta_{3 I}  \bar{\eta}_1^J   \, \langle  \psi_{1 I} A_{2\mu}   \bar{\psi}_3^J   \rangle \, \dot{x}_2^\mu  
~+~ \bar{\eta}_1^I \eta_{2 J}  \, \langle \psi_{1  I} \bar{\psi}_2^J \hat{A}_{3\mu} \rangle \, \dot{x}_3^\mu \right] \,  \Big\} 
\end{align}
We see that for even and odd $m$ the relative signs of the contributions change and we have to treat the cases of odd and even winding separately.

\paragraph{Odd winding number}
In the former situation the integrands are the same as in the singly wound case, in particular that of the corrected fermion diagram vanishes identically.
Therefore we can borrow the results of \cite{Bianchi:2013zda,Bianchi:2013rma,Griguolo:2013sma}, adapt the coupling and carry out the combinatorics associated to multiple winding and obtain
\begin{align}
\left.\raisebox{-0.8cm}{\includegraphics[width=2.cm]{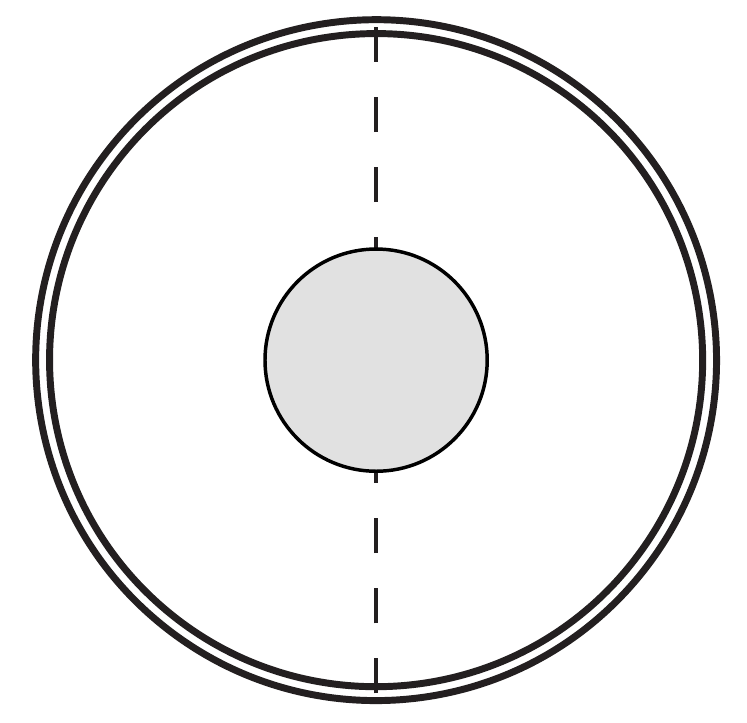}}\right|_{m\in 2\mathbb{Z}+1} &= 0 \nonumber\\
\left.\raisebox{-0.8cm}{\includegraphics[width=2.cm]{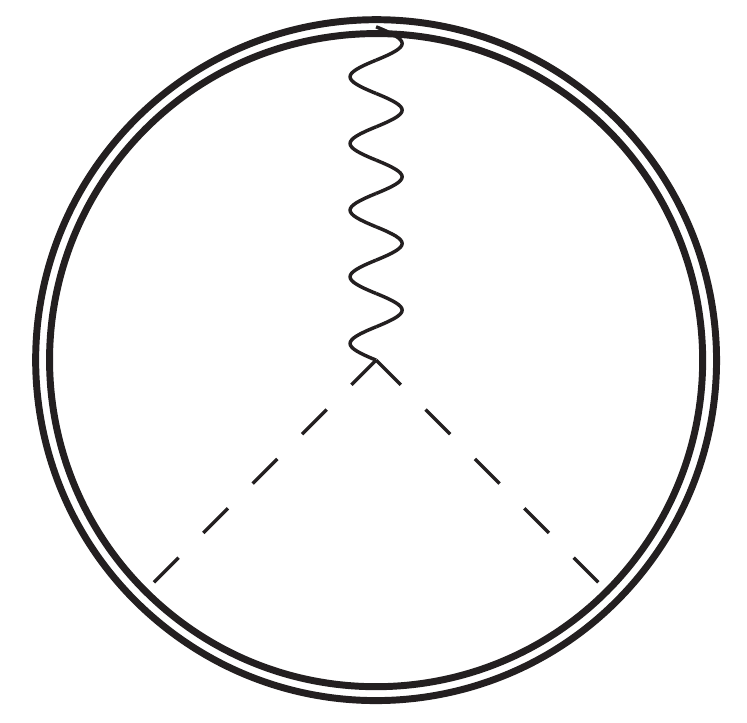}}\right|_{m\in 2\mathbb{Z}+1} &= -2\pi^2\, m^2\, \lambda_1\lambda_2 \left(\lambda_1 - (-1)^m \lambda_2\right) + {\cal O}(f)
\end{align}
In the last formula we have left the explicit dependence on $m$ for future reference, though it trivially produces the color structure $\lambda_1+\lambda_2$ for odd winding.

\paragraph{Even winding number}
At even winding the matrix in \eqref{eq:defhalf} changes the relative signs between the contributions of the two blocks. A similar phenomenon was observed in the computation of the expectation value of the latitude Wilson loop \cite{Bianchi:2014laa}. Unfortunately, the results of \cite{Bianchi:2013zda,Bianchi:2013rma,Griguolo:2013sma} are not thoroughly applicable to this case, as they exploited combining different terms in several steps.
Hence we perform a new computation for these contributions in this paper. We use the conventions and techniques of \cite{Bianchi:2013zda,Bianchi:2013rma,Griguolo:2013sma} to which we refer for more details. In particular we remark that we are working at framing 0 and regularizing intermediate divergences via dimensional reduction ($d=3-2\epsilon$), which proved suitable for these kinds of computation \cite{Bianchi:2013zda,Bianchi:2013rma,Griguolo:2013sma,Bianchi:2013pva}.
We start with the fermion exchange with a one-loop correction.
Notice the crucial different relative sign in \eqref{eq:fermioncorrected}, as $m$ is even, preventing them from cancelling each other. After introducing the explicit expressions for the relevant objects, we find the contribution
\begin{equation}
\left.\raisebox{-0.8cm}{\includegraphics[width=2.cm]{2Lfermion}}\right|_{m\in 2\mathbb{Z}} = 
\left(\frac{2\pi}{k}\right)^2 \,(\lambda_1-\lambda_2)N_1N_2\, \frac{\Gamma^2\left(\tfrac12-\epsilon\right)}{(4\pi)^{1-2\epsilon}}\, \int_0^{2\pi} d\tau_1 \int_0^{\tau_1} d\tau_2\, \frac{\cos \frac{\tau_{12}}{2}}{\left(\sin^2 \frac{\tau_{12}}{2}\right)^{1-2\epsilon}}
\end{equation}
Performing the integral we obtain
\begin{equation}
\left.\raisebox{-0.8cm}{\includegraphics[width=2.cm]{2Lfermion}}\right|_{m\in 2\mathbb{Z}} = 
\left( \lambda_1-\lambda_2\right) \lambda_1\lambda_2\, \frac{\Gamma^2\left(\tfrac12-\epsilon\right)}{(4\pi)^{1-2\epsilon}}
\left[-\frac{2}{\epsilon }-8 (1+\log 2)+{\cal O}\left(\epsilon\right)\right]
\end{equation}
This term looks particularly ugly. Not only it is divergent, but it also possesses a lower degree of transcendentality with respect to the expected order of two-loop contributions.

We now move to the mercedes diagram. From \eqref{eq:3expansion} and using the symmetries of the relevant objects we obtain
\begin{align}
\label{eq:beginningf}
\left. \raisebox{-0.8cm}{\includegraphics[width=2.cm]{2Lfermionmercedes}} \right|_{m\in 2\mathbb{Z}} &= -\lambda_1\lambda_2\left(\lambda_1-\lambda_2\right)\, (2\pi)^2\, \int d\tau_{1>2>3} \, 
\Big[ \left( \eta_1 \gamma_\lambda\gamma^\mu\gamma_\nu \bar\eta_2 \right) \, \varepsilon_{\mu\rho\sigma} \, \dot x_3^\rho\,  \Gamma^{\nu\lambda\sigma} + 
\nonumber\\&
+ \left( \eta_2 \gamma_\lambda\gamma^\mu\gamma_\nu \bar\eta_3 \right) \, \varepsilon_{\mu\rho\sigma} \, \dot x_1^\rho\, \Gamma^{\sigma\lambda\nu} 
- \left( \eta_1 \gamma_\lambda\gamma^\mu\gamma_\nu \bar\eta_3 \right) \, \varepsilon_{\mu\rho\sigma} \, \dot x_2^\rho\,  \Gamma^{\lambda\sigma\nu} \Big] 
\end{align}
where we have defined
\begin{equation}\label{eq:gammaintegrals}
\Gamma^{\mu\nu\rho} \equiv \,
\left(\frac{\Gamma\left( \tfrac12 - \epsilon \right)}{4\pi^{\tfrac32-\epsilon}}\right)^3\, \partial^{\mu}_1\, \partial^{\nu}_2\, \partial^{\rho}_3\, \int \frac{d^3 x}{[ (x-x_1)^2 (x-x_2)^2 (x-x_3)^2]^{\frac12 -\e}} 
\end{equation}
The strategy we choose here consists in the following: using the symmetries of the various pieces, we combine them extracting a part which coincides with the contribution at odd winding, plus the remainder
\begin{equation}
R = 2\,\lambda_1\lambda_2\left(\lambda_1-\lambda_2\right)\, (2\pi)^2\, \int d\tau_{1>2>3} \, \left( \eta_1 \gamma_\lambda\gamma^\mu\gamma_\nu \bar\eta_3 \right) \, \varepsilon_{\mu\rho\sigma} \, \dot x_2^\rho\,  \Gamma^{\lambda\sigma\nu}
\end{equation}
This evaluates more explicitly
\begin{align}
R &= 2\,\lambda_1\lambda_2\left(\lambda_1-\lambda_2\right)\, (2\pi)^2\, \int d\tau_{1>2>3} \, \Big[ -\left( \eta_1 \gamma^{\mu} \bar\eta_3 \right)\, \varepsilon_{\mu\rho\sigma}\, \dot x_{2}^{\rho}\, \Gamma^{\nu\sigma}_{\phantom{\sigma\nu}\nu} + \nonumber\\&
-i\, \left( \eta_1 \bar\eta_3 \right)\, \dot x_{2\sigma}\, \left( \Gamma^{\sigma\nu}_{\phantom{\sigma\nu}\nu} - \Gamma^{\nu}_{\phantom{\nu}\nu\sigma} \right)
+ \left( \eta_1 \gamma_\nu \bar\eta_3 \right)\, \varepsilon_{\mu\rho\sigma}\, \dot x_2^\rho\,  \left( \Gamma^{\nu\sigma\mu} + \Gamma^{\mu\sigma\nu} \right) \Big] = \nonumber\\&
= 2\,\lambda_1\lambda_2\left(\lambda_1-\lambda_2\right) \left( C_1 + C_2 + U \right)
\end{align}
The evaluation of these integrals is not immediate, but can be approached with the techniques developed in \cite{Bianchi:2013zda,Bianchi:2013rma,Griguolo:2013sma}.
In particular, for the terms where the $\Gamma$ integrals have contracted derivatives one can always produce a $\delta$ function which localizes the internal integration and one is left with a multiple integral over the circular contour.
The latter are nontrivial but can be reduced to multiple infinite sums which can in turn be expanded in series of $\epsilon$. The results read
\begin{align}
C_1 &= \frac{\pi ^{2 \epsilon }}{8} e^{2 \gamma_E  \epsilon } \left(-\frac{\pi ^2}{\epsilon }+2 \pi ^2-14 \zeta_3\right) + {\cal O}(\epsilon) \nonumber\\
C_2 &= \frac{\pi ^{2 \epsilon }}{4} e^{2 \gamma_E \epsilon } \left[\pi ^2 \left(\frac{1}{\epsilon }-1+4 \log 2\right)+\left( \frac{1}{\epsilon} + 4+12 \log 2\right)\right] + {\cal O}(\epsilon)
\end{align}
The term with an uncontracted integral is the most difficult and we tackle it as in \cite{Griguolo:2013sma}, adding and subtracting an appropriate integral which renders that contribution finite and is easier to compute, as it is proportional to $C_1$.
\begin{equation}\label{eq:subtraction}
U = \frac12 (\eta_1 \g^0 \bar{\eta}_3) \, 
\left[ (x_{12}^2 + x_{23}^2 ) h_{\mu} h_{\nu} \mathbf{V}^{\mu\nu} -2 (1-2\epsilon)\,  x_{2\mu} \Gamma^{\nu\mu}_{\phantom{\mu\nu}\nu} \right] + (1-2\epsilon)\, C_1 = U_{f} + (1-2\epsilon)\, C_1
\end{equation}
where $h^{\mu} = \delta_0^{\mu}$ and
\begin{align}
\label{eq:Vmn}
\mathbf{V}^{\mu\nu}
= -\left(\frac{\Gamma(\frac{3}{2}-\epsilon)}{2\pi^{3/2-\epsilon}}\right)^{3}
\int d^{3-2\epsilon}w
\frac{w^{\mu}w^{\nu}}{(x_{1w}^{2})^{3/2-\epsilon}(x_{2w}^{2})^{3/2-\epsilon}(x_{3w}^{2})^{3/2-\epsilon}}
\end{align}
The $U_f$ contribution is indeed finite and can be evaluated at $\epsilon=0$ where, after simplification, it takes the form
\begin{equation}
U_f = -\frac{1}{128}\,\int_0^{2\pi}d\tau_1\int_0^{\tau_1}d\tau_2\, \frac{\tau_{12}}{\sin \frac{\tau_1}{4} \cos \frac{\tau_2}{4} \cos \frac{\tau_{12}}{4}} \left[ \frac{\cos \frac{\tau_{12}}{4}}{\sin \frac{\tau_1}{4} \cos \frac{\tau_2}{4}}\left( \sin \frac{\tau_1}{2} + \sin \frac{\tau_2}{2} \right) - 4 \cos \frac{\tau_{12}}{4} \right]
\end{equation}
An analytical evaluation of the integral gives
\begin{equation}
U_f = \frac14 \left( \pi^2 - 4\, \pi^2\,\log 2 + 14\, \zeta_3 \right)
\end{equation}
Then, the net effect of the subtraction of \eqref{eq:subtraction} is multiplying $C_1$ by an extra $2(1-\epsilon)$ factor.
Taking into account the various pieces, we see that the contribution of the corrected fermion exchange is totally cancelled by $C_2$ and the total reads  
\begin{equation}
\left.\raisebox{-0.8cm}{\includegraphics[width=2.cm]{2Lfermionm}} + \raisebox{-0.8cm}{\includegraphics[width=2.cm]{2Lfermionmercedesm}}\right|_{m\in 2\mathbb{Z}} = - \frac{\pi^2}{2}\, m^2\, \lambda_1\lambda_2\left(\lambda_1-\lambda_2\right) + {\cal O}(f)
\end{equation}

\paragraph{1/2-BPS Wilson loop at multiple winding}
Summarising, the final result for the 1/2-BPS Wilson loop with winding $m$ and 0 framing reads
\begin{equation}
\langle W^{1/2}_m \rangle_0 = \left(\lambda _1 - (-1)^m \lambda _2\right) \left[
1-\frac{\pi^2}{6} m^2 \left(\lambda _1^2+\lambda _2^2-\lambda _1 \lambda _2 \left(\frac32+\frac72 (-1)^m + 6 m^2\right)\right)\right] + {\cal O}\left(k^{-4}\right)
\end{equation}
Comparing this to the result from localization at framing one we can derive some speculative conclusions concerning the perturbative origin of the framing dependence of the diagrams.

We recall that for single winding, up to two loops, the results differ by the phase $e^{i\pi(\lambda_1-\lambda_2)}$ as for pure Chern-Simons with $U(N_1|N_2)$ gauge group, as previously pointed out \cite{Bianchi:2013zda,Bianchi:2013rma,Griguolo:2013sma}. Diagrammatically this term comes partly from the combination of the pure Chern-Simons graphs. These contribute with the combined coupling $\lambda_1^3+\lambda_2^3$ which does not entirely reconstruct the framing contribution originating from the second order expansion of the exponential, proportional to $(\lambda_1+\lambda_2)(\lambda_1-\lambda_2)^2$. An additional term proportional to $(\lambda_1+\lambda_2)\lambda_1\lambda_2$ is necessary to generate the phase factor and hence we conclude that the fermion diagrams contribute to framing.

At winding $m$, subtracting the bosonic diagrams at framing one to the localization result and comparing to the framing 0 expression, we find the residual framing dependent fermionic contribution
\begin{equation}
\langle W^{1/2}_m \rangle_1 - \langle W^{1/2}_m \rangle_0 \bigg|_{\rm fermion} = -\frac{\pi ^2}{2} \lambda _1 \lambda _2 \left(\lambda _1 - (-1)^m \lambda _2\right) m^2 \left(2 m^2 - \frac32 (1-(-1)^m)\right) + {\cal O}\left(k^{-4}\right)
\end{equation}
We first note that the difference starts at two loops. Namely, the one-loop contribution to framing is only generated by the gluon exchange diagrams but not by the fermion exchanges.
At odd winding this statement is obvious since the fermion exchanges from the two blocks have the same color factor but opposite sign thanks to fermion statistics and hence cancel each other. This is similar to what happens for the exchange of a corrected fermion at two loops. Nonetheless, for even winding number the contributions from the two blocks add instead of cancelling out. At framing 0 such a contribution can be shown to be divergent, but subleading in the dimensional regularization parameter. At non-trivial framing the analysis could be more intricate. However in the limit of small displacement between the framing contours one expects to recover a similar divergent behaviour as for framing 0 and hence to be forced to introduce dimensional regularization. Then a problem of order of limits of the regulators may arise. 

At framing 0 and in dimensional regularization we remark that we also discarded a tadpole integral coming from the insertions of scalar bilinears. These contributions are also subtle to treat in the presence of framing.
A posteriori, the comparison with the localization result seems to suggest that these matter diagrams at one loop do not contribute at framing 1, since there are no terms with mixed gauge group couplings at one loop.

We now move to the two-loop part, which is more speculative.
Since the fermion diagrams contribute with different powers of the winding number we can separate their putative framing dependent part (at framing 1).
We recall that some fermionic diagrams were already discarded in the framing 0 analysis of this section, because entirely vanishing. For instance the mixed gluon and fermion exchange was neglected. One should ascertain whether this contributes to framing.
Such a diagram can be symmetrized by adding and subtracting a crossed contribution. Then the symmetric part factorizes into the product of a 1-loop gluon exchange and a 1-loop fermion exchange. We know that the former is framing dependent but the latter should give a vanishing contribution, according to the discussion on the one-loop dependence above. Hence this part is likely not to contribute to framing. For the remaining crossed integral, in pure Chern-Simons theory this diagram is framing independent and hence vanishing. We are not able to prove the same for the mixed exchange without a more accurate analysis and we only speculatively assume this is the case, relying on the fact that the same mechanism preventing crossed gauge propagators from developing a framing dependence can also work in this situation.

We now analyse the framing dependence of the fermionic diagrams which we have already considered in the framing 0 computation.
In particular, thanks to the fact that only the double fermion exchange can give rise to the $m^4$ power, we can immediately isolate its piece. Thus we find that the fermionic contributions due to framing have to read
\begin{align}
\left.\raisebox{-0.8cm}{\includegraphics[width=2.cm]{2Lfermion1m}} + \raisebox{-0.8cm}{\includegraphics[width=2.cm]{2Lfermion2m}}\right|^{f=1}_{f=0} &= -\frac{\pi^2}{2}\, m^2(2m^2+1)\, \lambda_1\lambda_2 \left(\lambda_1 - (-1)^m \lambda_2\right)\\
\left.\raisebox{-0.8cm}{\includegraphics[width=2.cm]{2Lfermionmercedesm}} + \raisebox{-0.8cm}{\includegraphics[width=2.cm]{2Lfermionm}}\right|^{f=1}_{f=0} &= \pi^2\, \frac{5-3(-1)^m}{4}\, m^2\, \lambda_1\lambda_2 \left(\lambda_1 - (-1)^m \lambda_2\right)  
\end{align}
where with the notation above we understand the difference between the result at framing 1 and that at framing 0.
Curiously enough this means that the contribution coming exclusively from framing 1 of the fermionic diagrams is precisely the opposite of their framing 0 result and cancel it individually.
In other words the fermionic diagrams seem to separately vanish when evaluated at framing 1.
On the total sum of them, this statement is effectively imposed by the cohomological equivalence between the 1/2-BPS Wilson loop and the combination of the 1/6-BPS ones \eqref{eq:equiv}, at framing 1, in the realm of localization. Indeed the fact that the 1/2-BPS Wilson loop can be computed in terms of 1/6-BPS objects means that effectively the sum of fermionic diagrams does not contribute at framing 1.
With multiple winding we are able to refine this statement at the level of the individual two-loop fermionic diagrams.
It would be interesting to test this indirect expectation against a genuine perturbative computation at framing 1 of the fermionic diagrams and also to understand whether this pattern continues at higher loops.

\section{Conclusions}

In this paper we have analysed the circular BPS Wilson loops for general framing and winding number in planar ABJM theory at weak coupling.
We have elucidated how to separate the combinatorics involved in multiple winding from the dynamics of the theory, by means of recursive relations.
In particular, we have derived an algorithm in order to reduce the contour integrals arising at multiple winding in terms of those which are relevant for the single winding computation.
Applying this technique we have first computed the 1/6-BPS Wilson loop perturbatively for general winding up to three loops. In particular we studied its dependence on the framing number, which for multiple winding is not captured by a phase.
At framing 1 the result we obtained coincides with the prediction from supersymmetric localization on $S^3$ and thus constitutes a robust test thereof.
In the limit where the contribution of the second gauge group becomes negligible we recover the pure Chern-Simons result for the unknot at general winding and framing number.
Finally we have computed the perturbative 1/2-BPS Wilson loop at two loops at 0 framing and generic winding. Comparing it to the localization result we have inferred that the contribution of the individual fermion diagrams at framing 1 vanishes.

\section*{Acknowledgements}

We thank Luca Griguolo, Matias Leoni, Andrea Mauri, Silvia Penati and Domenico Seminara for very useful discussions and especially Rodolfo Panerai for discussions and computational support. 
The work of MB was supported in part by the Science and Technology Facilities Council Consolidated Grant ST/L000415/1 \emph{String theory, gauge theory \& duality}.

\vfill
\newpage

\appendix

\section{Conventions and Feynman rules}\label{sec:conventions}

In this appendix we spell out the conventions used for the computation of the diagrams.

We work in euclidean three--dimensional space with coordinates $x^\mu = (x^0, x^1, x^2)$. We choose a set of gamma matrices satisfying Clifford algebra $\{ \g^\mu , \g^\nu \} = 2 \d^{\mu\nu} \mathbb{I}$ as
\beq
(\g^\mu)_\a^{\; \, \b} = \{ -\s^3, \s^1, \s^2 \}
\eeq
Useful identities are
\bea 
&&  \g^\mu \g^\nu = \d^{\mu \nu} \mathbb{I} - i \varepsilon^{\mu\nu\rho} \g^\rho
\non \\
&& \g^\mu \g^\nu \g^\rho = \d^{\mu\nu} \g^\rho - \d^{\mu\rho} \g^\nu+  \d^{\nu\rho} \g^\mu  - i \varepsilon^{\mu\nu\rho} \mathbb{I}
\non \\
&& 
\g^\mu \g^\nu \g^\rho \g^\s -  \g^\s \g^\rho \g^\nu \g^\mu = -2i \left( \d^{\mu\nu} \varepsilon^{\rho\s \eta}  + \d^{\rho \s}  \varepsilon^{\mu\nu\eta} + \d^{\nu\eta} \varepsilon^{\rho \mu \s} +
\d^{\mu\eta} \varepsilon^{\nu\rho\s}  \right) \g^\eta
 \\
&& 
\non \\
&&
\Tr (\g^\mu \g^\nu) = 2 \d^{\mu\nu}
\non \\
&&
\Tr (\g^\mu \g^\nu \g^\rho) = -2i \varepsilon^{\mu\nu\rho}
\eea 
Spinor indices are lowered and raised as $(\g^\mu)^\a_{\; \, \b} = \varepsilon^{\a \g}  (\g^\mu)_\g^{\; \, \d} \varepsilon_{\b \d}$, where $\varepsilon^{12} = - \varepsilon_{12} = 1$. 
When writing spinor products we conventionally choose the spinor indices of chiral fermions to be always up, while the ones of antichirals to be always down.

\vskip 20pt
\noindent
The euclidean action  of $U(N_1)_k \times U(N_2)_{-k}$ ABJM theory \cite{ABJM} reads
 \bea
\label{action}
S &=& \frac{k}{4\pi}\int d^3x\,\varepsilon^{\mu\nu\rho} \Big\{ -i\Tr \left( A_\mu\partial_\nu A_\rho+\frac{2}{3} i A_\mu A_\nu A_\rho \right)
+i \Tr \left(\hat{A}_\mu\partial_\nu 
\hat{A}_\rho+\frac{2}{3} i \hat{A}_\mu \hat{A}_\nu \hat{A}_\rho \right) 
\non \\
&~& \qquad \qquad \qquad \qquad  + \Tr \Big[ \frac{1}{\xi}  (\pa_\mu A^\mu)^2 -\frac{1}{\xi} ( \pa_\mu \hat{A}^\mu )^2 + \pa_\mu \bar{c} D^\mu c  
  - \pa_\mu \bar{\hat{c}} D^\mu \hat{c} \Big] \Big\}
\non \\
&~& + \int d^3x \, \Tr \Big[ D_\mu C_I D^\mu \bar{C}^I + i \bar{\psi}^I \g^\mu D_\mu \psi_I \Big] + S_{int} 
\non  
\eea
with covariant derivatives defined as
\bea
\label{covariant}
D_\mu C_I &=& \pa_\mu C_I + i A_\mu C_I - i C_I \hat{A}_\mu
\quad ; \quad 
D_\mu \bar{C}^I = \pa_\mu \bar{C}^I - i \bar{C}^I A_\mu + i \hat{A}_\mu \bar{C}^I  
\non \\
D_\mu \bar{\psi}^I  &=& \pa_\mu \bar{\psi}^I + i A_\mu \bar{\psi}^I - i \bar{\psi}^I \hat{A}_\mu
\quad ; \quad
D_\mu \psi_I = \pa_\mu \psi_I - i \psi_I A_\mu + i \hat{A}_\mu \psi_I  
\eea
Gauge fields are in the adjoint representation of the corresponding gauge group, $A_\mu = A_\mu^a T^a$, $\hat A_\mu = \hat A_\mu^a \hat T^a$ with $T^a$ ($\hat T^a$) a set of $U(N_1)$ ($U(N_2)$) hermitian matrices satisfying $\Tr (T^a T^b) = \d^{ab}$ ($\Tr (\hat T^a \hat T^b) = \d^{ab}$). Scalars $C_I$ ($\bar{C}^I$) and the corresponding fermions transform in the (anti)bifundamental representation of the gauge group and carry a fundamental index of the $SU(4)$ R-symmetry group.  

With these assignments the Feynman rules are:
\begin{itemize}
\item Vector propagators in Landau gauge
\bea
\label{treevector}
&& \langle A_\mu^a (x) A_\nu^b(y) \rangle^{(0)} =  \d^{ab}   \, \left( \frac{2\pi i}{k} \right) \frac{\G(\frac32-\e)}{2\pi^{\frac32 -\e}} \varepsilon_{\mu\nu\rho} \frac{(x-y)^\rho}{[(x-y)^2]^{\frac32 -\e} }
\non \\
&& \langle \hat{A}_\mu^a (x) \hat{A}_\nu^b(y) \rangle^{(0)} =  -\d^{ab}   \, \left( \frac{2\pi i}{k} \right) \frac{\G(\frac32-\e)}{2\pi^{\frac32 -\e}} \varepsilon_{\mu\nu\rho} \frac{(x-y)^\rho}{[(x-y)^2]^{\frac32 -\e} }
\eea
\item Scalar propagator
\beq
\label{scalar}
\langle (C_I)_i^{\; \hat{j}} (x) (\bar{C}^J)_{\hat{k}}^l(\; y) \rangle^{(0)}  = \d_I^J \d_i^l \d_{\hat{k}}^{\hat{j}} \, \frac{\G(\frac12 -\e)}{4\pi^{\frac32-\e}} 
\, \frac{1}{[(x-y)^2]^{\frac12 -\e}}
\eeq
\item Fermion propagator
\beq
\label{treefermion}
\langle (\psi_I^\a)_{\hat{i}}^{\; j}  (x) (\bar{\psi}^J_\b )_k^{\; \hat{l}}(y) \rangle^{(0)} = - i \, \d_I^J \d_{\hat{i}}^{\hat{l}} \d_{k}^{j} \, 
\frac{\G(\frac32 - \e)}{2\pi^{\frac32 -\e}} \,  \frac{(\g^\mu)^\a_{\; \, \b} \,  (x-y)_\mu}{[(x-y)^2]^{\frac32 - \e}}
\eeq
\item Gauge cubic vertex
\beq
\label{gaugecubic}
-i \frac{k}{12\pi} \varepsilon^{\mu\nu\rho} \int d^3x \, f^{abc} A_\mu^a A_\nu^b A_\rho^c
\eeq
\item Gauge--fermion cubic vertex
\beq
\label{gaugefermion}
-\int d^3x \, \Tr \Big[ \bar{\psi}^I \g^\mu \psi_I A_\mu - \bar{\psi}^I \g^\mu \hat{A}_\mu \psi_I  \Big]
\eeq 
\end{itemize}

\noindent
For two--loop calculations we also need the one--loop vector propagators  
\bea
\label{1vector}
&& \langle A_\mu^a (x) A_\nu^b(y) \rangle^{(1)} = \d^{ab}   \left( \frac{2\pi }{k} \right)^2 N_2 \frac{\G^2(\frac12-\e)}{4\pi^{3 -2\e}} 
\left[ \frac{\d_{\mu\nu}}{ [(x- y)^2]^{1-2\e}} - \pa_\mu \pa_\nu \frac{[(x-y)^2]^\e}{4\e(1+2\e)} \right]  \non  \\
&& \langle \hat{A}_\mu^a (x) \hat{A}_\nu^b(y) \rangle^{(1)} = \d^{ab}   \left( \frac{2\pi }{k} \right)^2 N_1 \frac{\G^2(\frac12-\e)}{4\pi^{3 -2\e}} 
\left[ \frac{\d_{\mu\nu}}{ [(x- y)^2]^{1-2\e}} - \pa_\mu \pa_\nu \frac{[(x-y)^2]^\e}{4\e(1+2\e)} \right] \non \\
\eea
and the one--loop fermion propagator  
\beq
\label{1fermion}
\langle (\psi_I^\a)_{\hat{i}}^{\; j}  (x) (\bar{\psi}^J_\b)_k^{\; \hat{l}}(y) \rangle^{(1)} =  i \,\left( \frac{2\pi}{k} \right) \,  \d_I^J \d_{\hat{i}}^{\hat{l}} \d_{k}^{j} \,  \, \d^\a_{\; \, \b}
\, (N_1-N_2) \frac{\G^2(\frac12 - \e)}{16 \pi^{3-2\e}} \, \frac{1}{[(x-y)^2]^{1 - 2\e}}  
\eeq	 

\newpage

\subsection{Integrals}
\label{app:integrals}

We give here the relevant integrals which have been symbolized by pictures in the previous sections.
Since all integrals are finite or ultimately regulated by framing we evaluate them at $\epsilon=0$. 
For arbitrary $m$ we have
\begin{align}
\raisebox{-0.8cm}{\includegraphics[width=2.cm]{1Lgauge}} &\equiv \int_{0}^{2\pi}d\tau_1 \int_{0}^{\tau_1}d\tau_2\, g(\tau_1,\tau_2) \\
\raisebox{-0.8cm}{\includegraphics[width=2.cm]{2Lgauge1}} &+ \raisebox{-0.8cm}{\includegraphics[width=2.cm]{2Lscalar}} \equiv \int_{0}^{2\pi}d\tau_1 \int_{0}^{\tau_1}d\tau_2\, \\
\raisebox{-0.8cm}{\includegraphics[width=2.cm]{2LgaugeMB}} &\equiv -\int_{0}^{2\pi}d\tau_1 \int_{0}^{\tau_1}d\tau_2\, \int_{0}^{\tau_2}d\tau_3\, \frac{1}{16\pi}\, \dot{x}_1^\s \, \dot{x}_2^\eta \, \dot{x}_3^\zeta \, \nonumber\\& 
~~~~\times \e^{\mu\nu\rho} \e_{\s\mu\xi} \e_{\eta\nu\tau} \e_{\zeta\rho \kappa} 
\,  \int d^3x  \,  
\frac{  (x-x_1)^\xi (x-x_2)^\tau (x-x_3)^\kappa}{|x-x_1|^{3} |x-x_2|^{3} |x-x_3|^{3} } \nonumber\\&
\equiv \int_{0}^{2\pi}d\tau_1 \int_{0}^{\tau_1}d\tau_2\, \int_{0}^{\tau_2}d\tau_3\, f(\tau_1,\tau_2,\tau_3) \label{eq:mercedes}  \\
\raisebox{-0.8cm}{\includegraphics[width=2.cm]{2Lgauge}} &\equiv \int_{0}^{2\pi}d\tau_1 \int_{0}^{\tau_1}d\tau_2\, \int_{0}^{\tau_2}d\tau_3\, \int_{0}^{\tau_3}d\tau_4\, g(\tau_1,\tau_2)\, g(\tau_3,\tau_4) \\
\raisebox{-0.8cm}{\includegraphics[width=2.cm]{2Lgauge2}} &\equiv \int_{0}^{2\pi}d\tau_1 \int_{0}^{\tau_1}d\tau_2\, \int_{0}^{\tau_2}d\tau_3\, \int_{0}^{\tau_3}d\tau_4\, g(\tau_1,\tau_4)\, g(\tau_2,\tau_3) \\
\raisebox{-0.8cm}{\includegraphics[width=2.cm]{2Lcross}} &\equiv \int_{0}^{2\pi}d\tau_1 \int_{0}^{\tau_1}d\tau_2\, \int_{0}^{\tau_2}d\tau_3\, \int_{0}^{\tau_3}d\tau_4\, g(\tau_1,\tau_3)\, g(\tau_2,\tau_4)
\end{align}
The function $g$ is defined as
\begin{equation}
g(\tau_1,\tau_2) \equiv i\, \dot x_1^{\mu}\, \dot x_2^{\nu}\, \varepsilon_{\mu\nu\rho}\, \frac{(x_1-x_2)^{\rho}}{|x_1 - x_2|^3}
\end{equation}
and satisfies the properties
\begin{equation}
g(\tau_1,\tau_2) = g(\tau_1+2\pi,\tau_2) = g(\tau_1,\tau_2+2\pi) = g(\tau_2,\tau_1)
\end{equation}
stating that it is symmetric and periodic in any argument.
Note that for a planar contour $g$ vanishes identically.

For the circle at winding 1, i.e. $0<\tau_3<\tau_2<\tau_1<2\pi$ the mercedes integral can be simplified
\begin{equation}
\raisebox{-0.8cm}{\includegraphics[width=2.cm]{2LgaugeMB}} = -\int_{0}^{2\pi}d\tau_1 \int_{0}^{\tau_1}d\tau_2\, \int_{0}^{\tau_2}d\tau_3\, \frac{1}{64\, \sin\frac{\tau_{13}}{4} \cos \frac{\tau_{12}}{4} \cos \frac{\tau_{23}}{4}} = -\frac{\pi^2}{6}
\end{equation}
The integrand is antisymmetric in the exchange of any two variables.

For the three-loop computation several other integrals are needed.
Using the same structures and diagrammatics as in the two-loop computation, we define the following contributions with five insertion points
\begin{equation}
\raisebox{-0.6cm}{\includegraphics[width=1.5cm]{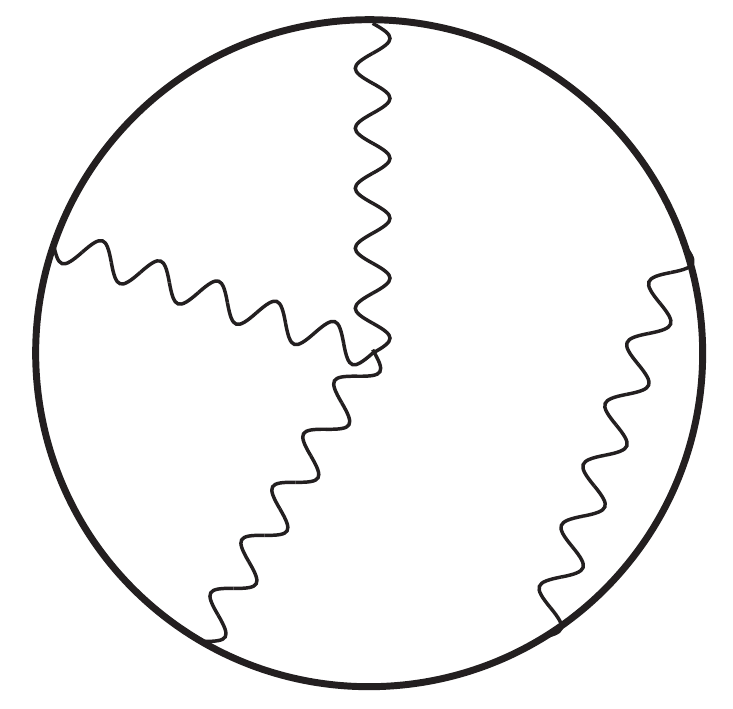}} \equiv a \quad \raisebox{-0.6cm}{\includegraphics[width=1.5cm]{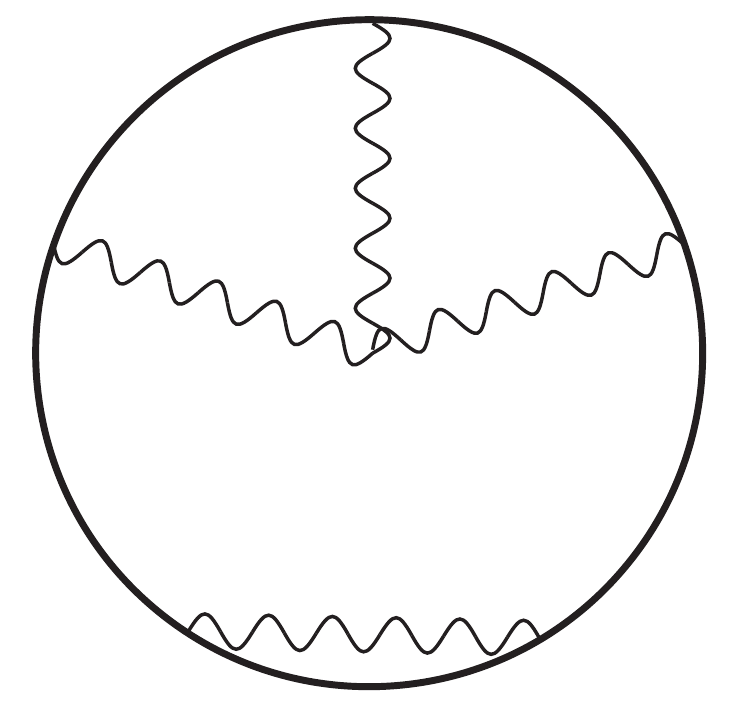}} \equiv b \quad \raisebox{-0.6cm}{\includegraphics[width=1.5cm]{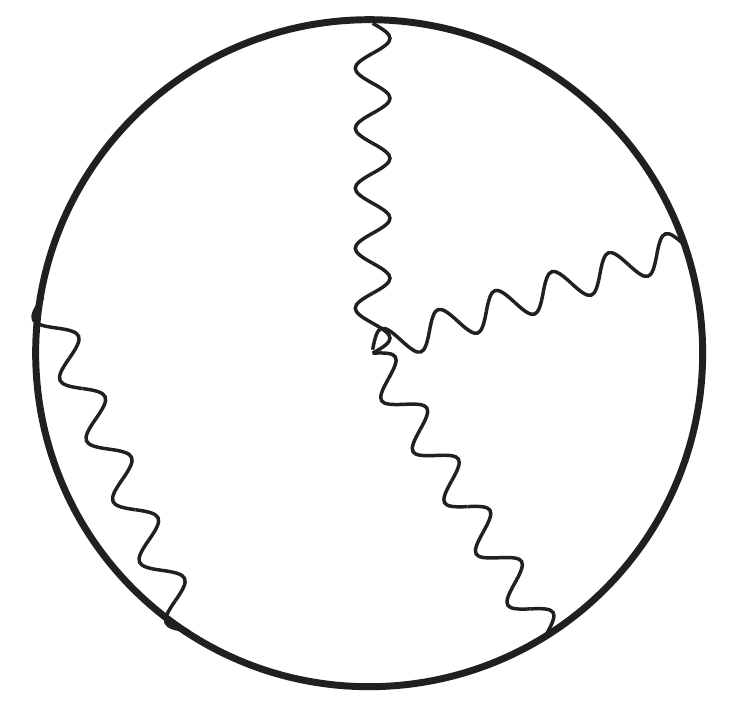}} \equiv c \quad \raisebox{-0.6cm}{\includegraphics[width=1.5cm]{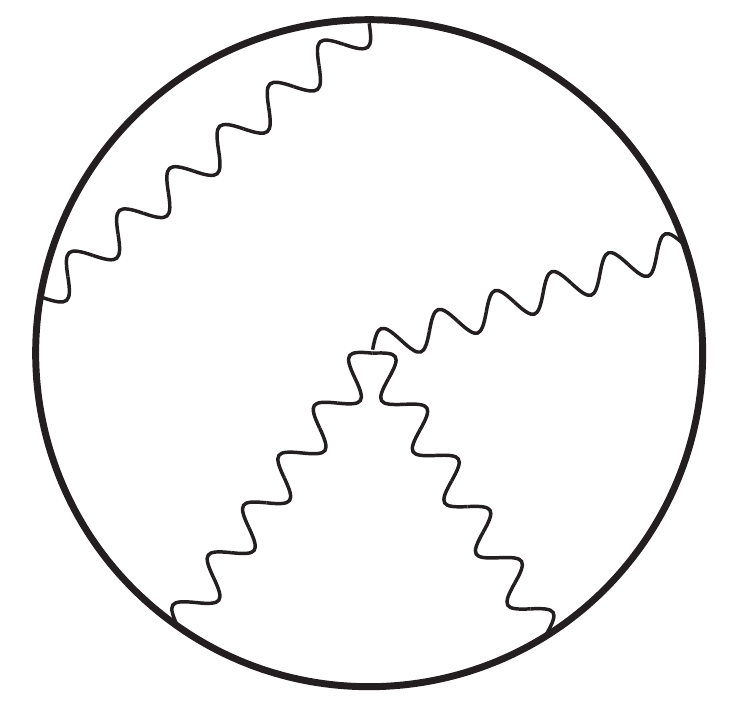}} \equiv d \quad
\raisebox{-0.6cm}{\includegraphics[width=1.5cm]{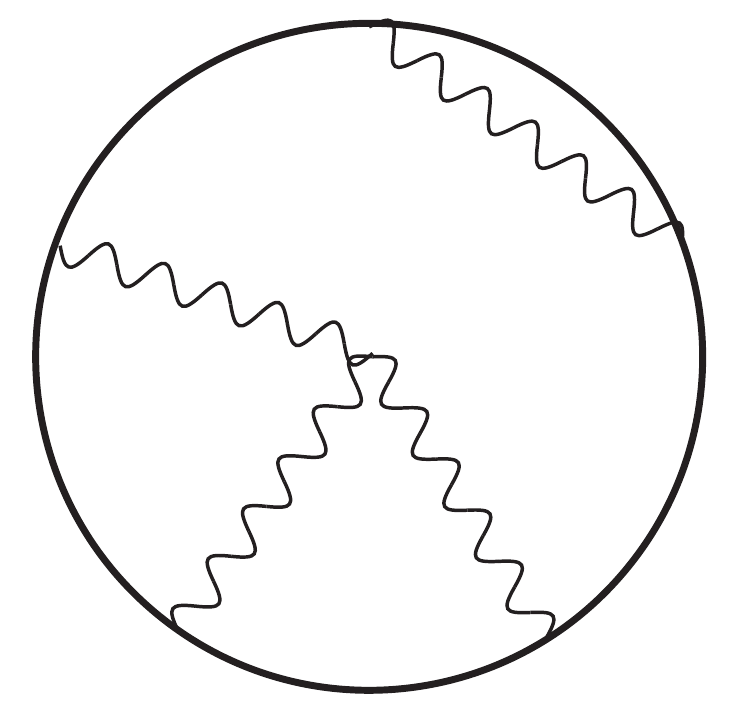}} \equiv e
\end{equation}
For instance the first integral reads explicitly
\begin{equation}
\raisebox{-0.8cm}{\includegraphics[width=2.cm]{3La5}} = \int_{0}^{2\pi}d\tau_1 \int_{0}^{\tau_1}d\tau_2\, \int_{0}^{\tau_2}d\tau_3\int_{0}^{\tau_3}d\tau_4\, \int_{0}^{\tau_4}d\tau_5\, f(\tau_1,\tau_2,\tau_3)\, g(\tau_4,\tau_5)
\end{equation}
where $f$ and $g$ are the same functions defined above.
The other integrals are obtained by a permutation of indices of the integrand, according to the ordering displayed in the pictures.
The nonplanar topologies do not give rise to framing dependent contributions, as pointed out in \cite{Alvarez:1991sx}, and hence vanish.
With six insertion points the relevant integrals read 
\begin{align}
&\raisebox{-0.6cm}{\includegraphics[width=1.5cm]{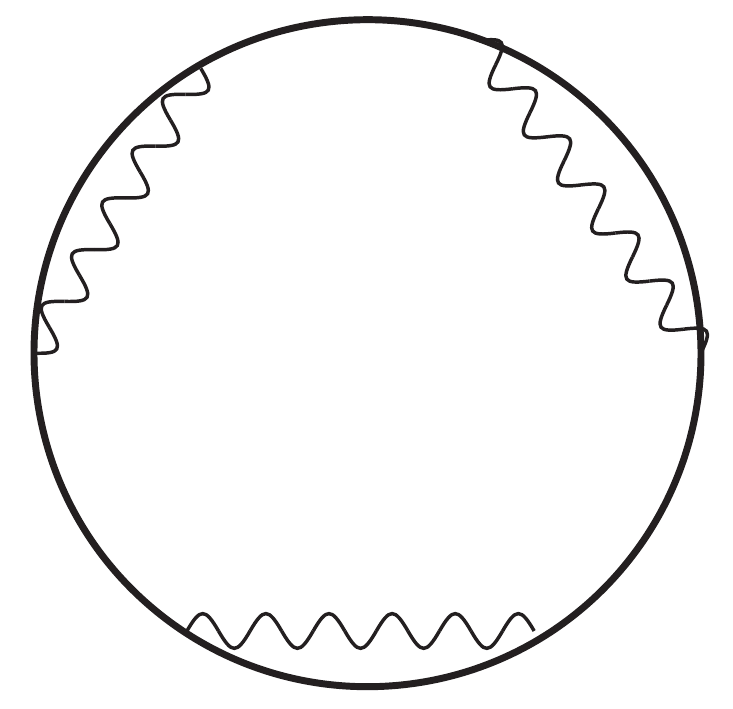}} \equiv A & \quad & \raisebox{-0.6cm}{\includegraphics[width=1.5cm]{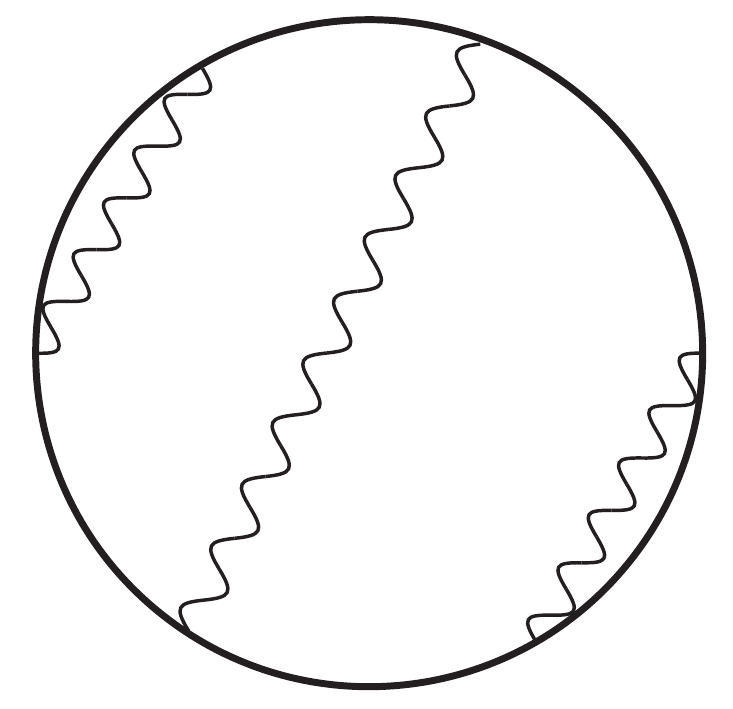}} \equiv B & \quad & \raisebox{-0.6cm}{\includegraphics[width=1.5cm]{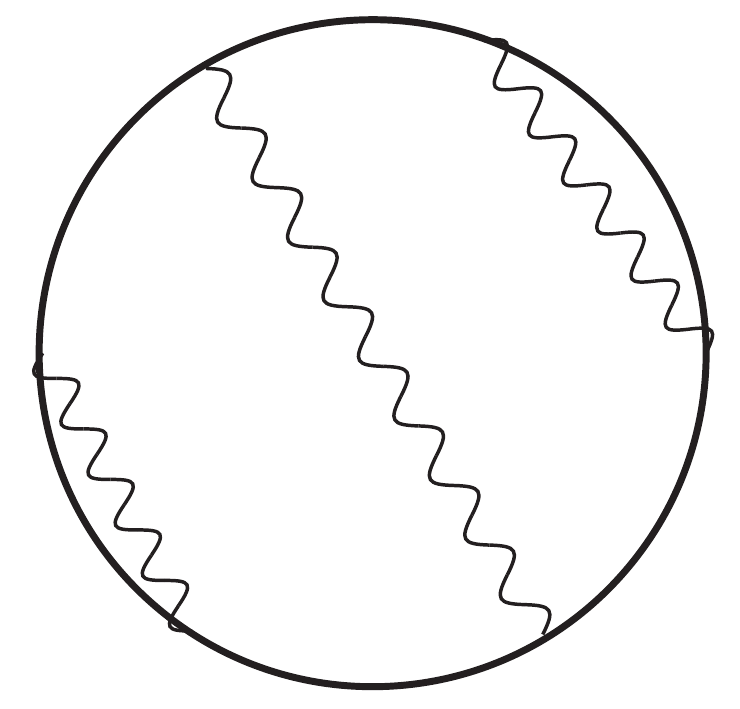}} \equiv C & \quad & \raisebox{-0.6cm}{\includegraphics[width=1.5cm]{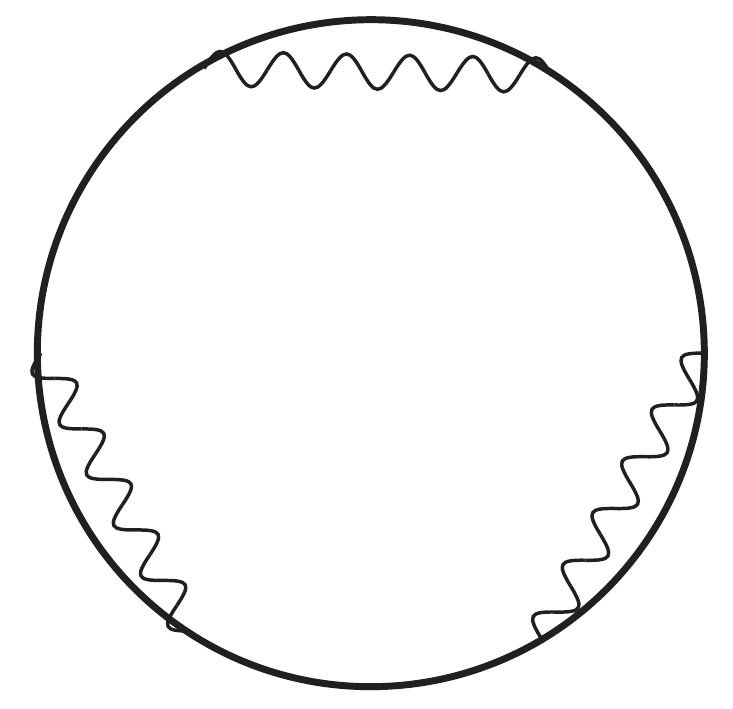}} \equiv D & \quad &
\raisebox{-0.6cm}{\includegraphics[width=1.5cm]{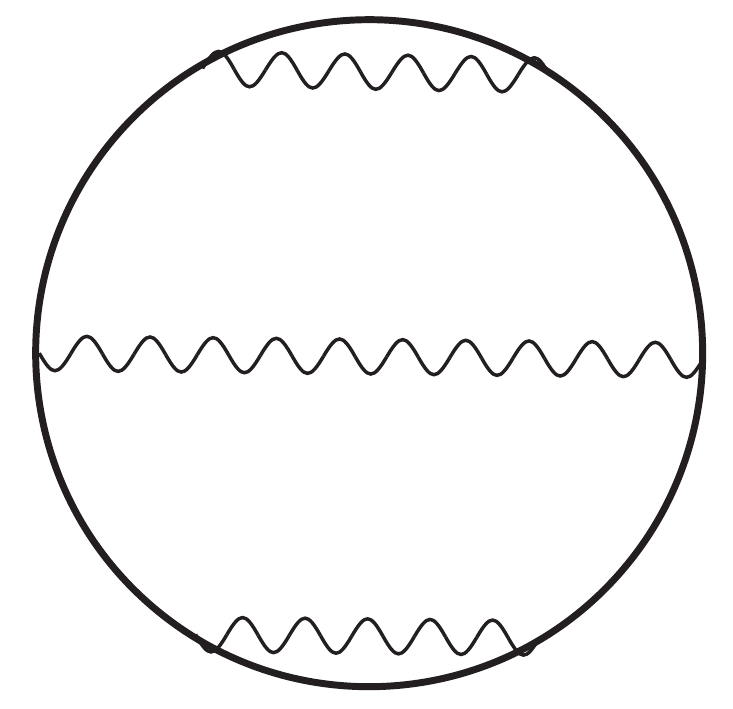}} \equiv E & \nonumber\\&
\raisebox{-0.6cm}{\includegraphics[width=1.5cm]{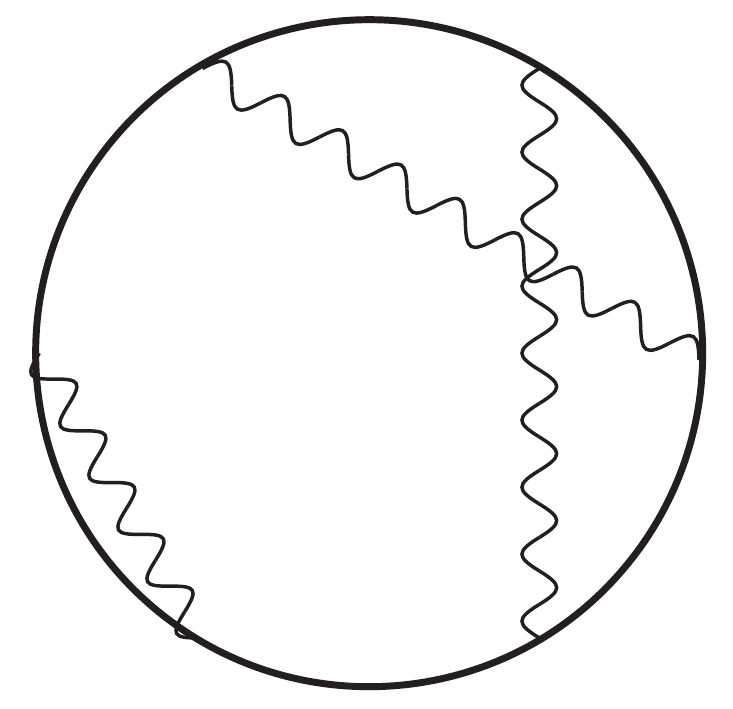}} \equiv f & \quad & \raisebox{-0.6cm}{\includegraphics[width=1.5cm]{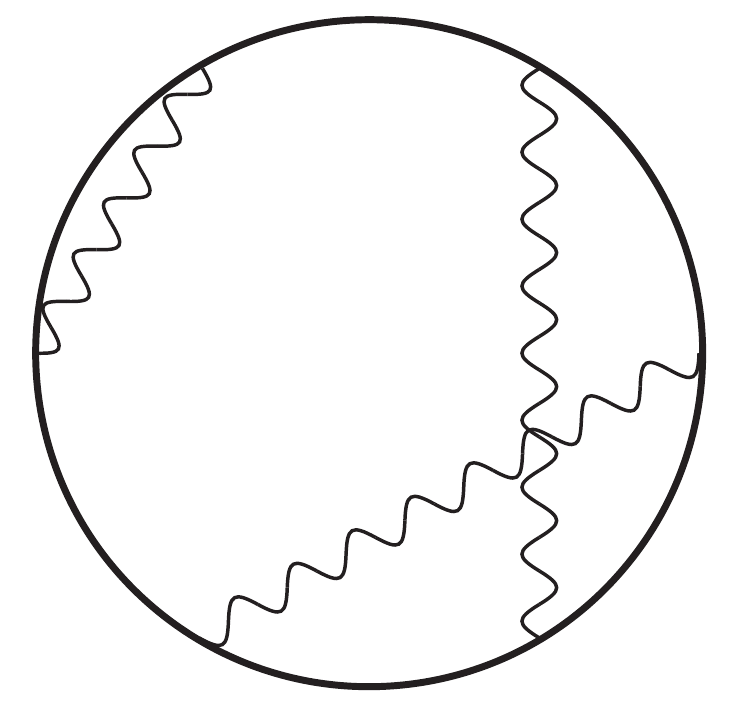}} \equiv g & \quad & \raisebox{-0.6cm}{\includegraphics[width=1.5cm]{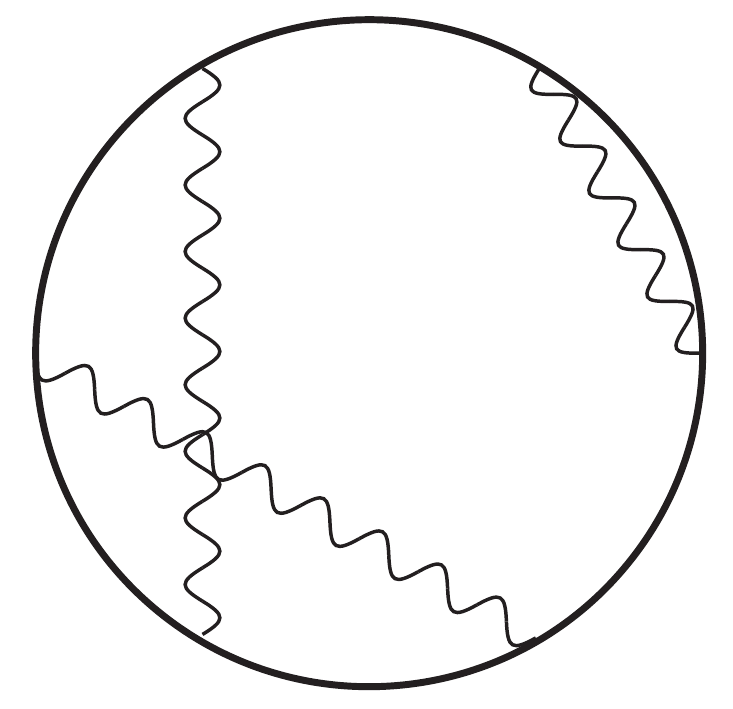}} \equiv h & \quad & \raisebox{-0.6cm}{\includegraphics[width=1.5cm]{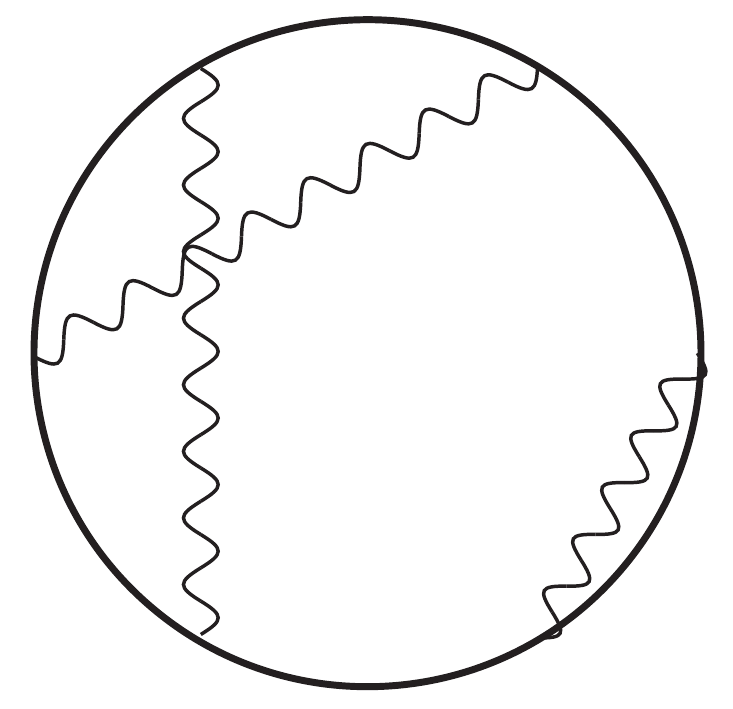}} \equiv i & \quad & \raisebox{-0.6cm}{\includegraphics[width=1.5cm]{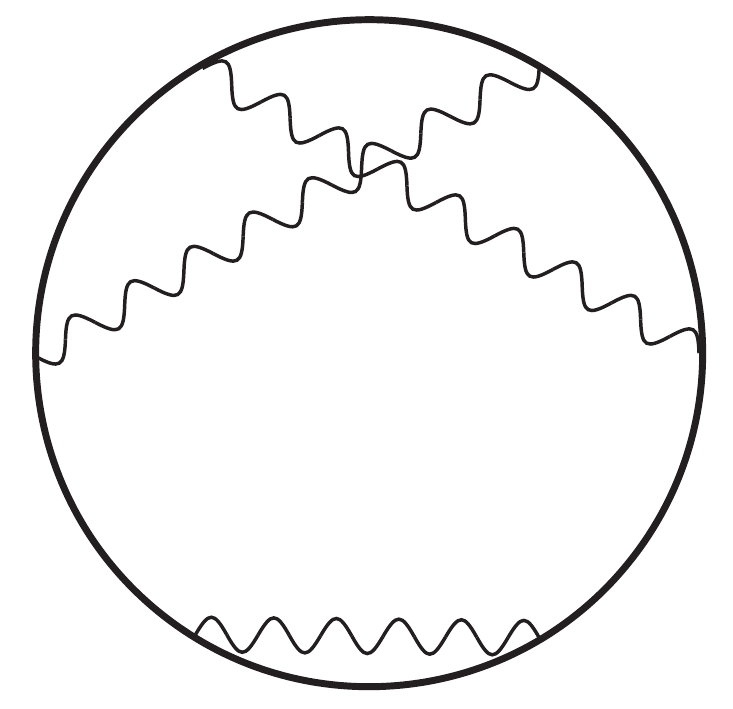}} \equiv j &
\end{align}
Again, the first integral has the explicit form
\begin{equation}
\raisebox{-0.8cm}{\includegraphics[width=2.cm]{3LA}} = \int_{0}^{2\pi}d\tau_1 \int_{0}^{\tau_1}d\tau_2\, \int_{0}^{\tau_2}d\tau_3\int_{0}^{\tau_3}d\tau_4\, \int_{0}^{\tau_4}d\tau_5\int_{0}^{\tau_5}d\tau_6\, g(\tau_1,\tau_2)\,g(\tau_3,\tau_4)\, g(\tau_5,\tau_6)
\end{equation}
and the others follow from relabelling of indices.
Other nonplanar topologies are possible with two and three crossings, which are framing independent and vanish.

\section{The recursion relations for 3-loop pure Chern-Simons}
\label{app:3loop}

In this appendix we spell out the systems of recursion relations which have been solved to derive the results \eqref{eq:3L5} and \eqref{eq:3L6}.
The three-loop single winding integrals are defined in the previous appendix.

We start with the five insertions integral. We consider \eqref{eq:system} with $n=5$ and integrand
\begin{equation}
G_5(1)=\raisebox{-0.8cm}{\includegraphics[width=2.cm]{3La5}} + \raisebox{-0.8cm}{\includegraphics[width=2.cm]{3Lb5}} + \raisebox{-0.8cm}{\includegraphics[width=2.cm]{3Lc5}} + \raisebox{-0.8cm}{\includegraphics[width=2.cm]{3Ld5}} +
\raisebox{-0.8cm}{\includegraphics[width=2.cm]{3Le5}} 
\end{equation}
From it we obtain the system of equations
\begin{equation}
\left\{\begin{array}{l}
G_5(m)=G_{5,2}(m-1)+G_{5,3}(m-1)+G_{5,4}(m-1)+\\~~~~+(m-1) (3 a+b+c+3 d+2 e)+a+b+c+d+e+G_5(m-1)\\
G_{5,2}(m)+2 d+e=G_{5,2}(m-1)+a (4 m-2)+4 m (b+c+d+e)\\
G_{5,3}(m)+2 a+2 d+e=G_{5,3}(m-1)+G_{5,3,2}(m-1)+4 m (a+b+c+d+e)\\
G_{5,4}(m)=G_{5,4}(m-1)+G_{5,4,2}(m-1)+G_{5,4,3}(m-1)+\\~~~~+4 (m-1) (a+b+c+d+e)+3 a+b+c+3 d+2 e\\
G_{5,3,2}(m) = G_{5,3,2}(m-1)+4 m (a+b+c+d+e)\\
G_{5,4,2}(m) = G_{5,4,2}(m-1)+4 m (a+b+c+d+e)\\
G_{5,4,3}(m)=G_{5,4,3}(m-1)+G_{5,4,3,2}(m-1)+4 m (a+b+c+d+e)\\
G_{5,4,3,2}(m) = G_{5,4,3,2}(m-1)+4 (a+b+c+d+e)=\\
G_5(1) = a+b+c+d+e\\
G_{5,2}(1)=G_{5,3}(1) = 2 a+4 b+4 c+2 d+3 e\\
G_{5,4}(1) = 3 a+b+c+3 d+2 e\\
G_{5,3,2}(1)=G_{5,4,2}(1)=G_{5,4,3}(1)=G_{5,4,3,2}(1)=4 (a+b+c+d+e)
\end{array}\right.
\end{equation}
whose solution reads
\begin{equation}
\left\{\begin{array}{l}
G_5(m)= \frac{1}{3} m^2 \left(2 m^2+1\right) (a+b+c+d+e)\\
G_{5,2}(m)= m (2 a m+2 b (m+1)+2 c (m+1)+2 d m+2 e m+e)\\
G_{5,3}(m)= \frac{1}{3} m \left(2 a \left(m^2+3 m-1\right)+2 b \left(m^2+3 m+2\right)+\right. \\~~~~\left.+2 c m^2+6 c m+4 c+2 d m^2+6 d m-2 d+2 e m^2+6 e m+e\right)\\
G_{5,4}(m)= m \left(2 a m^2+a+b \left(2 m^2-1\right)+2 c m^2-c+2 d m^2+d+2 e m^2\right)\\
G_{5,3,2}(m)= 2 m (m+1) (a+b+c+d+e)\\
G_{5,4,2}(m)= 2 m (m+1) (a+b+c+d+e)\\
G_{5,4,3}(m)= 4 m^2 (a+b+c+d+e)\\
G_{5,4,3,2}(m)= 4 m (a+b+c+d+e)
\end{array}\right.
\end{equation}
and the first equation is \eqref{eq:3L5}.
For the integrals over six insertion points the system of recursion relations is
\begin{equation}
\left\{\begin{array}{l}
G_6(m)= A+ B+ C+ D+ E+2 (m-1) (3  A+2  B+2  C+ D+ E+ f+ g +\\~~~~+ h+ i+ j)
+G_6(m-1)+G_{6,2}(m-1)+G_{6,3}(m-1)+G_{6,4}(m-1)+G_{6,5}(m-1)\\ 
G_{6,2}(m)=3  A+7  B+7  C+11  D+7  E+6  f+4  g+4  h+6  i+4  j+\\~~~~+2 (m-1) (9  A+7  B+7  C+7  D+5  E+5  f+6  g+6  h+5  i+4  j)+G_{6,2}(m-1)\\ 
G_{6,3}(m)=12  A+8  B+8  C+4  D+12  E+4  f+8  g+8  h+4  i+8  j+\\~~~~+2 (m-1) (12  A+9  B+15  C+10  D+10  E+12  f+9  g+13  h+8  i+9  j)+\\~~~~+G_{6,3}(m-1)+G_{6,3,2}(m-1)\\ 
G_{6,4}(m)=3  A+7  B+7  C+11  D+7  E+6  f+4  g+4  h+6  i+4  j+\\~~~~+2 (m-1) (12  A+15  B+9  C+10  D+10  E+8  f+13  g+9  h+12  i+9  j)+\\~~~~+G_{6,4}(m-1)+G_{6,4,2}(m-1)+G_{6,4,3}(m-1)\\ 
G_{6,5}(m)=6  A+4  B+4  C+2  D+2  E+2  f+2  g+2  h+2  i+2  j+\\~~~~+2 (m-1) (9  A+7  B+7  C+7  D+5  E+5  f+6  g+6  h+5  i+4  j)+\\~~~~+G_{6,5}(m-1)+G_{6,5,2}(m-1)+G_{6,5,3}(m-1)+G_{6,5,4}(m-1)\\ 
G_{6,3,2}(m)=24  A+30  B+18  C+20  D+20  E+16  f+26  g+18  h+24  i+18  j+\\~~~~+8 (m-1) (7  A+6  B+6  C+5  D+4  E+5  f+6  g+6  h+5  i+4  j)+G_{6,3,2}(m-1)\\ 
G_{6,4,2}(m)=6  A+18  B+18  C+42  D+54  E+30  f+12  g+12  h+30  i+48  j+\\~~~~+(m-1) (44  A+52  B+52  C+68  D+76  E+60  f+48  g+48  h+60  i+72  j)+\\~~~~+G_{6,4,2}(m-1)\\ 
G_{6,4,3}(m)=24  A+18  B+30  C+20  D+20  E+24  f+18  g+26  h+16  i+18  j+\\~~~~+(m-1) (68  A+68  B+68  C+52  D+52  E+60  f+68  g+68  h+60  i+52  j)+\\~~~~+G_{6,4,3}(m-1)+G_{6,4,3,2}(m-1)\\ 
G_{6,5,2}(m)=24  A+18  B+30  C+20  D+20  E+24  f+18  g+26  h+16  i+18  j+\\~~~~+8 (m-1) (7  A+6  B+6  C+5  D+4  E+5  f+6  g+6  h+5  i+4  j)+G_{6,5,2}(m-1)\\ 
G_{6,5,3}(m)=24  A+30  B+18  C+20  D+20  E+16  f+26  g+18  h+24  i+18  j+\\~~~~+(m-1) (44  A+52  B+52  C+68  D+76  E+60  f+48  g+48  h+60  i+72  j)+\\~~~~+G_{6,5,3}(m-1)+G_{6,5,3,2}(m-1)\\ 
G_{6,5,4}(m)=18  A+14  B+14  C+14  D+10  E+10  f+12  g+12  h+10  i+8  j+\\~~~~+8 (m-1) (7  A+6  B+6  C+5  D+4  E+5  f+6  g+6  h+5  i+4  j)+\\~~~~+G_{6,5,4}(m-1)+G_{6,5,4,2}(m-1)+G_{6,5,4,3}(m-1)\\ 
G_{6,4,3,2}(m)=44  A+52  B+52  C+68  D+76  E+60  f+48  g+48  h+60  i+72  j+\\~~~~+120 (m-1) ( A+ B+ C+ D+ E+ f+ g+ h+ i+ j)+G_{6,4,3,2}(m-1)\\ 
G_{6,5,3,2}(m)=68  A+68  B+68  C+52  D+52  E+60  f+68  g+68  h+60  i+52  j+\\~~~~+120 (m-1) ( A+ B+ C+ D+ E+ f+ g+ h+ i+ j)+G_{6,5,3,2}(m-1)\\ 
G_{6,5,4,2}(m)=44  A+52  B+52  C+68  D+76  E+60  f+48  g+48  h+60  i+72  j+\\~~~~+120 (m-1) ( A+ B+ C+ D+ E+ f+ g+ h+ i+ j)+G_{6,5,4,2}(m-1)\\ 
G_{6,5,4,3}(m)=56  A+48  B+48  C+40  D+32  E+40  f+48  g+48  h+40  i+32  j+\\~~~~+120 (m-1) ( A+ B+ C+ D+ E+ f+ g+ h+ i+ j)+\\~~~~+G_{6,5,4,3}(m-1)+G_{6,5,4,3,2}(m-1)\\
G_{6,5,4,3,2}(m) = 120 (2 m-1) ( A+ B+ C+ D+ E+ f+ g+ h+ i+ j)+G_{6,5,4,3,2}(m-1)
\end{array}\right. \nonumber
\end{equation}
This is derived from \eqref{eq:system}, setting $n=6$ and plugging the integrand
\begin{equation}
G_6(1)=\raisebox{-0.8cm}{\includegraphics[width=2.cm]{3LA}} + \raisebox{-0.8cm}{\includegraphics[width=2.cm]{3LB}} + \raisebox{-0.8cm}{\includegraphics[width=2.cm]{3LC}} + \raisebox{-0.8cm}{\includegraphics[width=2.cm]{3LD}} +
\raisebox{-0.8cm}{\includegraphics[width=2.cm]{3LE}} 
\end{equation}
produced by the sum of the planar topologies.
As recalled in section \ref{sec:contours}, dealing with contours to reduce all integrals to the first winding also produces nonplanar topologies.
Of these we have already discarded those with two and three crossing since are individually vanishing.
Nonplanar topologies with one crossing $f-j$ are not necessarily 0 separately, as they contain a contractible gauge propagator which introduces a framing dependence.
Nonetheless the sum of them factorises
\begin{equation}
f+g+h+i+j = \raisebox{-0.8cm}{\includegraphics[width=2.cm]{1Lgauge}} \times \raisebox{-0.8cm}{\includegraphics[width=2.cm]{2Lcross}} = 0
\end{equation}
and so vanishes as the crossed diagram does so.
Using this result and the vanishing of the other nonplanar topologies one can symmetrize the integrand of the planar topologies under the exchange of any integration variable, then symmetrize the contour of integration with a $3!$ combinatorial factor and find \cite{Alvarez:1991sx}
\begin{equation}
A+B+C+D+E = \frac16 \raisebox{-0.8cm}{\includegraphics[width=2.cm]{1Lgauge}}^3 = -i\, \frac{\pi^3}{6}\, f^3\, \lambda_1^3
\end{equation}

The initial conditions read
\begin{equation}
\left\{\begin{array}{l}
G_6(1) = A+ B+ C+ D+ E\\
G_{6,2}(1) = 3  A+7  B+7  C+11  D+7  E+6  f+4  g+4  h+6  i+4  j\\ 
G_{6,3}(1)=4 (3  A+2  B+2  C+ D+3  E+ f+2  g+2  h+ i+2  j)\\
G_{6,4}(1) = 3  A+7  B+7  C+11  D+7  E+6  f+4  g+4  h+6  i+4  j\\ 
G_{6,5}(1)=2 (3  A+2  B+2  C+ D+ E+ f+ g+ h+ i+ j)\\ 
G_{6,3,2}(1)=2 (12  A+15  B+9  C+10  D+10  E+8  f+13  g+9  h+12  i+9  j)\\ 
G_{6,4,2}(1)=6 ( A+3  B+3  C+7  D+9  E+5  f+2  g+2  h+5  i+8  j)\\ 
G_{6,4,3}(1)=2 (12  A+9  B+15  C+10  D+10  E+12  f+9  g+13  h+8  i+9  j)\\ 
G_{6,5,2}(1)=2 (12  A+9  B+15  C+10  D+10  E+12  f+9  g+13  h+8  i+9  j)\\ 
G_{6,5,3}(1)=2 (12  A+15  B+9  C+10  D+10  E+8  f+13  g+9  h+12  i+9  j)\\ 
G_{6,5,4}(1)=2 (9  A+7  B+7  C+7  D+5  E+5  f+6  g+6  h+5  i+4  j)\\
G_{6,4,3,2}(1) = 44  A+52  B+52  C+68  D+76  E+60  f+48  g+48  h+60  i+72  j\\
G_{6,5,3,2}(1) = 68  A+68  B+68  C+52  D+52  E+60  f+68  g+68  h+60  i+52  j\\
G_{6,5,4,2}(1) = 44  A+52  B+52  C+68  D+76  E+60  f+48  g+48  h+60  i+72  j\\ 
G_{6,5,4,3}(1)=8 (7  A+6  B+6  C+5  D+4  E+5  f+6  g+6  h+5  i+4  j)\\ 
G_{6,5,4,3,2}(1)=120 ( A+ B+ C+ D+ E+ f+ g+ h+ i+j)
\end{array}\right.
\end{equation}
The solution to the system reads
\begin{equation}
\left\{\begin{array}{l}
G_6(m)= \frac{1}{3} m^2 \left( \left(m^2+2\right) m^2 (A+B+C+D+E) + (f + g + h + i + j ) (m^4-1)\right)\\
G_{6,2}(m)= m (m(9A+7  B+7D+5 E+5f+6g+6h+5i+4j)+\\~~~~+2  E+7  C m+ 4D -6 A+ f-2  g-2  h+ i)\\
G_{6,3}(m)= \frac{2}{3} m \left(m^2 (14A+12B+12C+10D+8E+10f+12g+12h+10i+8j)+\right.\\~~~~\left.+m(6  E-6A-3  g-3  h+3  j)+4  E-4  D+10  A-4  f+3  g+3  h-4  i+ j\right)\\
G_{6,4}(m)= \frac{1}{3} m \left(15 m^3 (A+BC+D+E+f+15g+15h+15i+15j)+\right.\\~~~~+ m^2(-12A-4B -4  C +4  D+12  E-8  g-8  h+8  j)+\\~~~~+ m(12A+6B+6  C+6  D+3  g+3  h-3  j)+\\~~~~\left.+4  C+8  D-6  E-6  A+ 4 B+3  f+2  g+2  h+3  i-8  j\right)\\ 
G_{6,5}(m)= \frac{2}{3} m \left(3 m^4 ( A +3B+3C +3  D+3  E+3  f+3  g+3  h+3  i+3  j)+\right.\\~~~~\left.+m^2( 6 A + 4B + 4C +2  D+2  g+2  h-2  j)-2  D- B- C-2  g-2  h+2  j\right)\\ 
G_{6,3,2}(m)= 2 m (m(14A+12B+12 C+10  D+8  E+10f+12g+10i+8j)+\\~~~~-3  C+2  E+3  B-2  A-2  f+ g+12 m  h-3  h+2  i+ j)\\ 
G_{6,4,2}(m)= 2 m (m(11A+13B+13C+17  D+19  E+15f+12g+12h+15i+18j)+\\~~~~-4  C+4  D+8  E-8 A-4 B-6  g-6  h+6  j)\\ 
G_{6,4,3}(m)= 2 m \left(10m^2(A+B+C+D+E+f+g+h+i+j)+\right.\\~~~~\left.+m(2  E-2A- g- h+ j)+5  C-2  E+4  A - B +2  f+4  h-2  i-2  j\right)\\ 
G_{6,5,2}(m)= 2 m (m(14A+12B+12C+10D+8  E+10f+12g+12h+10i+8j)+\\~~~~+3  C+2  E-3  B - 2  A +2  f-3  g+ h-2  i+ j)\\ 
G_{6,5,3}(m)= 2 m \left(10 m^2 (A+B+C+D+E+f+g+h+i+j)+\right.\\~~~~\left.+m(2E-2A-g-h+j)- C-2  E+5  B +4  A-2  f+4  g+2  i-2  j\right)\\ 
G_{6,5,4}(m)= 2 m^2 \left(5 m^2 (A + B + C + D+ E + f+ g+ h+ i+ j)+\right.\\~~~~\left.+2 (C+D+B)+ 4A+ g+ h- j\right)\\ 
G_{6,4,3,2}(m)= 4 m ((15 m+2) (D+E) + (15 m-2)  C + A (15 m-4)+\\~~~~+ B (15 m-2) + 15 m  f + (15 m-3) (g+h) + (15 m+3) (i+j))\\ 
G_{6,5,3,2}(m)= 4 m ((15 m+2) (A+B+C+g+h) (15 m-2) (D+E+j) +15 m ( f+i))\\ 
G_{6,5,4,2}(m)= 4 m ((15 m+2) D + (15 m-2) (C+B) + E (15m+4)+\\~~~~ + A (15 m-4)+ 15 m  f + (15 m-3) (g + h) + (15 m+3) (i+j))\\ 
G_{6,5,4,3}(m)= 8 m \left(5 m^2 (A+ B + C +D + E+ f + g+ h+ i+ j)+\right.\\~~~~\left.+ B+ C- E+ 2A+ g+ h- j\right)\\ 
G_{6,5,4,3,2}(m)= 120 m^2 ( A+ B+ C+ D+ E+ f+ g+ h+ i+ j)
\end{array}\right.
\end{equation}
where the first equation is \eqref{eq:3L6}.

\section{Weak coupling expansion of the matrix model}
\label{app:mmexp}

In this appendix we provide an explicit expansion of the matrix model average \eqref{eq:mmwl} for the ABJM 1/6-BPS Wilson loop winding $m$ times the great circle of $S^3$. To eight loops it reads
\begin{align}\label{eq:localization8loops}
\langle W_m \rangle &= 1+i  \pi m^2 \lambda _1 + 
\pi^2\left[\left(-\frac{1}{3} m^2-\frac{m^4}{3}\right) \lambda _1^2+m^2 \lambda _1 \lambda _2\right] + \nonumber\\&
+ i\pi^3\left[ \left(-\frac{1}{18} m^2-\frac{2}{9} m^4-\frac{1}{18} m^6\right) \lambda _1^3+\left(\frac{1}{3} m^2+\frac{2}{3} m^4\right) \lambda _1^2 \lambda _2-\frac{1}{2} m^2 \lambda _1 \lambda _2^2 \right] + \nonumber\\&
+ \pi^4\left[\left(\frac{13 m^4}{180}+\frac{m^6}{18}+\frac{m^8}{180}\right) \lambda _1^4+\left(-\frac{5}{12} m^4-\frac{m^6}{6}\right) \lambda _1^3 \lambda _2 \right. + \nonumber\\&
+\left.\left(-\frac{1}{6} m^2+\frac{2 m^4}{3}\right) \lambda _1^2 \lambda _2^2-\frac{1}{6} m^2 \lambda _1 \lambda _2^3 \right] + \nonumber\\&
+ i\pi^5\left[\left(-\frac{1}{675}   m^2  +\frac{1}{90}   m^4  +\frac{73   m^6  }{2700}+\frac{1}{135}   m^8  +\frac{  m^{10}  }{2700}\right) \lambda _1^5 \right.+ \nonumber\\&
+\left(\frac{1}{60}   m^2  -\frac{4}{45}   m^4  -\frac{7}{45}   m^6  -\frac{1}{45}   m^8  \right) \lambda _1^4 \lambda _2 + \nonumber\\&
+\left(-\frac{1}{8}   m^2  +\frac{1}{6}   m^4  +\frac{1}{4}   m^6  \right) \lambda _1^3 \lambda _2^2 
+\left.\left(\frac{5}{18}   m^2  -\frac{4}{9}   m^4  \right) \lambda _1^2 \lambda _2^3+\frac{1}{24}  m^2   \lambda _1 \lambda _2^4 \right] + \nonumber\\&
+ \pi^6 \left[ \left(\frac{m^4  }{4050}-\frac{17 m^6  }{2268}-\frac{13 m^8  }{2700}-\frac{m^{10}  }{1620}-\frac{m^{12}  }{56700}\right) \lambda _1^6 \right. + \nonumber\\&
+\left(-\frac{1}{216} m^4  +\frac{11 m^6  }{180}+\frac{m^8  }{36}+\frac{m^{10}  }{540}\right) \lambda _1^5 \lambda _2 + \nonumber\\&
+\left(\frac{m^2  }{120}+\frac{13 m^4  }{180}-\frac{29 m^6  }{180}-\frac{2 m^8  }{45}\right) \lambda _1^4 \lambda _2^2 + \nonumber\\&
+\left(\frac{13 m^2  }{72}-\frac{5 m^4  }{24}+\frac{m^6  }{4}\right) \lambda _1^3 \lambda _2^3 
+\left.\left(\frac{13 m^2  }{72}-\frac{2 m^4  }{9}\right) \lambda _1^2 \lambda _2^4+\frac{1}{120} m^2   \lambda _1 \lambda _2^5 \right] + \nonumber\\&
+ i \pi^7 \left[\left(-\frac{ m^2  }{13230}+\frac{ m^4  }{3150}-\frac{223  m^6  }{226800}-\frac{188  m^8  }{99225}-\frac{19  m^{10}  }{37800}-\frac{ m^{12}  }{28350}-\frac{ m^{14}  }{1587600}\right) \lambda _1^7 \right. + \nonumber\\&
+\left(\frac{1}{840}  m^2  -\frac{83  m^4  }{18900}+\frac{1}{105}  m^6  +\frac{74  m^8  }{4725}+\frac{11  m^{10}  }{3780}+\frac{ m^{12}  }{9450}\right) \lambda _1^6 \lambda _2 + \nonumber\\&
+\left(-\frac{1}{144}  m^2  +\frac{19}{540}  m^4  -\frac{1}{45}  m^6  -\frac{2}{45}  m^8  -\frac{1}{216}  m^{10}  \right) \lambda _1^5 \lambda _2^2 + \nonumber\\&
+\left(\frac{4}{45}  m^2  -\frac{7}{540}  m^4  +\frac{17}{540}  m^6  +\frac{8}{135}  m^8  \right) \lambda _1^4 \lambda _2^3 + \nonumber\\&
+\left(-\frac{7}{24}  m^2  +\frac{13}{36}  m^4  -\frac{3}{16}  m^6  \right) \lambda _1^3 \lambda _2^4
+\left.\left(-\frac{29}{360}  m^2  +\frac{4}{45}  m^4  \right) \lambda _1^2 \lambda _2^5-\frac{1}{720} m^2   \lambda _1 \lambda _2^6 \right] + \nonumber\\&
+ \pi^8 \left[\left(\frac{23 m^4}{178605}-\frac{17 m^6  }{34020}+\frac{1229 m^8  }{272160}+\frac{3643 m^{10}  }{1428840}+\frac{47 m^{12}  }{136080}+\frac{m^{14}  }{68040}+\frac{m^{16}  }{5715360}\right) \frac{\lambda _1^8}{10} \right. + \nonumber\\&
+\left(-\frac{m^4  }{3600}+\frac{2 m^6  }{2025}-\frac{4241 m^8  }{907200}-\frac{23 m^{10}  }{10800}-\frac{13 m^{12}  }{64800}-\frac{m^{14}  }{226800}\right) \lambda _1^7 \lambda _2 + \nonumber\\&
+\left(-\frac{m^2  }{5040}+\frac{13 m^4  }{9450}-\frac{23 m^6  }{1890}+\frac{641 m^8  }{37800}+\frac{47 m^{10}  }{7560}+\frac{m^{12}  }{3150}\right) \lambda _1^6 \lambda _2^2 + \nonumber\\&
+\left(-\frac{61 m^2  }{2160}-\frac{179 m^4  }{3240}+\frac{m^6  }{36}-\frac{37 m^8  }{1080}-\frac{5 m^{10}  }{648}\right) \lambda _1^5 \lambda _2^3 + \nonumber\\&
+\left(-\frac{37}{144} m^2  +\frac{1159 m^4  }{4320}-\frac{53 m^6  }{432}+\frac{8 m^8  }{135}\right) \lambda _1^4 \lambda _2^4 + \nonumber\\&
+\left(-\frac{37}{144} m^2  +\frac{29 m^4  }{96}-\frac{9 m^6  }{80}\right) \lambda _1^3 \lambda _2^5
+ \left.\left(-\frac{61 m^2  }{2160}+\frac{4 m^4  }{135}\right) \lambda _1^2 \lambda _2^6-\frac{m^2   \lambda _1 \lambda _2^7}{5040}\right] + \dots
\end{align}

\bibliographystyle{JHEP-2}

\bibliography{biblio}

\end{document}